\documentclass[english,11pt,aps,prd,a4paper,preprintnumbers,floatfix,nofootinbib,showpacs,superscriptaddress, notitlepage]{revtex4-1} 
\usepackage[mode=buildnew]{standalone}
\usepackage[usenames,dvipsnames]{color} 
\usepackage{graphicx}
\usepackage{bm}
\usepackage{dcolumn}
\usepackage[colorlinks=true,citecolor=darkred,urlcolor=darkred, pdfborder={0 0 0}]{hyperref}
\usepackage{setspace}
\usepackage{caption}
\usepackage{subcaption}
\usepackage{slashed}
\usepackage[normalem]{ulem}
\usepackage{float}
\usepackage[usenames,dvipsnames]{xcolor}
\usepackage{microtype}
\usepackage{lipsum}
\usepackage{amsmath}
\usepackage{amssymb}
\usepackage{mathrsfs}
\usepackage{placeins}
\usepackage{lettrine}
\usepackage{tikz}
\usetikzlibrary{shapes.misc}
\input Eileen.fd

\definecolor{darkred}{rgb}{0.6,0,0}

\definecolor{linkcolor}{rgb}{0,0,0.5}
\usepackage[T1]{fontenc} 
\usepackage[compat=1.1.0]{tikz-feynhand}
\usepackage{tikz-feynman}
\tikzfeynmanset{compat=1.1.0}
\usepackage{feynmp}
\usepackage{tikzsymbols}
\usepackage{array}
\usepackage{pifont}

\definecolor{linkcolor}{rgb}{0,0,0.5}
\usepackage{orcidlink}
\usepackage{multirow}
\begin{document}
\title{\boldmath \color{Blue}
The Simplest Dirac Scoto-Seesaw Realization
}

\author{Sin Kyu Kang}
\email{skkang@seoultech.ac.kr}
\affiliation{Seoul National University of Science and Technology, Seoul 01811, Republic of Korea}
\author{Ranjeet Kumar}
\email{kumarranjeet.drk@gmail.com}
\affiliation{Institute for Convergence of Basic Studies, Seoul National University of Science and Technology, Seoul 01811, Republic of Korea}
\author{Hemant Kumar Prajapati}
\email{hemant19@iiserb.ac.in}
\affiliation{Department of Physics, Indian Institute of Science Education and Research - Bhopal \\ Bhopal Bypass Road, Bhauri, Bhopal 462066, India}
\begin{abstract}
  \vspace{1cm} 

  \noindent
We present a simple Dirac scoto-seesaw framework based on the anomaly-free $U(1)_{B-L}$ charge assignment $(-4,-4,5)$ for $\nu_R$. This chiral charge assignment naturally accounts for the observed neutrino mass-squared differences, with $\Delta m^2_{\rm atm}$ generated at tree level and $\Delta m^2_{\rm sol}$ arising radiatively.
After the spontaneous breaking of gauged $U(1)_{B-L}$, a residual $Z_6$ symmetry stabilizes the dark matter candidate. We investigate two minimal realizations of the framework, finding that both normal and inverted orderings are viable in one case, whereas only normal ordering survives in the other, with distinctive features for neutrino observables.
Moreover, the chiral nature of the $U(1)_{B-L}$ charges suppresses the dilepton branching fraction of $Z'$, resulting in weaker ATLAS mass bounds than in the conventional vector $B-L$ scenario, thereby easing constraints on the dark sector.
We explore the dark matter phenomenology of the singlet scalar and fermionic dark matter candidates. While singlet scalar DM is often severely constrained, the presence of the $Z'$ portal together with annihilation and co-annihilation channels substantially broadens the allowed parameter space.
Thus, the framework offers a predictive scenario for neutrino and dark matter phenomenology that can be probed in future experiments.

\end{abstract}
\maketitle
%--------------------------------------------------------------------------------------------------------------------------------------------------------------------------------------------------------
%

\section{Introduction}
\label{sec:intro}
%%%%%%%
Although the Standard Model (SM) is a remarkably successful theory, it fails to accommodate the explanation of neutrino masses and the nature of dark matter (DM). The discovery of neutrino oscillations has firmly established that neutrinos are massive particles~\cite{Kamiokande-II:1990wrs,Kamiokande-II:1992hns,Super-Kamiokande:1998kpq,Cleveland:1998nv,SNO:2002tuh}. However, the mechanism responsible for neutrino mass generation, as well as the fundamental nature of neutrinos, whether Majorana or Dirac,  remains elusive.
In parallel, the existence of DM has been inferred from a wide range of astrophysical observations and further supported by cosmological measurements~\cite{Zwicky:1933gu,Rubin:1970zza,Rubin:1980zd,Planck:2018vyg}. 
Since the SM lacks a viable DM candidate and cannot explain neutrino masses, these shortcomings hint toward physics beyond the Standard Model (BSM).
 Thus, extensions of the SM that can simultaneously explain neutrino masses and DM within a unified framework are particularly appealing.
In this context, the scotogenic model~\cite{Ma:2006km} provides an elegant framework linking neutrino mass generation to DM phenomenology. Motivated by this idea, numerous scotogenic extensions have been proposed \cite{Guo:2020qin,Dasgupta:2021ggp,Borah:2022enh,Borah:2022phw,ChuliaCentelles:2022ogm,Bharadwaj:2024crt,CentellesChulia:2024iom,Nomura:2024zca,Singh:2025jtn,Kumar:2025aek,Avila:2025qsc,Kang:2026osw}. 
%%%%%%

Nevertheless, these models often fail to provide the explanation of the existence of two different mass scales, namely, the atmospheric mass-squared difference $\Delta m^2_{\rm atm}$, and the solar mass-squared difference $\Delta m^2_{\rm sol}$, observed in neutrino oscillation experiments ~\cite{Kamiokande-II:1990wrs,Kamiokande-II:1992hns,Super-Kamiokande:1998kpq,Cleveland:1998nv,SNO:2002tuh}. 
Interestingly, an alternative approach was recently proposed in Ref. \cite{Rojas:2018wym}, where the observed neutrino mass-squared differences can be naturally accommodated while preserving the key features of the scotogenic model~\cite{Ma:2008cu}. This is achieved through a hybrid mass generation mechanism, known as the `scoto-seesaw' mechanism, in which the atmospheric mass scale arises from a tree-level seesaw contribution, while the solar scale is generated radiatively \cite{Barreiros:2020gxu,Mandal:2021yph,Barreiros:2022aqu, Ganguly:2022qxj,Ganguly:2023jml,VanDong:2023xbd,Leite:2023gzl,Kumar:2023moh,Luong:2025pjj,Nasri:2026nbf}.
Furthermore, this hybrid mass generation framework can be obtained from the breaking of the $A_4$ flavor symmetry~\cite{Kumar:2024zfb,Borah:2024gql,Kumar:2025cte,Kumar:2025zvv}.
Consequently, it not only intertwines DM phenomenology with neutrino mass generation but also naturally explains the observed neutrino flavor structure and the emergence of the two distinct neutrino mass scales.

Motivated by this, in this work, we propose a simple and minimal framework realizing the Dirac scoto-seesaw mechanism. 
An important feature of our framework is the anomaly-free chiral $(-4,-4,5)$ charge assignment for the right handed neutrinos $\nu_{R_i}$ under the gauged $U(1)_{B-L}$ symmetry. This charge assignment leads to distinct roles for the three $\nu_R$ states.
The two $\nu_R$ carrying charge $-4$ participate in a Dirac type-I seesaw mediated by $N\equiv(N_L,N_R)$, whereas the $\nu_R$ with charge $5$ contributes only through a scotogenic loop involving $f\equiv(f_L,f_R)$ and the scalars $(\eta,S_1,S_2)$. Consequently, the two neutrino mass scales originate from distinct mass generation mechanisms dictated by the chiral $U(1)_{B-L}$ charges.
 In addition, the $U(1)_{B-L}$ symmetry is spontaneously broken by the vacuum expectation value (VEV) of the SM singlet scalar $\chi$. This breaking leaves a residual $Z_6$ symmetry that not only guarantees the Dirac nature of neutrinos but also acts as the dark symmetry and stabilizes the DM candidate. Only the particles $(f,\eta,S_1,S_2)$ running in the loop are odd under the residual $Z_6$ symmetry, whereas all other particles are even.

Within this model, $\Delta m^2_{\rm sol}$ is generated radiatively through the scotogenic mechanism, while $\Delta m^2_{\rm atm}$ arises from the tree level Dirac seesaw. We consider two scenarios: case-I with one generation of the tree level mediator $N$, and case-II with two generations, while a single generation of the loop mediator $f$ is present in both cases. Case-I accommodates two massive neutrinos and is compatible with both normal ordering (NO) and inverted ordering (IO), whereas case-II yields three massive neutrinos and allows only NO. The model leads to predictive ranges for $m_{\beta}$ and $\sum m_i$, with similar lower bounds in the NO scenario for both cases but a broader upper range in case-II. We also calculate the minimum of the $\chi^2$ function and provide the corresponding best-fit (BF) values of the model parameters and neutrino observables.
%%%

Furthermore, the gauged $U(1)_{B-L}$ symmetry predicts the existence of a neutral gauge boson $Z'$, which can be probed at high energy collider experiments such as ATLAS and CMS~\cite{CMS:2018mgb,ATLAS:2019erb}. Using recent ATLAS dilepton search results, we show that the corresponding mass limits in this framework are comparatively weaker than those in the conventional vector $B-L$ scenario. This difference arises from the chiral charge assignment $(-4,-4,5)$ of $\nu_{R_i}$, which enhances the invisible decay width of the $Z'$ boson and consequently suppresses its branching fraction into dileptons.
%%%
In the dark sector, the lightest neutral particle in the $Z_6$ odd sector, among $(f, \zeta_p)$, serves as a viable DM candidate.\footnote{$\zeta_p$ ($p=1,2,3$) denotes the mass eigenstates arising from the mixing of the $Z_6$ odd scalar fields $(\eta, S_1, S_2)$.} 
%%%
Notably, the scalar candidate $\zeta_p$, contains both $SU(2)_{L}$ doublet and singlet components.
%%%
While the doublet-dominated scenario is known to satisfy DM constraints via efficient co-annihilation channels \cite{Deshpande:1977rw,Barbieri:2006dq,Ma:2006km,Batra:2022pej,CentellesChulia:2022vpz,Kumar:2023moh,CentellesChulia:2024iom,Kumar:2024zfb,Kumar:2025cte,Kumar:2025zvv}, the pure singlet scenario is typically much more constrained. In the latter case, direct detection bounds often restrict the viable parameter space to a narrow region around the Higgs resonance \cite{Yaguna:2008hd,Profumo:2010kp,Feng:2014vea,DiMauro:2023tho,Yu:2024xsy}. In contrast, the present framework introduces additional scalar and $Z'$-mediated channels, significantly altering this picture; in particular, we focus on their combined impact in opening up the $SU(2)_{L}$ singlet-dominated parameter space.
%%%

The remainder of the paper is structured as follows. In Sec.~\ref{sec:model}, we present the model framework and discuss the generation of neutrino masses via the hybrid Dirac scoto-seesaw mechanism. We perform a detailed numerical study of the neutrino sector and its phenomenological consequences in Sec.~\ref{sec:nusec}. The constraints from $Z'$ decay and collider have been discussed in Sec.~\ref{sec:Zprime}. Furthermore, we present the dark sector phenomenology in Sec.~\ref{sec:dm}. Finally, we provide our concluding remarks in Sec.~\ref{sec:conc}.

\section{The Minimal Dirac Scoto-Seesaw Model}\label{sec:model}

\begin{figure}[!h]
    \centering
    \tikzset{
  cross/.style={
    draw,
    cross out,
    minimum size=6pt,
    inner sep=0pt
  }
}
\tikzset{
DM/.style={draw=black,line width=0.8pt, postaction={decorate},
        decoration={}},
fermion/.style={draw=cyan, postaction={decorate},
        decoration={markings,mark=at position .55 with {\arrow[draw=black]{>}}}},
nucleon/.style={draw=black, thick},
    vertex/.style={draw=black, fill=black, circle, inner sep=0pt, minimum size=1mm},
nucleonN/.style={draw=cyan,line width=1pt, postaction={decorate},
        decoration={markings,
            mark=at position .3 with {\arrow[draw=black]{>}},
            mark=at position .7 with {\arrow[draw=black]{>}}
        }
    },
    fermionbar/.style={draw=cyan, postaction={decorate},
        decoration={markings,mark=at position .55 with {\arrow[draw=black]{<}}}},
    fermionnoarrow/.style={draw=black},
    gluon/.style={decorate, draw=red,
        decoration={coil, amplitude=4pt, segment length=5pt}},
        photon/.style={draw=red, decorate, decoration={snake}, draw},
            scalar6/.style={dashed,draw=black,line width=0.8pt, postaction={decorate},
        decoration={markings,mark=at position .55 with {\arrow[draw=black]{<}} }},
    scalar/.style={dashed,draw=black,line width=0.8pt, postaction={decorate},
        decoration={markings,mark=at position .2 with {\arrow[draw=black]{<}} }},
            scalar2/.style={dashed,draw=black,line width=0.01pt, postaction={decorate},
        decoration={markings,mark=at position .7 with {\arrow[draw=black]{<}} }},
                    scalar3/.style={dashed,draw=black,line width=0.01pt, postaction={decorate},
        decoration={markings,mark=at position .9 with {\arrow[draw=black]{<}} }},
       scalarN/.style={dashed,draw=black, postaction={decorate},
       }, 
    scalarbar/.style={dashed,draw=black, postaction={decorate},
        decoration={markings,mark=at position .55 with {\arrow[draw=black]{>}}}},
majorana/.style={draw=cyan, postaction={decorate},
        decoration={markings,mark=at position .5 with {\arrow[draw=black]{><}}}},
    majoranabar/.style={draw=orange, postaction={decorate},
        decoration={markings,mark=at position .55 with {\arrow[draw=black]{><}}}}, 
crossmark/.style={postaction={decorate},
    decoration={markings,
      mark=at position 0.55 with {\node[cross out,draw,minimum size=8pt,inner sep=0pt] {};}}},
cross/.style={draw=red,line width=1.5pt,,minimum size=7pt,inner sep=0pt,shape=cross out}         
}
\def\iimg{{\bf i}}

\hspace{-7cm}
\begin{tikzpicture}

%%%%%%%%%%%%%%%%%%%%%%%%%%%%%%%%%%%%
%%%%%%%%%%%%%%%%%%%%%%%%%%%%%%%%%%%%
%%%%%%%%%%%%%%%%%%%%%%%%%%%%%%%%%%%%

%Type-1 Seesaw

     \draw[scalarbar, line width=1.5pt] (0,0)--(0,-2);
     \draw[fermion, line width=1.5pt] (0,-2)--(1.5,-2);
     \draw[fermionbar, line width=1.5pt] (0,-2)--(-1.5,-2);
%%%      
      \draw[fermion, line width=1.5pt] (1.5,-2)--(3.0,-2);
       \draw[fermion, line width=1.5pt] (3.0,-2)--(4.5,-2);
%%%        
%%%        
      \draw[scalar6, line width=1.5pt] (3,0)--(3,-2);
%     \draw[scalar, line width=1.5pt] (3,0)--(2,1);
%     \draw[scalar, line width=1.5pt] (3,0)--(4,1);
%%%        
%%%      
%%%      
% \draw[fermion,crossmark] (0,0) -- (2,0); 
%%%      
     \node[cross] at (1.5,-2) {};
   \node at (0.0,0.4){$\langle H \rangle$};
   \node at (3.0,0.4){$\langle \chi \rangle$};
    
      \node at (-0.8,-2.4){$L_{i}$};
     \node at (3.8,-2.4){$\nu_{R_{j}}$}; 
                    \node at (0.8,-2.4){$N_{R}$};
   \node at (2.4,-2.4){$N_{L}$};

%\end{tikzpicture}

% Scotogenic
\hspace{8cm}
%%%%%%%%%%
%\begin{tikzpicture}

      \draw[fermion, line width=1.5pt] (0,-2)--(2,-2);
      \draw[fermionbar, line width=1.5pt] (0,-2)--(-2,-2);
       \draw[fermion, line width=1.5pt] (2,-2)--(4.0,-2);
        \draw[fermion, line width=1.5pt] (4.0,-2)--(6,-2);
%%%        
        \node[cross] at (2,-2) {};
%%%        
         \draw[scalar, line width=1.5pt] (0,-2) arc (180:0:2.0);
           \draw[scalar2, line width=1.5pt] (0,-2) arc (180:0:2.0);
         \draw[scalar3, line width=1.5pt] (0,-2) arc (180:0:2.0);
%%%   
   \draw[scalarbar, line width=1.5pt] (-0.1,1.0)--(2,0);
     \draw[scalarbar, line width=1.5pt] (4.0,1.0)--(2,0.0);
          \draw[scalarbar, line width=1.5pt] (5.5,-0.2)--(3.7,-1.0);
%%%     
%%%     
%%%     
            \node at (-0.8,-2.4){$L_{i}$};
      \node at (5.0,-2.4){$\nu_{R_3}$}; 
%%%      
%%%      
                  \node at (1.0,-2.4){$f_{R}$};
      \node at (3.0,-2.4){$f_{L}$}; 
%%%    
   \node at (0.0,-0.9){$\eta$}; 
    \node at (4.3,-1.3){$S_1$};
    \node at (3.5,-0.2){$S_2$};
      \node at (-0.6,1.3){$ \langle H \rangle $};    
       \node at (4.3,1.3){$ \langle \chi \rangle  $};   
        \node at (5.8,0.1){$ \langle \chi \rangle  $};
          
\end{tikzpicture}
    \caption{Neutrino mass generation from Dirac scoto-seesaw mechanism, where $i=1,2,3$ and $j=1,2$.}
    \label{fig:feyn_numass}
\end{figure}
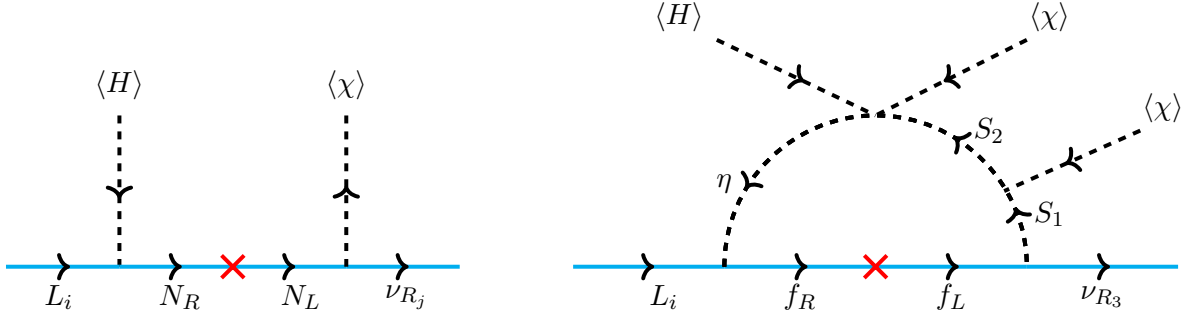

We first outline the particle content and the underlying symmetry structure governing the model framework.
In particular, the model is constructed within the simplest realization of the Dirac scoto-seesaw mechanism for neutrino mass generation. An appealing feature of the model is its natural explanation of the two observed neutrino mass scales. To generate the two different scales, we have utilized the chiral anomaly-free charge assignment ($-4,-4,5$) of right handed neutrinos $\nu_{R_i}$ under the gauged $U(1)_{B-L}$ symmetry. 
The atmospheric mass scale $\Delta m^2_{\rm atm}$ is generated at tree level through the participation of the two $\nu_{R_{1,2}}$ carrying charge $-4$. In contrast, the comparatively smaller solar mass scale $\Delta m^2_{\rm sol}$ arises radiatively at one loop from the $\nu_{R_3}$ carrying charge $5$. 
The schematic realization of this neutrino mass generation mechanism is illustrated in Fig.~\ref{fig:feyn_numass}.
In addition to the right handed neutrinos $\nu_{R_i}$, we introduce BSM Dirac fermion pairs $N \equiv (N_L, N_R)$ and $f \equiv (f_L, f_R)$, which are singlets under the SM gauge group. In the scalar sector, besides the SM Higgs doublet $H$, the model contains an additional $SU(2)_L$ scalar doublet $\eta$ as well as scalar singlets, $S_1$, $S_2$, and $\chi$. 
Under $U(1)_{B-L}$, the fermions $(L_i,e_{R_i},N)$ carry charge $-1$, while $f$ carries charge $-1/2$. 
The BSM scalars $\eta,S_1,S_2$ and $\chi$ have been assigned charges $1/2,-11/2,-5/2$ and $3$, respectively, while SM Higgs $H$ remains neutral under $U(1)_{B-L}$. The particle content along with their charge assignments under the various symmetries of the model is summarized in Table~\ref{tab:field_content}. 
\begin{table}[h]
\centering
\renewcommand{\arraystretch}{1.3}
\begin{tabular}{|c|c||c|c| c|}
\hline
  \quad  \quad \quad & ~~Fields~~ 
 & ~~$SU(2)_L \otimes U(1)_Y$~
 & $U(1)_{B-L} \to $
 $\bold{Z_6}$  \\ 
\hline\hline

\multirow{5}{*}{\rotatebox{90}{Fermions }} 
 & $L_i$   
 & $(2,-1)$ 
 & $-1$ 
 $\to$ $\boldsymbol{\omega^4}$  \\

  & $e_{R_i}$   
 & $(1,-2)$ 
 & $-1$ 
 $\to$ $\boldsymbol{\omega^4}$  \\

 & $\nu_{R_{i}}$ 
 & $(1,0)$ 
 & $(-4,-4,5)$ 
$\to$ $\boldsymbol{(\omega^4,\omega^4,\omega^4)}$ \\

 & $N_{L_p}$, $N_{R_p}$  
 & $(1,0)$ 
 & $-1$ 
$\to$ $\boldsymbol{\omega^4}$ \\

 %& $N_{R}$  & $(1,0)$  & $-1$ $\to$ $\boldsymbol{\omega^4}$ \\

& $f_{L}$, $f_{R}$ 
 & $(1,0)$ 
 & $-1/2$ 
 $\to$ $\boldsymbol{\omega^5}$ \\

 %& $f_{R}$  & $(1,0)$ & $-1/2$  $\to$ $\boldsymbol{\omega^5}$ \\
 
\hline\hline

\multirow{4}{*}{\rotatebox{90}{Scalars}} 
 & $H$ 
 & $(2,1)$ 
 & $0$ 
$\to$ $\boldsymbol{1}$ \\

 & $\chi$ 
 & $(1,0)$ 
 & $3$ 
 $\to$ $\boldsymbol{1}$ \\

 & $\eta$ 
 & $(2,1)$ 
 & $1/2$ 
 $\to$ $\boldsymbol{\omega}$ \\

 & $S_1$, $S_2$  
 & $(1,0)$ 
 & $(-11/2,-5/2)$ 
$\to$ $(\boldsymbol{\omega},\boldsymbol{\omega})$ \\

%& $S_2$  & $(1,0)$ & $-5/2$ $\to$ $\boldsymbol{\omega}$ \\

\hline
\end{tabular}
\caption{Field contents and their transformation properties under 
$SU(2)_L \otimes U(1)_Y$ and $U(1)_{B-L}$. Here,  $Z_6$ is the residual symmetry, which serves as dark symmetry, and $i=1,2,3$. For case-I (two massive neutrinos) $p=1$ and for case-II (three massive neutrinos) $p=2$. Only one generation of $f_L,f_R$ has been introduced.}
\label{tab:field_content}
\end{table}
%%%%

The electroweak symmetry is spontaneously broken by the VEV of $H$, while the VEV of $\chi$ induces the breaking of the $U(1)_{B-L}$ symmetry. The other scalar fields do not develop VEVs, allowing the lightest among them to provide a viable DM candidate. The breaking of $U(1)_{B-L}$  symmetry leads to a residual $Z_6$ symmetry. The residual $Z_6$ symmetry here serves a dual purpose: it provides the stability to the DM candidate and forbids the Majorana mass terms, ensuring the Dirac nature of neutrinos. 
Under the $Z_6$, only the particles running in the loop ($\eta, S_1,S_2, f$) carry odd charges, while all remaining particles are even.\footnote{Since the $B-L$ charged scalars do not couple to quarks, the quark sector remains unaffected by the symmetry breaking and therefore transforms trivially under the residual $Z_6$ symmetry.} As a result, the lightest $Z_6$ odd particle cannot decay into lighter states, naturally providing a stable DM candidate. 
Moreover, the specific charge assignment forbids Majorana mass terms such as $\bar{L}_i^c L_j$ and $\bar{\nu}_{R_i}^c \nu_{R_j}$ ($i,j=1,2,3$), since they are not invariant under $Z_6$. On the other hand, the effective Dirac operators $\bar{L}_i \tilde{H} \nu_{R_k} \chi$ and $\bar{L}_i \tilde{H} \nu_{R_{3}} \chi^{*} \chi^{*}$ ($k=1,2$) are consistent with the $Z_6$ symmetry and hence allowed, leading naturally to Dirac neutrino masses. 

Furthermore, with the adopted $U(1)_{B-L}$ charge assignments, the fermion $N$ contributes exclusively at tree level, whereas $f$ appears only in loop level neutrino mass generation, as shown in Fig.~\ref{fig:feyn_numass}. 
In our analysis, we explore two different realizations of the framework. The first scenario (case-I) consists of a single generation of $N$, representing the minimal tree level setup, whereas the second scenario (case-II) includes two generations of $N$. In both cases, however, the loop sector is kept minimal by considering only a single generation of $f$.
The minimal structure of case-I results in one massless neutrino, which is consistent with the current neutrino oscillation data \cite{Kamiokande-II:1990wrs,Kamiokande-II:1992hns,Super-Kamiokande:1998kpq,Cleveland:1998nv,SNO:2002tuh}. Meanwhile, case-II gives rise to masses for all three neutrinos. In the dark sector, our framework accommodates both scalar and fermionic DM candidates. The scalar sector includes $S_1$, $S_2$, and the neutral part of $\eta$, while the fermionic sector contains $f$. The lightest among these states can naturally serve as a viable DM candidate.
%%%%%%%%%%%%%%%%%%%%

%%%%%%%%%%%%%%%%%%%%%%%%%%%%%%%%%%%%%%%%
\subsection{Yukawa Sector and Neutrino Mass Generation}
%%%%%

Having discussed the particle content and their transformation properties, we now examine the Yukawa sector responsible for the mass generation of the neutrinos.
Following the charge assignment given in Table \ref{tab:field_content}, the relevant Yukawa Lagrangian of the model can be formulated as,
%%%%
\begin{equation}
\begin{split}
    -\mathcal{L}_{Y} &=  Y^{ij}_{e}\overline{L}_i H e_{R_j} + Y^{ip}_{\nu} \overline{L}_i \tilde{H}N_{R_p} + M^{qp} \,\overline{N}_{L_q} N_{R_p} + Y^{q \alpha}_{\chi}  \overline{N}_{L_q} \chi\nu_{R_{\alpha}} \\
    & + Y^{i}_{f} \overline{L}_i \tilde{\eta}f_{R} + M_{f}\, \overline{f}_{L}f_{R} + y_{S} \, \overline{f}_{L}S_1 \nu_{R_{3}}  +\text{H.c.} ,
\end{split}    
\end{equation}
%%%%%%%%
where, $i,j=1,2,3$, $\alpha=1,2$, and $p,q=1$ ($p,q=1,2$) for case-I (case-II). We adopt a basis where the charged lepton mass matrix is diagonal, so that the observed leptonic mixing is generated exclusively through the neutrino sector. 
 The neutrino masses and mixing are generated from the interplay between a tree level seesaw mechanism and a loop induced scotogenic contribution. After spontaneous symmetry breaking (SSB), scalars $H$, $\chi$ acquire VEV, $\langle H \rangle = v_H/\sqrt{2}$,  $\langle \chi \rangle = v_{\chi}/\sqrt{2}$ and neutrino masses are generated.

At the tree level, the mass matrix relevant for neutrino mass generation, written in the basis $(\overline{\nu}_{L_i},\overline{N}_{L_q} )$ and $({\nu_{R_{\alpha}}},{N_{R_p}} )^T$, is given by
\begin{align}
    \mathcal{M}_{\nu} = \begin{pmatrix}
         0 & \mathcal{M_{D}} \\
         \mathcal{M'_{D}} & \mathcal{M}
     \end{pmatrix},
\end{align}
where, $\mathcal{M_{D}} \equiv Y^{ip}_{\nu}v_H/\sqrt{2} $, $\mathcal{M'_{D}} \equiv Y^{q \alpha}_{\chi}v_{\chi}/\sqrt{2} $, and $\mathcal{M} \equiv M^{qp}$. 
Now, in the seesaw limit, $M^{qp} \gg Y^{ip}_{\nu}v_H, Y^{q\alpha}_{\chi}v_{\chi} $, the light Dirac neutrino mass can be written as follows 
\begin{align}
    -(m_{\nu}^{i \alpha})^{(\text{tree})} = \mathcal{M_{D}} \mathcal{M}^{-1} \mathcal{M'_{D}} = \frac{v_H v_{\chi}}{2} Y^{ip}_{\nu} (M^{qp})^{-1} Y^{q\alpha}_{\chi} \ .
\end{align}
For one generation of fermion $(N_L, N_R)$ (case-I), $M^{qp}$ representing fermion mass is a number and for two generations (case-II), it is $2 \times 2 $ matrix. The structure of Yukawas, $Y^{ip}_{\nu}$, $Y^{q\alpha}_{\chi}$ for case-I (case-II) are $3 \times 1$ ($3 \times 2$) and  $1 \times 2$ ($2 \times 2$), respectively. In case-I, only one neutrino acquires mass at tree level, whereas in case-II, two neutrinos become massive at tree level.
%%%%%%%%%
The remaining neutrino mass is then generated radiatively through the scoto-loop mechanism, whose contribution is given by
%%%
\begin{equation} \label{eq:loopnumass}
    (m_{\nu}^{i})^{(\text{loop})} = \frac{1}{16 \pi^{2}} Y^{i}_{f}y_{S} \sum_k \mathcal{O}^{T}[1,k]\mathcal{O}^{T}[2,k]M_{f} \left(  \frac{M_{\zeta_{k}}^{2}}{M_{\zeta_{k}}^{2}-M_{f}^2} \right) \ln{\left( \frac{M^2_{\zeta_k}}{M^2_f} \right)} \equiv  Y^{i}_{f}y_{S} \mathcal{F} \ ,
\end{equation}
where $y_S$, $M_f$ is a single entry number, $Y^{i}_f$ is a $3\times 1$ matrix\footnote{Note that only the non-zero Yukawa entries relevant for neutrino mass generation are written. Consequently, the tree-level and loop-level Yukawa structures appear as $3\times 2$ and $3\times 1$, respectively. In the full $(\nu_{R_1},\nu_{R_2},\nu_{R_3})^T$ basis, however, both are $3\times 3$ matrices: the third column vanishes at tree level since $\nu_{R_3}$ does not contribute, while the first and second columns vanish at loop level since $\nu_{R_{1,2}}$ do not contribute.}, $M_{\zeta_k}$ are masses of scalars in their mass eigenstates and $\mathcal{O}$ is their rotation matrix for $k=1,2,3$ (see App.~\ref{sec:scalar}). Further, we defined the loop contribution in terms of $\mathcal{F}$ given as
\begin{align}
    \mathcal{F}= \frac{1}{16 \pi^{2}}  \sum_k \mathcal{O}^{T}[1,k]\mathcal{O}^{T}[2,k]M_{f} \left(  \frac{M_{\zeta_{k}}^{2}}{M_{\zeta_{k}}^{2}-M_{f}^2} \right) \ln{\left( \frac{M^2_{\zeta_k}}{M^2_f} \right)}.
\end{align}
%%%%%
Consequently, the two observed neutrino mass-squared differences originate from distinct contributions. In particular, $\Delta m^2_{\rm atm}$ is generated at tree level seesaw, while $\Delta m^2_{\rm sol}$ arises radiatively from the loop level scotogenic.
After incorporating both seesaw and scotogenic contributions, the final neutrino mass matrix is obtained as follows,
%%%%%%%%%%%%
\begin{equation} \label{eq:nutot}
    m_{\nu}^{(\text{TOT})} = m_{\nu}^{(\text{tree})} + m_{\nu}^{(\text{loop})}\,.
\end{equation}
Next, in the subsequent section, we discuss our numerical findings of the neutrino sector for both cases.

%%%%%%%%%%%%%%%%%%%%%%%%
\section{Numerical analysis of Neutrino Sector} \label{sec:nusec}
We now delve into the details of the numerical analysis of the neutrino sector. In our analysis, we employ the NuFIT 6.1 neutrino oscillation parameters at the $3\sigma$ level, given as follows ~\cite{Esteban:2024eli,Esteban:2026phq,Esteban:2026xyz}
%%%%%%%%%%%%%%%%
\begin{align}
\textbf{NO}: \quad  \Delta m^2_{31}&=[2.450, 2.576]\times 10^{-3}\ {\rm eV}^2,\
\Delta m^2_{21}=[7.236, 7.823]\times 10^{-5}\ {\rm eV}^2,\nonumber\\
\sin^2{\theta_{12}}&=[0.2893, 0.3295],\ 
\sin^2{\theta_{23}}=[0.435, 0.584],\ 
\sin^2{\theta_{13}}=[0.02064, 0.02418] .
\label{eq:mixNO}
\end{align} 
\begin{align}
\textbf{IO}: \quad  &\Delta m^2_{32}=-[2.421, 2.547]\times 10^{-3}\ {\rm eV}^2,\
\Delta m^2_{21}=[7.236, 7.823]\times 10^{-5}\ {\rm eV}^2,\nonumber\\
&\sin^2{\theta_{12}}=[0.2893, 0.3295],\ 
\sin^2{\theta_{23}}=[0.439, 0.584],\ 
\sin^2{\theta_{13}}=[0.02093, 0.02441] .
\label{eq:mixIO}
\end{align} 
%%%%%%%%
Here, $\Delta m^2_{21}$ corresponds to the solar mass-squared difference $\Delta m^2_{\rm sol}$, whereas $\Delta m^2_{31}$ ($\Delta m^2_{32}$) is associated with the atmospheric mass-squared difference $\Delta m^2_{\rm atm}$, for NO (IO).
To discuss the phenomenological implications of the model, we numerically diagonalize the combined neutrino mass matrix (seesaw $+$ scoto) obtained in Eq.~\eqref{eq:nutot}. This enables us to extract the neutrino masses and mixing parameters, which are subsequently used to analyze the numerical predictions of the model for the two distinct scenarios: case-I and case-II. 
Our numerical analysis reveals that both NO and IO are compatible with the neutrino oscillation data in case-I, while in case-II the IO scenario is ruled out, leaving only NO as the viable possibility. In what follows, we now proceed to present the detailed numerical results and phenomenological features of both case-I and case-II.
\subsection{Case-I: Two Massive Neutrinos}

We begin by analyzing the case-I, where only two neutrinos are massive and one remains massless. The numerical analysis is performed by varying the relevant input parameters within the ranges listed in Table \ref{tab:scan1}. For our numerical analysis, we consider the tree level couplings to be real, whereas couplings contributing at the loop level are taken to be complex. Assuming the scalar potential parameters to be real, the predicted Dirac CP-violating phase $\delta_{\rm CP}$ originates entirely through the loop level couplings.  
\begin{table}[!h]
\centering
\renewcommand{\arraystretch}{1.3}
\begin{tabular}{|c|c||c|c|c| c|}
\hline
  ~Yukawas~
 & ~~Ranges~~
 & ~Masses (in GeV)~ &
 ~~Ranges~~  \\ 
\hline\hline
  $Y^{i 1}_{\nu}$   
 & ~$ (10^{-6}- 10^{-2}) $~ 
 &  $M^{11}$  & $(10^{9}- 10^{10})$ \\
  $Y^{1 \alpha}_{\chi}$ 
 & $(10^{-6}- 10^{-2}) $ 
 &  $M_{f}$
 & $(10^{6}- 10^{9})$  \\ 
  $Y^{i}_{f}$, $y_S$ 
 & $|y| e^{i\phi}$
 &  $M_{\zeta_{i}}$ & $(10^{1}- 10^{5})$\\ 
\hline
\end{tabular}
\caption{Ranges of the input parameters used for the numerical scan for case-I, where $i=1,2,3$, $\alpha =1,2$ and $|y| = (10^{-6}-10^{-4})$, $\phi \in [0,2\pi)$. The singlet VEV $v_{\chi} \approx (10^{2}-10^{7}) $ GeV.}
\label{tab:scan1}
\end{table}
In the case-I scenario, the lightest neutrino mass remains massless, i.e., $m_1=0$ ($m_3=0$) for NO (IO). Combined with the two experimentally measured mass-squared differences, this condition enables a complete determination of the neutrino masses.
For NO, $m_3>m_2>m_1$ and $m_1 = 0$, while for IO, $m_2>m_1>m_3$ and $m_3 = 0$. Now, utilizing the best-fit (BF) values from NuFIT  data~\cite{Esteban:2026phq}, we can extract the explicit masses as follows 
\begin{align} \label{eq:numass}
   \textbf{NO:} \quad  &m_3=5.011 \times 10^{-2}\ {\rm eV}, \quad  m_2= 8.682 \times 10^{-3}\ {\rm eV} \,. \nonumber
\\
   \textbf{IO:} \quad  &m_1=4.907 \times 10^{-2}\ {\rm eV}, \quad  m_2= 4.983 \times 10^{-2}\ {\rm eV} \,.
\end{align}
This feature has important implications for the sum of neutrino masses ($\sum m_i$) and the effective mass of beta decay ($|m_{\beta}|$), as we discuss next.

The numerical analysis is carried out by performing a random scan over the input parameter space specified in Table~\ref{tab:scan1}. Further, to analyze the observables in the neutrino sector, we calculate the $\chi^2$, making use of the NuFIT data provided in Eqs.~\eqref{eq:mixNO} and \eqref{eq:mixIO}. 
The $\chi^2$ function is defined as:
\begin{align}
    \chi^2 = \sum_i \frac{(x_i - \bar{x}_i)^2}{\sigma_i^2}, 
\end{align}
where $x_i = (\Delta m^2_{31}\ \text{or} \ \Delta m^2_{32},  \Delta m^2_{21},\sin^2\theta_{12}, \sin^2\theta_{23}, \sin^2\theta_{13})$, and  $\bar{x}_i$ and $\sigma_i$ are the corresponding BF values and $1 \sigma$ uncertainties obtained from NuFIT  data~\cite{Esteban:2026phq}.
Since, for neutrino masses and mixing, only the loop factor $\mathcal{F}$ defined in Eq.~\eqref{eq:loopnumass} plays a role, not the individual masses $M_f$ and $M_{\zeta_i}$, we provide the BF value of $\mathcal{F}$.
The BF values of the model parameters and neutrino observables are obtained at $\chi^2_{\rm min}=1.475$ and $\chi^2_{\rm min}=9.033$ for the NO and IO cases, respectively.
The corresponding values are summarized in Table \ref{tab:chi2min_NO2m} and Table \ref{tab:chi2min_IO2m}, respectively.
\begin{table}[!h]
\centering
\renewcommand{\arraystretch}{1.3}
\begin{tabular}{|c|c||c|c||c|c|}
\hline
  ~Parameters~
 & BF ($\times 10^{-4}$)
 & ~Parameters~ &
 ~~BF~~ &  ~Observables~ & ~~BF~~ \\ 
\hline\hline
%\multirow{5}{*}{\rotatebox{90}{Fermions }} 
  $Y^{11}_{\nu}$   
 & ~$ 1.07 $~ 
 &  $Y^{1}_{f} (\times 10^{-6})$ & $-9.29 -  9.69 i$ & \colorbox{green!20}{$\boldsymbol{\theta_{12}}$} & \colorbox{green!20}{$\boldsymbol{33.96}^{\circ}$} \\
 \hline
$Y^{21}_{\nu}$ 
 & $2.68$ 
 & $Y^{2}_{f} (\times 10^{-5})$ & $5.20 +  1.93 i$ & \colorbox{green!20}{$\boldsymbol{\theta_{13}}$} & \colorbox{green!20}{~$\boldsymbol{8.68}^{\circ}$~} \\ \hline
  $Y^{31}_{\nu}$ 
 & $0.016$
 & $Y^{3}_{f} (\times 10^{-5})$ & $-3.05 -  5.19 i$
  & \colorbox{green!20}{$\boldsymbol{\theta_{23}}$} &\colorbox{green!20}{$\boldsymbol{43.70}^{\circ}$} \\ \hline
 $Y^{11}_{\chi}$ 
 & $6.98$ 
  & $Y_S (\times 10^{-4})$ 
 & $2.58 -7.5 i$ & \colorbox{green!20}{$\boldsymbol{\delta_{\rm CP}}$} & \colorbox{green!20}{\quad $ \boldsymbol{0.88 \pi}$ \quad } \\ \hline
 $Y^{12}_{\chi}$  & $0.014$
 & $M^{11}$ (in GeV)
  & $1.98 \times 10^{9}$ & \colorbox{green!20}{$\boldsymbol{\Delta m^2_{\rm sol}}$ (in $\text{eV}^2$)}  & \colorbox{green!20}{$\boldsymbol{7.47 \times 10^{-5}} $} \\ \hline
$v_{\chi}$ (in GeV)
   & $8.49 \times 10^{5}$   & $\mathcal{F}$ (in GeV) 
 & $7.53 \times 10^{-4}$ & \colorbox{green!20}{$\boldsymbol{\Delta m^2_{\rm atm}}$ (in $\text{eV}^2$)} & \colorbox{green!20}{$\boldsymbol{2.50  \times 10^{-3}}$} \\ \hline
 $-$ 
 & $-$ 
 & $-$ 
   & $-$ & \colorbox{green!20}{ $\boldsymbol{\sum m_{i}}$ (in $\text{meV}$) }& \colorbox{green!20}{\quad  ~$\boldsymbol{58.65}$ \quad} \\ \hline
 $-$ 
 & $-$ 
 & $-$
   & $-$ & \colorbox{green!20}{ $\boldsymbol{| m_{\beta}|}$ (in $\text{meV}$) }& \colorbox{green!20}{\quad  ~$\boldsymbol{8.93}$ \quad} \\ \hline
\hline
\end{tabular}
\caption{The obtained BF values of input parameters and observables for the NO scenario corresponding to $\chi^2_{\rm min}=1.475$ for case-I.}
\label{tab:chi2min_NO2m}
\end{table}
%%%%%%%%%%%%%%%%%
\begin{table}[!h]
\centering
\renewcommand{\arraystretch}{1.3}
\begin{tabular}{|c|c||c|c||c|c|}
\hline
  ~Parameters~
 & BF ($\times 10^{-4}$)
 & ~Parameters~ &
 ~~BF~~ &  ~Observables~ & ~~BF~~ \\ 
\hline\hline
%\multirow{5}{*}{\rotatebox{90}{Fermions }} 
  $Y^{11}_{\nu}$   
 & ~$ 5.18 $~ 
 &  $Y^{1}_{f} (\times 10^{-6})$ & $3.61 + 5.62 i$ & \colorbox{green!20}{$\boldsymbol{\theta_{12}}$} & \colorbox{green!20}{$\boldsymbol{33.83}^{\circ}$} \\
 \hline
  $Y^{21}_{\nu}$ 
 & $0.55$
 & $Y^{2}_{f} (\times 10^{-5})$ & $2.60 -  3.01 i$ & \colorbox{green!20}{$\boldsymbol{\theta_{13}}$} & \colorbox{green!20}{~$\boldsymbol{8.64}^{\circ}$~} \\ \hline
 $Y^{31}_{\nu}$ 
 & $0.79$ 
 & $Y^{3}_{f} (\times 10^{-5})$ & $-4.13 -  1.11 i$
  & \colorbox{green!20}{$\boldsymbol{\theta_{23}}$} &\colorbox{green!20}{$\boldsymbol{47.41}^{\circ}$} \\ \hline
 $Y^{11}_{\chi}$ 
 & $15.56$  
  & $Y_S (\times 10^{-5})$ 
 & $-0.49 +1.23 i$ & \colorbox{green!20}{$\boldsymbol{\delta_{\rm CP}}$} & \colorbox{green!20}{\quad $ \boldsymbol{1.5 \pi}$ \quad } \\ \hline
 $Y^{12}_{\chi}$  & $5.65$ 
 & $M^{11}$ (in GeV)
  & $1.90 \times 10^{9}$ & \colorbox{green!20}{$\boldsymbol{\Delta m^2_{\rm sol}}$ (in $\text{eV}^2$)}  & \colorbox{green!20}{$\boldsymbol{7.48 \times 10^{-5}} $} \\ \hline
 $v_{\chi}$ (in GeV)
   & $8.73 \times 10^{5}$  & $\mathcal{F}$ (in GeV) 
 & $6.38 \times 10^{-2}$ & \colorbox{green!20}{$\boldsymbol{\Delta m^2_{\rm atm}}$ (in $\text{eV}^2$)} & \colorbox{green!20}{$\boldsymbol{2.42  \times 10^{-3}}$} \\ \hline
 $-$ 
 & $-$ 
 & $-$ 
   & $-$ & \colorbox{green!20}{ $\boldsymbol{\sum m_{i}}$ (in $\text{meV}$) }& \colorbox{green!20}{\quad  ~$\boldsymbol{99.25}$ \quad} \\ \hline
 $-$ 
 & $-$ 
 & $-$
   & $-$ & \colorbox{green!20}{ $\boldsymbol{| m_{\beta}|}$ (in $\text{meV}$) }& \colorbox{green!20}{\quad  ~$\boldsymbol{48.92}$ \quad} \\ \hline
\hline
\end{tabular}
\caption{Best fit values of input parameters and observables for the IO scenario corresponding to $\chi^2_{\rm min}=9.033$ for case-I.}
\label{tab:chi2min_IO2m}
\end{table}

\begin{figure}[!h]
    \centering
    \includegraphics[width=0.49\linewidth]{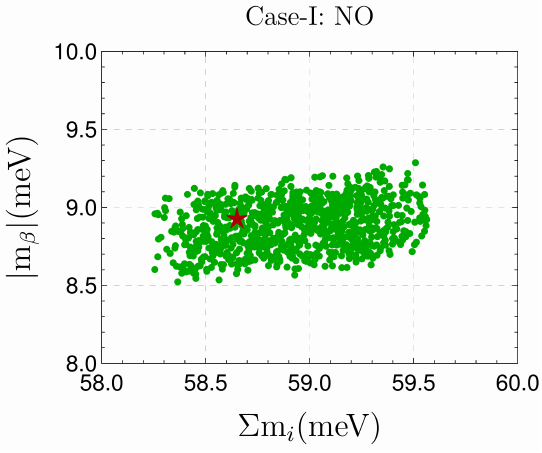}
   \includegraphics[width=0.49\linewidth]{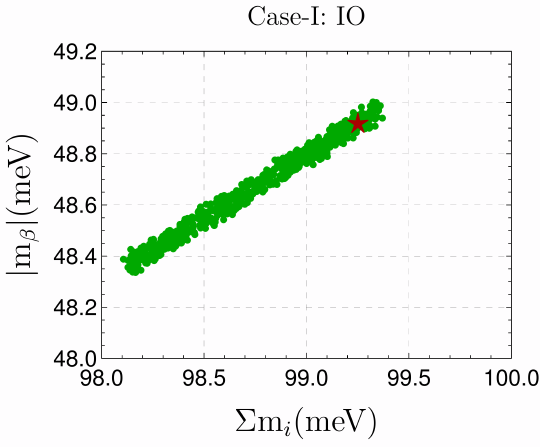}
    \caption{Correlation between $\sum m_i$ and $|m_{\beta}|$ predicted by the model for case-I. The allowed parameter space is shown by the green scattered points, while the red star indicates the BF values. The left (right) panel corresponds to the NO (IO) case.}
    \label{fig:2m_sumbeta}
\end{figure}

We now present the resulting predictions that lead to interesting correlations between various observables.
The correlations between the sum of neutrino masses ($\sum m_i$) and the effective mass of beta decay ($|m_{\beta}|$) are presented in Fig.~\ref{fig:2m_sumbeta}. The left and right panels correspond to the NO and IO cases, respectively. 
The effective mass of beta decay ($|m_{\beta}|$) is defined as follows
\begin{align} \label{eq:me}
 | m_{\beta}| \equiv \sqrt {\sum_{i}  |U_{ei} |^2 m^2_i}  = \sqrt{c^2_{12}c^2_{13}m^2_1 + s^2_{12}c^2_{13} m^2_2 +s^2_{13}m^2_3  }\;.
\end{align}
where $c_{ij}=\cos \theta_{ij}$ and $s_{ij}=\sin \theta_{ij}$. 
The model predictions are shown in green scattered points and the red star represents the corresponding BF values obtained in our analysis, as provided in Tables \ref{tab:chi2min_NO2m} and  \ref{tab:chi2min_IO2m}. 
From Fig.~\ref{fig:2m_sumbeta}, it is evident that the NO case (left panel) predicts comparatively smaller values of both $\sum m_i$ and $|m_{\beta}|$ than the IO case (right panel). This behavior emerges as a direct consequence of the presence of one massless neutrino. As can be understood from Eq.~\eqref{eq:numass}, in the NO scenario only $m_3$ has a large value, while $m_2$ remains relatively small. In contrast, for the IO scenario, both $m_1$ and $m_2$ have large values. Consequently, the predicted ranges of $\sum m_i$ and $|m_{\beta}|$ are significantly lower for NO compared to IO. The sum of neutrino masses $\sum m_i$ remains consistent with the Planck data \cite{Planck:2018vyg}. 
Moreover, the effective mass of beta decay $|m_{\beta}|$ is currently being probed by the forefront direct neutrino mass experiments KATRIN \cite{KATRIN:2024cdt} and Project-8 \cite{Project8:2022hun,Project8:2025aar}. Notably, KATRIN has recently released updated results establishing an upper limit on $|m_{\beta}|$,
\begin{align}
| m_{\beta}| &< 450~\text{meV},\quad 90\%\ \text{CL}~(\text{2024}).
\end{align}
Compared to KATRIN, the present sensitivity achieved by Project-8 is relatively weaker, and therefore the corresponding upper limit on $|m_{\beta}|$ is less restrictive \cite{Project8:2022hun}, given by
\begin{align}
| m_{\beta}| < 155~\text{eV} \quad (\text{2022}).
\end{align}
Nevertheless, the planned large scale upgrades of Project-8 are anticipated to significantly improve its sensitivity, potentially reaching the level of $|m_{\beta}| < 40~\text{meV}$ \cite{Project8:2022wqh}, comparable to the reach of KATRIN. Therefore, both the NO and IO predictions of the model satisfy the present experimental constraints. However, with the improved future sensitivities, the IO scenario is expected to become testable, while the NO scenario will likely remain consistent with the experimental limits. 

\begin{figure}[!h]
    \centering
     \includegraphics[width=0.49\linewidth]{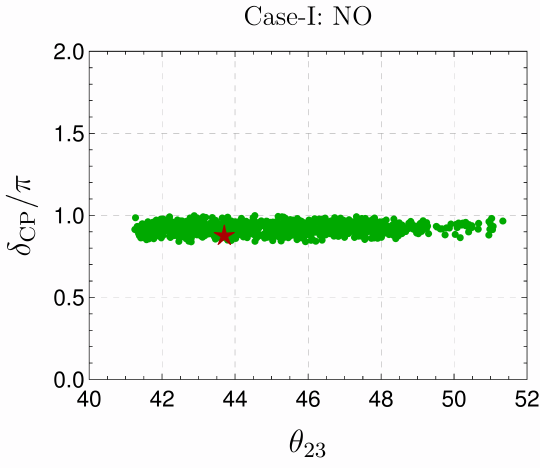}
    \includegraphics[width=0.49\linewidth]{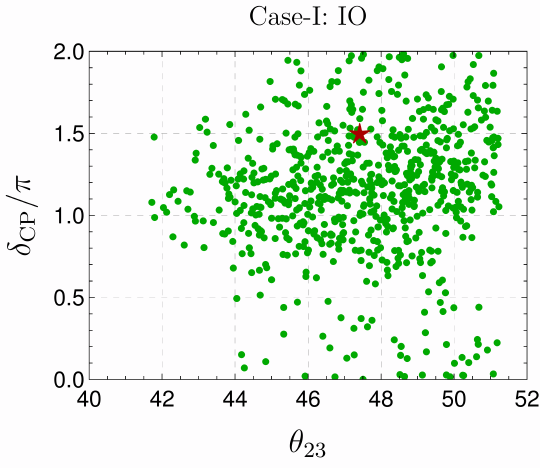}
    \caption{Predicted correlation between $\delta_{\rm CP}$ and $\theta_{23}$ for case-I. The color coding remains the same as in Fig. \ref{fig:2m_sumbeta}. The left (right) panel corresponds to the NO (IO) case.}
    \label{fig:2m_delcp}
\end{figure}

The possible correlation between the Dirac CP-violating phase $\delta_{\rm CP}$ and atmospheric mixing angle $\theta_{23}$ are shown in Fig. \ref{fig:2m_delcp}. 
The NO and IO scenarios are shown in the left and right panels, respectively, with the same color coding as adopted in Fig.~\ref{fig:2m_sumbeta}. 
\begin{figure}[!h]
    \centering
    \includegraphics[width=0.6\linewidth]{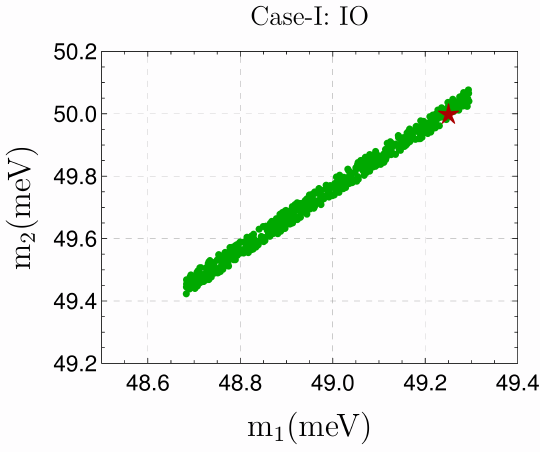}
    \caption{Correlation between neutrino masses obtained for case-I corresponding to the IO case. The color coding remains the same as in Fig. \ref{fig:2m_sumbeta}.}
    \label{fig:2m_IO}
\end{figure}
For the NO case, the allowed parameter space is found to be clustered around the CP-conserving region, with the BF value occurring at $\delta_{\rm CP}=0.88\pi$. In contrast, for the IO case, the allowed parameter space spans almost the entire range of $\delta_{\rm CP}$, while the BF point corresponds to maximal CP-violation with $\delta_{\rm CP}=1.5\pi$. Moreover, the BF value of $\theta_{23}$ lies in the lower octant for NO, whereas it resides in the upper octant in the IO scenario.
We also show the correlation between the neutrino masses $m_2$ and $m_1$ in Fig.~\ref{fig:2m_IO}. The color scheme remains identical to that of Fig.~\ref{fig:2m_sumbeta}. It can be seen that the BF values lie towards the higher mass values. 

\FloatBarrier
%%%%%%%%%%%%%%%%%%%

\subsection{Case-II: Three Massive Neutrinos}

Having discussed the case-I scenario with two massive neutrinos, we now proceed to the detailed discussion of case-II, where all three neutrinos are massive.
For this case, the numerical analysis has been done by performing a random scan over the input parameter space given in Table~\ref{tab:scan2}. 
\begin{table}[!h]
\centering
\renewcommand{\arraystretch}{1.3}
\begin{tabular}{|c|c||c|c|c| c|}
\hline
  ~Yukawas~
 & ~~Ranges~~
 & ~Masses (in GeV)~ &
 ~~Ranges~~  \\ 
\hline\hline
  $Y^{i \alpha}_{\nu}$   
 & ~$ (10^{-7}- 10^{-3}) $~ 
 &  $M^{\alpha \beta}$  & $(10^{10}- 10^{12})$ \\

  $Y^{\alpha \beta}_{\chi}$ 
 & $(10^{-7}- 10^{-3}) $ 
 &  $M_{f}$
 & $(10^{5}- 10^{6})$  \\ 

  $Y^{i}_{f}$, $y_S$ 
 & $|y| e^{i\phi}$
 
 &  $M_{\zeta_{i}}$ & $(10^{1}- 10^{5})$\\ 

\hline
\end{tabular}
\caption{Ranges of the input parameters used for the numerical scan for case-II, where $i=1,2,3$, $\alpha, \beta =1,2$ and $|y| = (10^{-6}-10^{-4})$, $\phi \in [0,2\pi)$. The singlet VEV $v_{\chi} \approx (10^{2}-10^{7}) $ GeV.}
\label{tab:scan2}
\end{table}
%%%%%%%%%%%%%%%%%%%%
Similar to the case-I, here we also consider the tree level couplings to be real and couplings contributing at the loop level are complex. 
In our analysis, we find that the IO scenario is disfavored for case-II. This conclusion follows from the correlation between the mass-squared differences $\Delta m^2_{\rm sol}$ and $\Delta m^2_{\rm atm}$ obtained in our random scan approach.
\begin{figure}[!h]
    \centering
    \includegraphics[width=0.6\linewidth]{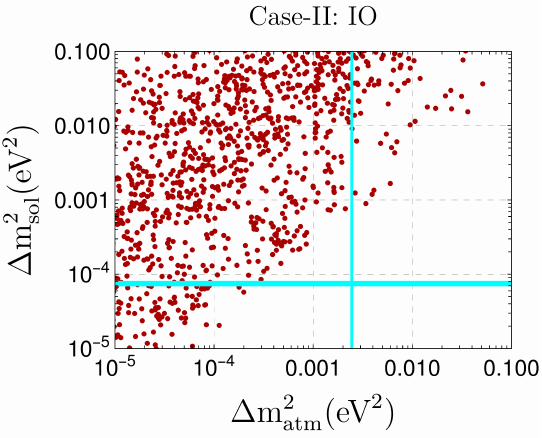}
    \caption{Correlation between the mass-squared differences for case-II in the IO scenario. The $3\sigma$ allowed ranges of both mass-squared differences cannot be simultaneously satisfied in this scenario.}
    \label{fig:m21vsm31IO}
\end{figure}
In this scenario, the model fails to simultaneously satisfy the constraints of both mass-squared differences, as illustrated in Fig.~\ref{fig:m21vsm31IO}. The scattered red points represent the parameter space allowed by the model. The horizontal and vertical bands shown in light cyan correspond to the $3\sigma$ allowed regions of $\Delta m^2_{\rm sol}$ and $\Delta m^2_{\rm atm}$, respectively. As it is evident from Fig.~\ref{fig:m21vsm31IO}, no overlap region simultaneously satisfies both mass-squared difference constraints. Thus, the IO scenario remains forbidden in case-II.

We therefore focus on the NO scenario in the following and examine its numerical predictions in detail. To analyze the model predictions for the NO scenario, we calculate the $\chi^2$, making use of the NuFIT data provided in Eq.~\eqref{eq:mixNO}. The BF values of the model are obtained at $\chi^2_{\rm min}=0.536$. The obtained BF values of Yukawa couplings $Y^{22}_{\nu}$, $Y^{22}_{\chi}$ are close to $Y^{11}_{\nu}$ and $Y^{11}_{\chi}$, respectively. The corresponding BF values of the model parameters and neutrino observables are listed in Table \ref{tab:chi2min_NO3m}. 
\begin{table}[!h]
\centering
\renewcommand{\arraystretch}{1.3}
\begin{tabular}{|c|c||c|c||c|c|}
\hline
  ~Parameters~
 & BF ($\times 10^{-5}$)
 & ~Parameters~ &
 ~~BF~~ &  ~Observables~ & ~~BF~~ \\ 
\hline\hline
%\multirow{5}{*}{\rotatebox{90}{Fermions }} 
  $Y^{11}_{\nu}, Y^{22}_{\nu}$   
 & ~$ 3.70 $~ 
 &  $Y^{1}_{f} (\times 10^{-6})$  & $-0.095-2.18 i$ & \colorbox{green!20}{$\boldsymbol{\theta_{12}}$} & \colorbox{green!20}{$\boldsymbol{33.91}^{\circ}$} \\
 \hline

  $Y^{12}_{\nu}$ 
 & $2.20$ 
 & $Y^{2}_{f} (\times 10^{-6})$ & $-9.03 +  3.07 i$ & \colorbox{green!20}{$\boldsymbol{\theta_{13}}$} & \colorbox{green!20}{~$\boldsymbol{8.62}^{\circ}$~} \\ \hline

 $Y^{21}_{\nu}$ 
 & $0.025$ 
 & $Y^{3}_{f} (\times 10^{-6})$ 
 & $9.99 -3.25 i$  & \colorbox{green!20}{$\boldsymbol{\theta_{23}}$} &\colorbox{green!20}{$\boldsymbol{42.93}^{\circ}$} \\ \hline

  $Y^{31}_{\nu}$ 
 & $0.17$ 
 & $Y_S (\times 10^{-6})$
 & $-22.14 -1.04 i$ & \colorbox{green!20}{$\boldsymbol{\delta_{\rm CP}}$} & \colorbox{green!20}{\quad $\boldsymbol{1.38 \pi}$ \quad } \\ \hline

 $Y^{32}_{\nu}$ 
 & $7.9$ 
 & $M^{11}$ (in GeV)
  & $3.84 \times 10^{11}$ & \colorbox{green!20}{$\boldsymbol{\Delta m^2_{\rm sol}}$ (in $\text{eV}^2$)}  & \colorbox{green!20}{$\boldsymbol{7.58 \times 10^{-5}} $} \\ \hline

 $Y^{11}_{\chi},Y^{22}_{\chi}$
 & $41.00$ 
 & $M^{22}$ (in GeV)
  & $5.76 \times 10^{10}$ & \colorbox{green!20}{$\boldsymbol{\Delta m^2_{\rm atm}}$(in $\text{eV}^2$)} & \colorbox{green!20}{$\boldsymbol{2.51  \times 10^{-3}}$} \\ \hline
  $Y^{12}_{\chi}$
 & $0.67$  
 & $\mathcal{F}$ (in GeV) 
 & $-0.157 $ & \colorbox{green!20}{ $\boldsymbol{\sum_i m_{i}}$ (in $\text{meV}$) }& \colorbox{green!20}{\quad  ~$\boldsymbol{59.67}$ \quad} \\ \hline
  $Y^{21}_{\chi}$
 & $1.20$
 & $v_{\chi}$ (in GeV)
   & $2.06 \times 10^{5}$ & \colorbox{green!20}{ $\boldsymbol{| m_{\beta}|}$ (in $\text{meV}$) }& \colorbox{green!20}{\quad  ~$\boldsymbol{8.95}$ \quad} \\ \hline
\hline
\end{tabular}
\caption{Best fit values of input parameters and observables for the NO scenario corresponding to $\chi^2_{\rm min}=0.536$ for case-II.}
\label{tab:chi2min_NO3m}
\end{table}
%

\begin{figure}[!h]
    \centering
    \includegraphics[width=0.49\linewidth]{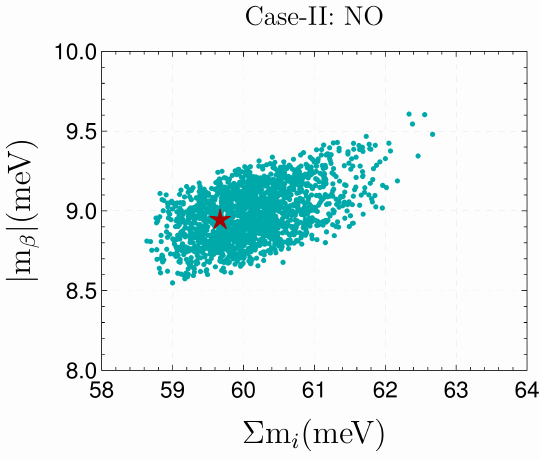}
    \includegraphics[width=0.49\linewidth]{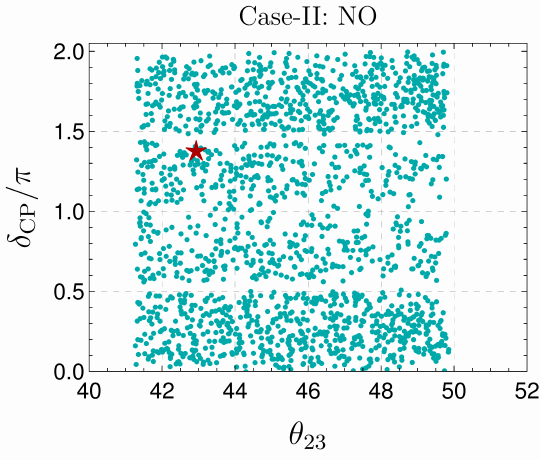}
    \caption{Model predictions for case-II in the NO scenario. The left panel corresponds to the correlation between $\sum m_i$ and $|m_{\beta}|$, and the right panel is for the $\delta_{\rm CP}$-$\theta_{23}$ plane. The allowed parameter space is shown by cyan scattered points and the red star represents the BF values.}
    \label{fig:3m_mbeta_NO}
\end{figure}

%%%%%%%%%%%%%%%%%%%%
Now, we present the key predictions obtained by random scan, emphasizing the correlations that emerge among various observables. 
In all the figures for NO in case-II, the model predictions are shown in scattered cyan points and the red star represents the BF values of the parameters following from Table \ref{tab:chi2min_NO3m}. 
Similar to case-I, we begin by presenting the correlation between $\sum m_i$ and $|m_{\beta}|$, as illustrated in the left panel of Fig.~\ref{fig:3m_mbeta_NO}. While the lower values of both observables remain close to those found in case-I, their upper limits are larger in case-II. This behavior can be attributed to the fact that all three neutrinos acquire nonzero masses in case-II, thereby increasing the allowed range of the total neutrino mass scale. These values of $\sum m_i$ and $|m_{\beta}|$ remain consistent with the current limits of Planck \cite{Planck:2018vyg} and KATRIN \cite{KATRIN:2024cdt}, respectively.
In the right panel of Fig.~\ref{fig:3m_mbeta_NO}, we present the model predictions in the $\delta_{\rm CP}-\theta_{23}$ plane. For this scenario, the allowed parameter space spans the entire range of $\delta_{\rm CP}$, while the BF point lies close to the maximally CP-violating value. This behavior clearly distinguishes case-II from case-I, where both the allowed region and the BF value of $\delta_{\rm CP}$ were concentrated near the CP-conserving regime.

\begin{figure}[!h]
    \centering
    \includegraphics[width=0.49\linewidth]{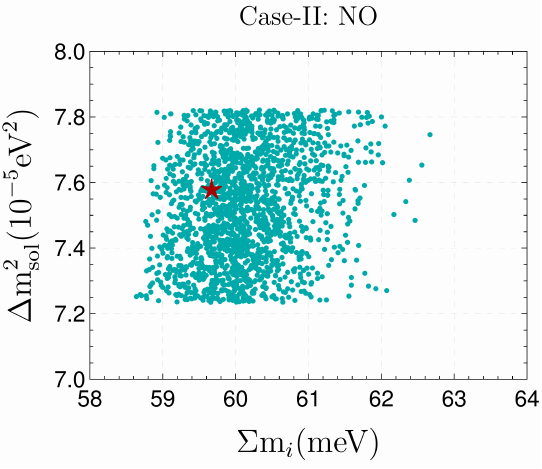}
    \includegraphics[width=0.49\linewidth]{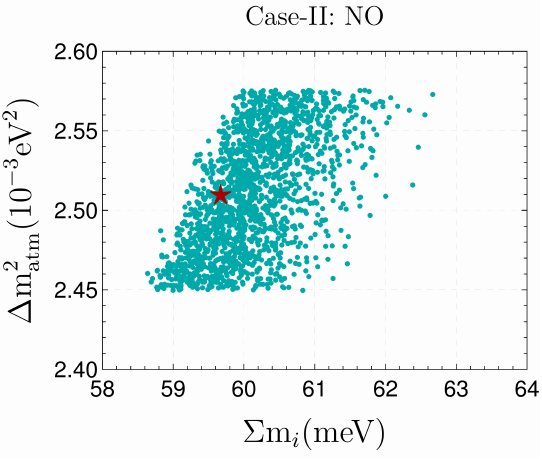}
    \caption{Correlations of mass-squared differences with $\sum m_i$ have been shown. The left and right panels correspond to the solar mass-squared difference $\Delta m^2_{\rm sol}$ and atmospheric mass-squared difference $\Delta m^2_{\rm atm}$, respectively. The color coding remains the same as in Fig. \ref{fig:3m_mbeta_NO}.}
    \label{fig:3m_sumvsm21_NO}
\end{figure}

\begin{figure}[!h]
    \centering
    \includegraphics[width=0.49\linewidth]{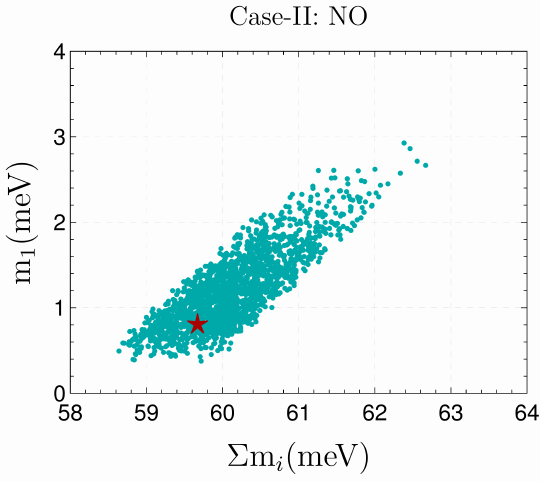}
    \includegraphics[width=0.49\linewidth]{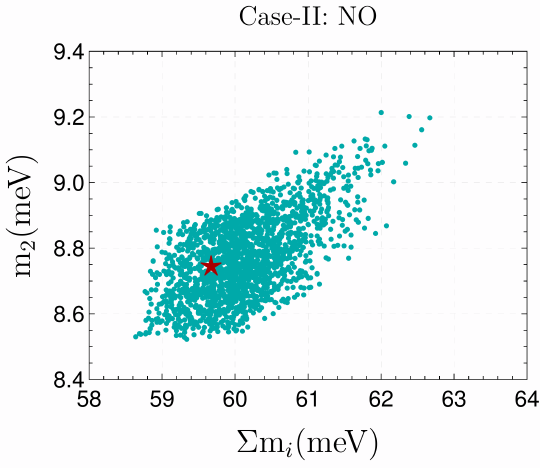}
    \caption{Correlation of neutrino masses $m_1$ and $m_2$ with $\sum m_i$ have been shown. The color coding remains the same as in Fig. \ref{fig:3m_mbeta_NO}.}
    \label{fig:3m_sumvsmi_NO}
\end{figure}

\begin{figure}[!h]
    \centering
    \includegraphics[width=0.49\linewidth]{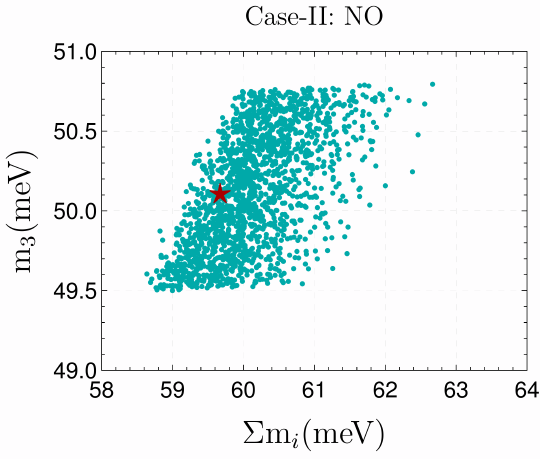}
    \includegraphics[width=0.49\linewidth]{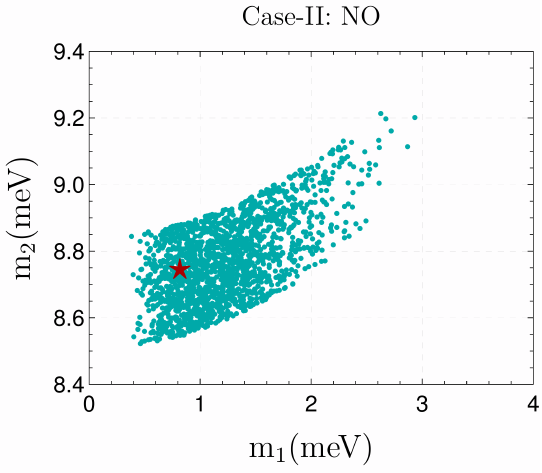}
    \caption{In the left panel correlation of neutrino mass $m_3$ with $\sum m_i$ has been shown. The right panel corresponds to the correlation between neutrino masses $m_2$ and $m_1$. The color coding remains the same as in Fig. \ref{fig:3m_mbeta_NO}.}
    \label{fig:3m_m1vsm2_NO}
\end{figure}

Next, we present the possible correlation of mass-squared differences with the sum of neutrino masses. This has been shown in the left (right) panel of Fig. \ref{fig:3m_sumvsm21_NO} corresponding to the $\Delta m^2_{\rm sol}$ ($\Delta m^2_{\rm atm}$) with $\sum m_i$. 
Similarly, the correlation of neutrino masses $m_1$ and $m_2$ with $\sum m_i$ has been shown in the left and right panels of Fig. \ref{fig:3m_sumvsmi_NO}. 
In addition, the correlations of neutrino mass $m_3$ with $\sum m_i$ and among masses $m_2$ and $m_1$ are shown in the left and right panels of Fig. \ref{fig:3m_m1vsm2_NO}. The obtained BF values of neutrino masses are: $m_1=0.82$ meV, $m_2=8.75$ meV, and $m_3=50.11$ meV. 

Thus, our numerical analysis demonstrates that the Dirac scoto-seesaw framework can successfully accommodate both the two massive neutrino (case-I) and three massive neutrino (case-II) scenarios. The two cases exhibit distinct phenomenological features and lead to different allowed parameter spaces for the relevant observables. In particular, the two scenarios can be distinguished through their predictions for neutrino mass observables, the Dirac CP phase, and the neutrino mass ordering. These features highlight the rich predictive nature of the framework and provide promising avenues for experimental scrutiny in future neutrino oscillation, beta decay, and cosmological observations. 
Moreover, our discussion is restricted to the scalar DM scenario, as the neutrino sector phenomenology is found to be nearly identical in the fermionic DM case. Consequently, presenting the latter separately would not provide any additional insight and is therefore omitted for brevity.
Having established the viability and phenomenological implications of the neutrino sector, we now turn to other important aspects of the framework, including the constraints from $Z'$ decays in Sec.~\ref{sec:Zprime} and the DM phenomenology in Sec.~\ref{sec:dm}.

\FloatBarrier
%%%%%%%%%%%%%%%%%%%%%
\section{Constraints from $Z'$ decay} \label{sec:Zprime}
%%%%%%%%%

%%%%%%%%%
In this section, we investigate collider constraints on the $Z^\prime$ boson, focusing on its production and detection prospects at the LHC ~\cite{CMS:2018mgb,ATLAS:2019erb}.
The production of the $Z'$ boson is possible if the SM quarks carry charges under the new $U(1)$ symmetry, or if they interact with the $Z'$ through mixing with the SM $Z$ boson.
For simplicity, we assume that the kinetic mixing between the SM hypercharge and $U(1)_{B-L}$ is negligible and set it to zero. In addition, since the Higgs boson is not charged under $U(1)_{B-L}$, there is no mass mixing between the $Z$ and $Z'$ bosons.
%%%
Nevertheless, as both quarks and leptons carry charges under $U(1)_{B-L}$, the $Z'$ couples directly to them. It can therefore be probed at the LHC through Drell-Yan production, followed by its decay into charged leptonic final states,
%%%
\begin{equation}
q \,\bar{q} \longrightarrow Z' \longrightarrow l^{+} l^{-},
\end{equation}
%%%
where $q$ is either a valence quark or a sea quark in the proton, and $l=e,\mu$.
%%%

\begin{figure}[!h]
    \centering
    \includegraphics[width=0.6\linewidth]{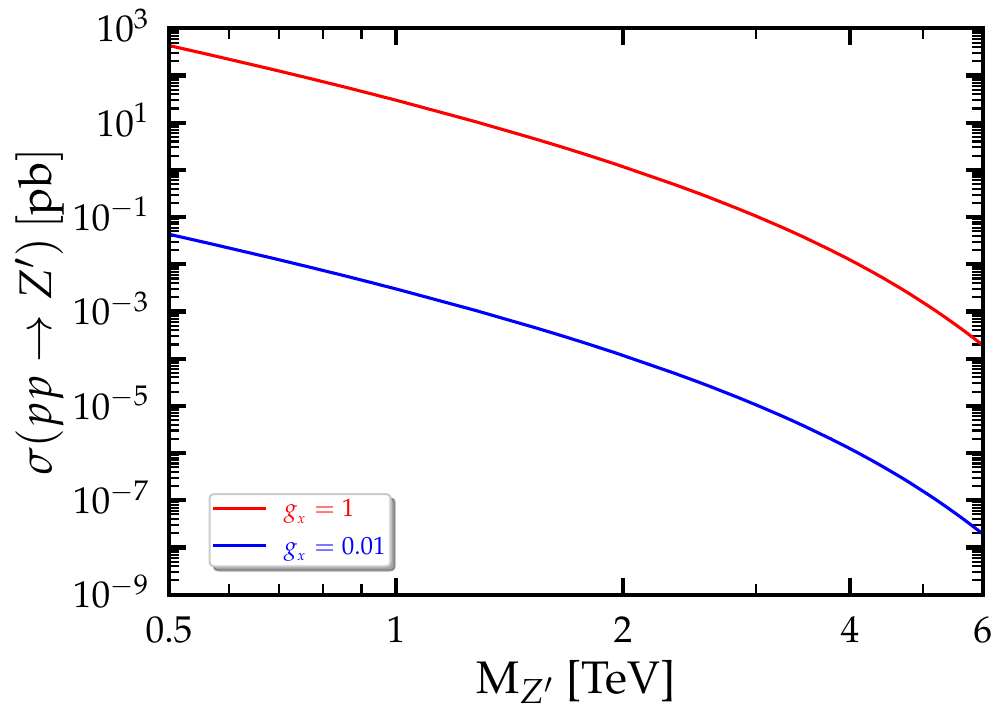}
    \caption{Production cross section of the $Z'$ boson in $pp$ collisions at $\sqrt{s}=13$ TeV as a function of the mass $M_{Z'}$. The red and blue lines correspond to benchmark gauge couplings of $g_x = 1$ and $g_x = 0.01$, respectively.}
    \label{fig:Sig_Zp}
\end{figure}

The production cross section of the $Z'$ boson, which quantifies the probability of its production in a collision, is primarily governed by the three parameters: $Z'$ mass $M_{Z'}$, its gauge coupling $g_x$, and the $U(1)_{B-L}$ charges of the quarks.
%%%%%
For the collider analysis, we employ the package SARAH \cite{Staub:2015kfa} to derive the interaction vertices and mass matrices. The numerical analysis is carried out using SPheno \cite{Porod:2003um,Goodsell:2017pdq}, while collider simulations are performed with MadGraph \cite{Alwall:2014hca}.
%%%%%
In Fig. \ref{fig:Sig_Zp}, we illustrate the $Z'$ production cross section in $pp$ collisions at $\sqrt{s}=13$ TeV for two benchmark values of gauge coupling $g_x$.
%%%%%
%%%%%
The red line corresponds to the case where $g_x = 1$. Because the production cross section $\sigma(pp \to Z')$ scales as $g_x^2$, the cross section for any other coupling value can be computed by simply rescaling the red curve. This scaling behavior is shown by the blue line, corresponding to $g_x = 0.01$.
%%%
Notably, the quark charges under the chiral $U(1)_{B-L}$ symmetry in this framework are identical to those in the vector $B-L$ model. 
Hence, the $Z^\prime$ production cross section is exactly the same as in the vector $B-L$ scenario, where the right handed neutrinos carry charges $(-1,-1,-1)$.

We now turn to the decay modes of the $Z'$ boson. The $Z'$ has both SM and BSM decay channels. In the SM sector, it undergoes two-body decays into all SM fermions, $Z' \to \bar{\psi}_{\text{SM}} \psi_{\text{SM}}$. In the BSM sector, the allowed two-body decay channels include $Z' \to \zeta_p \zeta_q^*$, $Z' \to \eta^{+} \eta^{-}$ and BSM fermions.
%%%%%%%%%
\begin{figure}[!h]
    \centering        \includegraphics[width=0.6\linewidth]{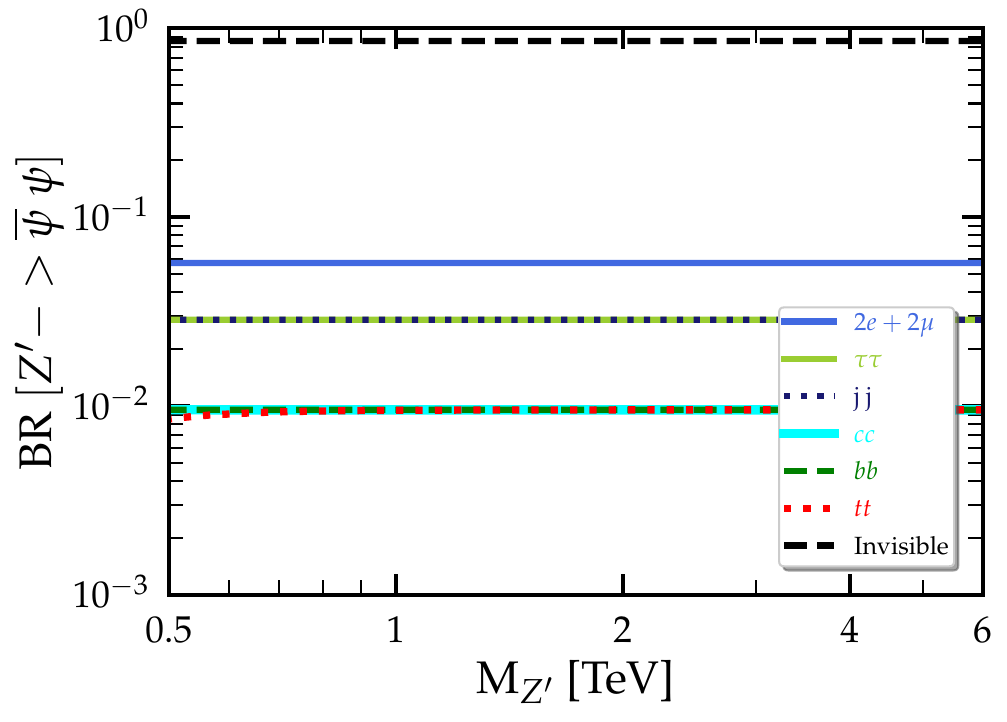}
    \caption{\centering Branching fractions of the $Z'$ boson as a function of its mass $M_{Z'}$.}
    \label{fig:Br_Zp1}
\end{figure}
%%%%%%%
Fig.~\ref{fig:Br_Zp1} shows the branching fractions of the $ Z' $ into visible and invisible SM decay channels. The visible channels include quarks and charged leptons, while the invisible channels correspond to neutrinos. In plotting Fig. \ref{fig:Br_Zp1}, we assume that decays into BSM particles are kinematically forbidden. The plotted branching fractions therefore represent the maximum possible values for the SM decay modes.
%%%
As shown in Fig. \ref{fig:Br_Zp1}, the invisible channel is dominant, accounting for  $\sim 86\%$ of the total width.
%%%
This large invisible branching fraction arises from the sizable $U(1)_{B-L}$ charges of the right handed neutrinos, $(-4,-4,5)$, in this model. In contrast, in the vector $B-L$ scenario, the invisible branching fraction is typically around $38\%$ \cite{Mandal:2023oyh,Prajapati:2024wuu}.
%%%
The total visible branching fraction is this model is approximately $14\%$. Among these, the dilepton final states ($Z' \to l^+ l^-$) contribute about $6\%$.
%%%
This is significantly smaller than in the vector $B-L$ scenario, where the dilepton branching fraction is typically around $25\%$ \cite{Mandal:2023oyh,Prajapati:2024wuu}.
It should be noted that if BSM decay channels are kinematically allowed in this model, the branching fractions into visible SM states will be further decrease.
%%%%%%%
\begin{figure}[!h]
    \centering        \includegraphics[width=0.6\linewidth]{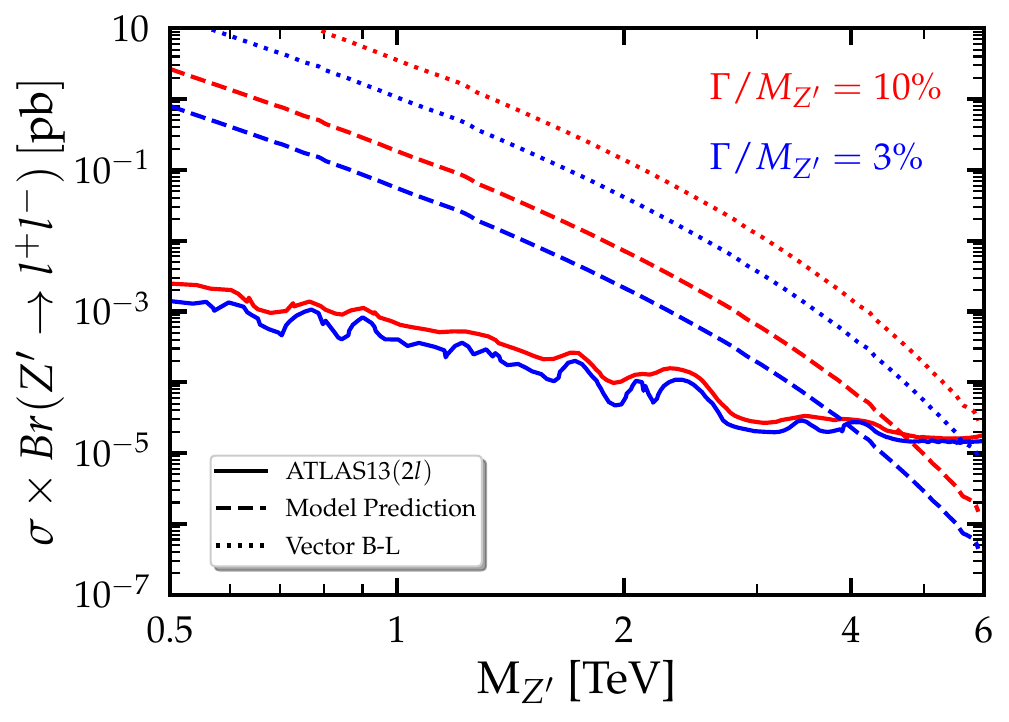}
\caption{Comparison of $\sigma(pp \to Z') \times \mathrm{Br}(Z' \to \ell^{+}\ell^{-})$ with ATLAS limits at $\sqrt{s} = 13$~TeV as a function of $M_{Z'}$. Solid lines denote ATLAS bounds for $\Gamma/M_{Z'} = 10\%$ (red) and $3\%$ (blue), while dashed lines show the present model predictions and dotted lines correspond to the vector $B-L$ case. The chosen $M_{Z'}$ total widths to mass ratio correspond to $g_x \approx 0.33~(0.18)$ in the present model and $g_x \approx 0.69~(0.38)$ in the vector $B-L$ scenario.}
    \label{fig:Br_Zp2}
\end{figure}
%%%%%%

Using the production cross section and the dileptonic branching fraction of the $Z'$, we can derive constraints on the $Z'$ mass.
%%%
In Fig.~\ref{fig:Br_Zp2}, we compare our model predictions against the ATLAS search results for $Z'$ resonances in the dilepton final state, using $pp$ collision data at $\sqrt{s} = 13$~TeV with an integrated luminosity of $139~\text{fb}^{-1}$ \cite{ATLAS:2019erb}. The solid red and blue lines represent the ATLAS upper limits on the cross section times branching ratio [$\sigma (pp \to Z') \times B (Z' \to l^{+} l^{-})$] for two different values of $Z'$ total decay width to mass ratios: $\Gamma/M_{Z'} = 10\%$ and $\Gamma/M_{Z'} = 3\%$, respectively.
%%%%
The dashed blue (red) line represents the predictions of this model for $\Gamma/M_{Z'} = 10\%$ ($3\%$). 
%%%%%
In the limit where $Z'$ decays only into SM fermions, the ratio of the partial decay width to the $Z'$ mass, $\Gamma(Z' \to \overline{\psi}\psi)/M_{Z'}$, depends solely on the gauge coupling $g_x$ and is given by \cite{Prajapati:2024wuu},
%%%%%
\begin{equation}\label{Eq:ZP_decay_Width}
    \frac{\Gamma(Z' \to \overline{\psi} \psi)}{M_{Z'}} \approx  \frac{n_c}{48 \pi} \left[(Q_{\psi_{L}} + Q_{\psi_{R}})^{2} + (Q_{\psi_{L}} - Q_{\psi_{R}})^{2}  \right] g_{x}^{2}\,,
\end{equation}
%%%%%
where $n_c$ is the color factor, taking the value $3$ for quarks and $1$ for leptons. The quantities $Q_{\psi_L}$ ($Q_{\psi_R}$) denote the $U(1)_{B-L}$ charges of the left handed (right handed) components $\psi_L$ ($\psi_R$) of the Dirac fermion $\psi$.
%%%%%
%%%%%
Summing over all fermionic final states, the total decay width to mass ratio $\Gamma/M_{Z'}$ in this model is approximately
\begin{equation}
\frac{\Gamma}{M_{Z'}} \approx 0.93\, g_x^2.
\end{equation}
Thus, $\Gamma/M_{Z'} = 10\%$ ($3\%$) corresponds to $g_x \approx 0.33$ ($0.18$). For comparison, we also show the predictions of the vector $B-L$ model (dotted lines), where $\Gamma/M_{Z'} \approx 0.21\, g_x^2$, leading to $g_x \approx 0.69$ ($0.38$) for the same benchmark widths.
Due to the larger $U(1)_{B-L}$ charges of the right handed neutrinos in the present model, the total decay width is enhanced compared to the vector $B-L$ case. Consequently, a smaller value of $g_x$ is required to obtain the same $\Gamma /M_{Z'}$, which in turn leads to a reduced $Z'$ production cross section. In addition, the branching fraction into dileptons is further suppressed. As a result, the limits on the $Z'$ mass are weaker in this scenario.
%%%
From the comparison with ATLAS data, we obtain lower bounds on the $Z'$ mass of $\sim 4.6$~TeV and $\sim 3.9$~TeV for $\Gamma/M_{Z'} = 10\%$ and $3\%$, respectively. These limits are significantly weaker than in the vector $B-L$ scenario, where the corresponding bounds are $\sim 6$~TeV and $\sim 5.6$~TeV.

Hence, the enhanced invisible decay width of the $Z'$ weakens the collider bounds relative to the conventional vector $B-L$ scenario. This feature originates from the chiral right handed neutrino charge assignment $(-4,-4,5)$.
%%%%
Here, we would like to emphasize that these benchmark points correspond to the most stringent constraints on the $Z'$ mass in this model, since all BSM decay channels are kinematically forbidden. The inclusion of additional BSM decay channels would increase the total decay width and reduce the branching fraction into dileptons, thereby further relaxing these mass limits.
%%%%
%
The results presented above are obtained for representative benchmark values and illustrate the qualitative behavior of the collider constraints on the $Z'$ boson. In the next section, we turn to the phenomenology of the dark sector, where these limits are incorporated to constrain the viable DM parameter space. Notably, the relatively weaker bounds on the $Z'$ mass compared to the vector $B-L$ scenario lead to a less constrained and hence more relaxed DM parameter space.

%%%%%%%%%%%%%%%%%%%%%%%%%%%%%%%%%%%%%%%%%%%%%%%%%%%%%%

\section{Phenomenology of Dark Sector} \label{sec:dm}

This section is devoted to the analysis of dark sector phenomenology of the model. 
As discussed in Sec.~\ref{sec:model}, the VEVs of the Higgs doublet $H$ and the singlet scalar $\chi $ spontaneously break the electroweak and $U(1)_{B-L}$ symmetries, respectively. The breaking of $U(1)_{B-L}$ leads to a residual $Z_6$ symmetry. Under this residual $Z_6$ symmetry, all the SM fermion fields, along with the BSM singlet fermions $N\equiv (N_{L}, N_{R})$ which participate in neutrino mass generation at tree level, are even. The Higgs doublet $H$ and the singlet scalar $\chi$ transform trivially under $Z_{6}$.
%%%
In contrast, the BSM particles running in the loop to generate neutrino masses, $\eta = (\eta^{+},\eta^0)^{T}$, $S_1$, $S_2$, $f\equiv (f_L,f_R)$, are odd under $Z_6$.
%%%
After spontaneous symmetry breaking, the $Z_{6}$ odd neutral scalars $(\eta^0, S_1, S_2)$ mix to form three physical scalar fields, $\zeta_{p}$ ($p=1,2,3$), as discussed in App. \ref{sec:scalar}.
Consequently, the $Z_6$ odd sector comprises three neutral complex scalars $\zeta_{p}$, one charged scalar $\eta^{\pm}$, and one fermion $f$.
The lightest neutral particle among these $Z_6$ odd fields serves as a viable DM candidate.
Therefore, our model accommodates both scalar and fermionic DM scenarios. In the following subsections, we consider the scalar and fermionic DM scenarios separately, analyzing the parameter space consistent with both collider and direct detection constraints.
%%

%%%%%%%%%%%%%%%%%%%%%%%%%%
\subsection{Singlet Scalar DM}

The $Z_6$ odd scalar sector consists of three neutral physical fields, $\zeta_p$ ($p=1,2,3$). 
%%%
The details regarding the scalar potential, mass spectrum, and mixing matrices are presented in App.~\ref{sec:scalar}.
%%%
We consider $\zeta_{1}$ to be the lightest particle in the $Z_6$ odd sector ($M_{\zeta_1}< M_{\zeta_{2/3}},~M_{\eta^{\pm}},~M_{f}$), thereby ensuring its stability. 
%%%
Due to the mixing between the $SU(2)_L$ scalar doublet $\eta_0$ and the scalar singlets $S_1$ and $S_2$, the fields $\zeta_1$ exhibit a joint doublet-singlet nature. 
Consequently, within certain parameter limits, our scalar DM candidate $\zeta_1$ can inherit the characteristics of a scalar doublet, a scalar singlet, or a mixture of both.
%%%
In the doublet dominated or in the mixed doublet-singlet scenario, the efficient co-annihilation channels allow the scalar DM to satisfy the observed relic density while evading collider and direct detection bounds over a wide mass range \cite{Deshpande:1977rw,Barbieri:2006dq,Ma:2006km,Batra:2022pej,CentellesChulia:2022vpz,Kumar:2023moh,CentellesChulia:2024iom,Kumar:2024zfb,Kumar:2025cte,Kumar:2025zvv}.
%%%%
Conversely, in the standard pure scalar singlet DM scenario, direct detection experiments severely constrain the parameter space, leaving only a narrow viable window near the Higgs resonance region ($60.5~\text{GeV} \lesssim M_{\text{DM}} \lesssim 62.5~\text{GeV}$) \cite{Yaguna:2008hd,Profumo:2010kp,Feng:2014vea,DiMauro:2023tho,Yu:2024xsy}.
%%%%%
However, this severe restriction applies strictly to the pure singlet case.
%%%%%%
In the present framework, we have additional scalar channels due to the rich scalar sector and also $Z'$ mediated channels.
%%%
These additional channels may contribute to the DM relic density without significantly increasing the direct detection cross section. 
%%%
It is therefore instructive to study how these additional scalar and gauge interactions reshape the singlet-dominated DM parameter space.
%%%%
Accordingly, here we focus on the singlet-dominated DM scenario.  
%%%%%
For this analysis, we take the hierarchy in the mass parameters $m_{\eta}^{2} \gg m_{s_1, s_2}^2$, ensuring that the DM candidate is either $S_1$ dominated, $S_2$ dominated, or a mixture of both singlets.
%%%%%%
To quantitatively analyze these scenarios, the DM observables, including the relic density and direct detection cross sections, are computed using the package micrOMEGAs \cite{Belanger:2014vza,Belanger:2020gnr}.
%%%
In what follows, we examine the effects of the $Z'$ mediated and scalar mediated channels separately to assess their individual impacts on the singlet dominated DM scenario.

\subsubsection{Singlet DM with pure $Z'$ interactions}
\begin{figure}[!h]
    \centering
    \includegraphics[width=0.425\linewidth]{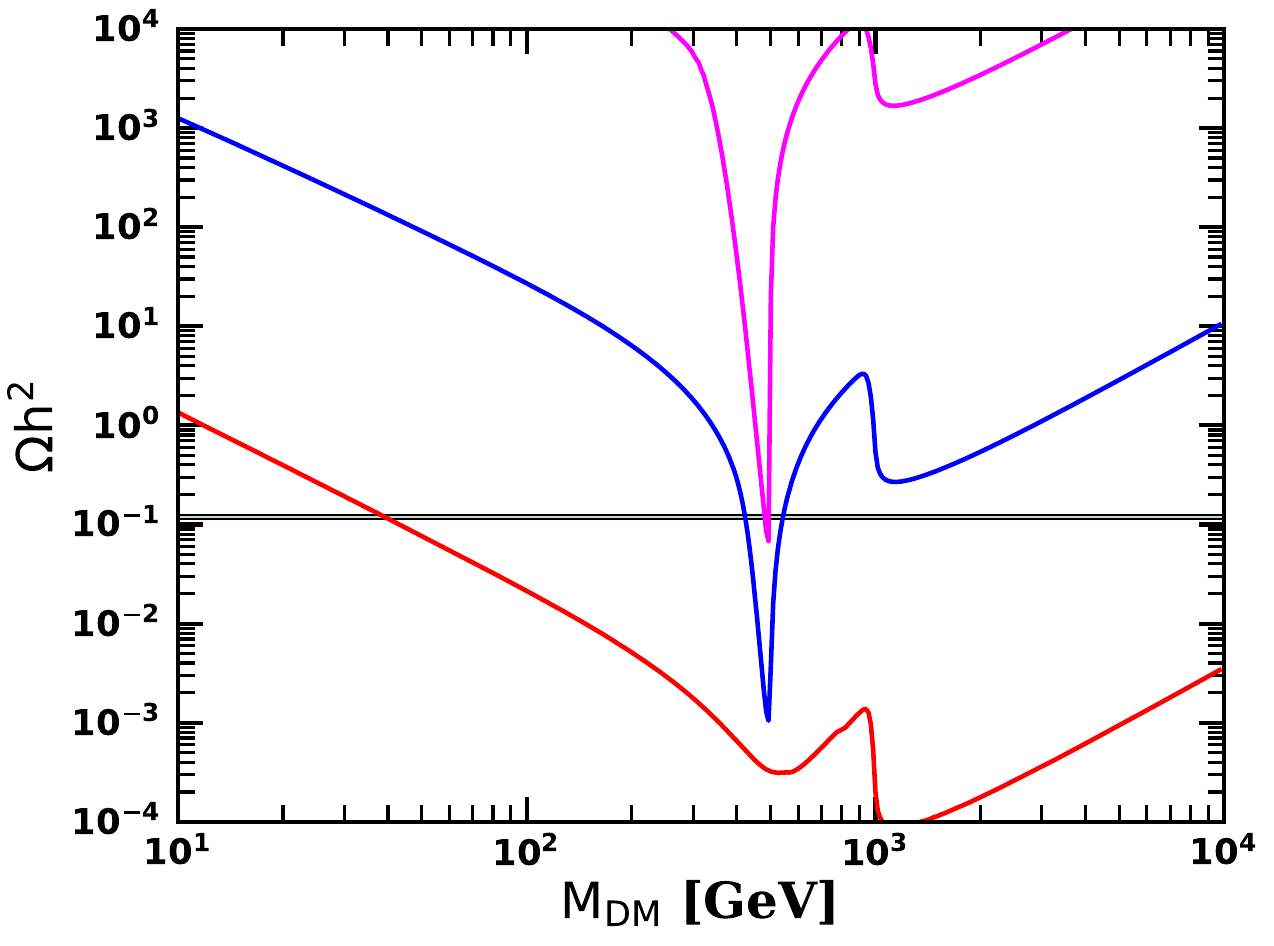}    \includegraphics[width=0.49\linewidth]{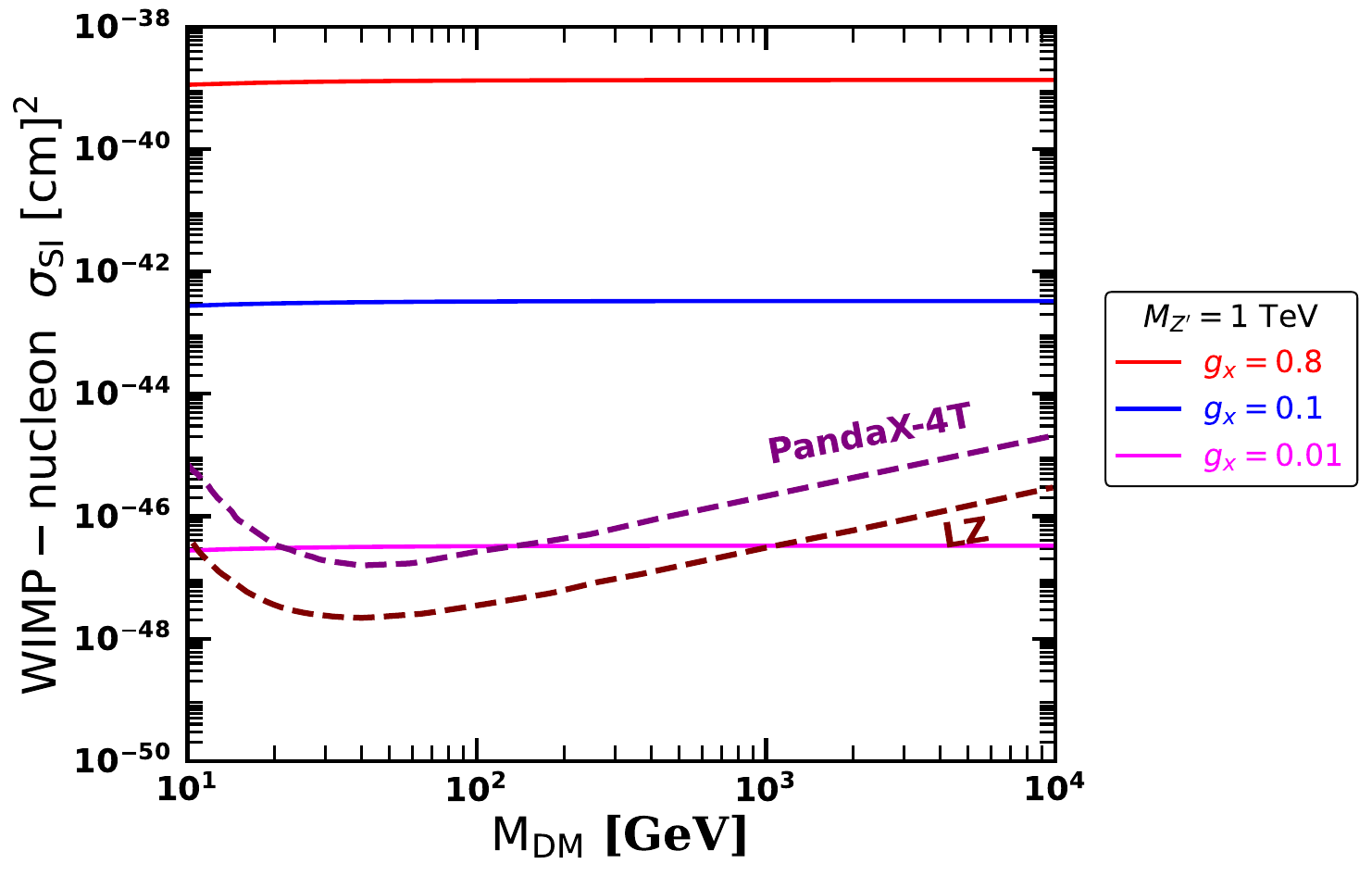}
    \includegraphics[width=0.425\linewidth]{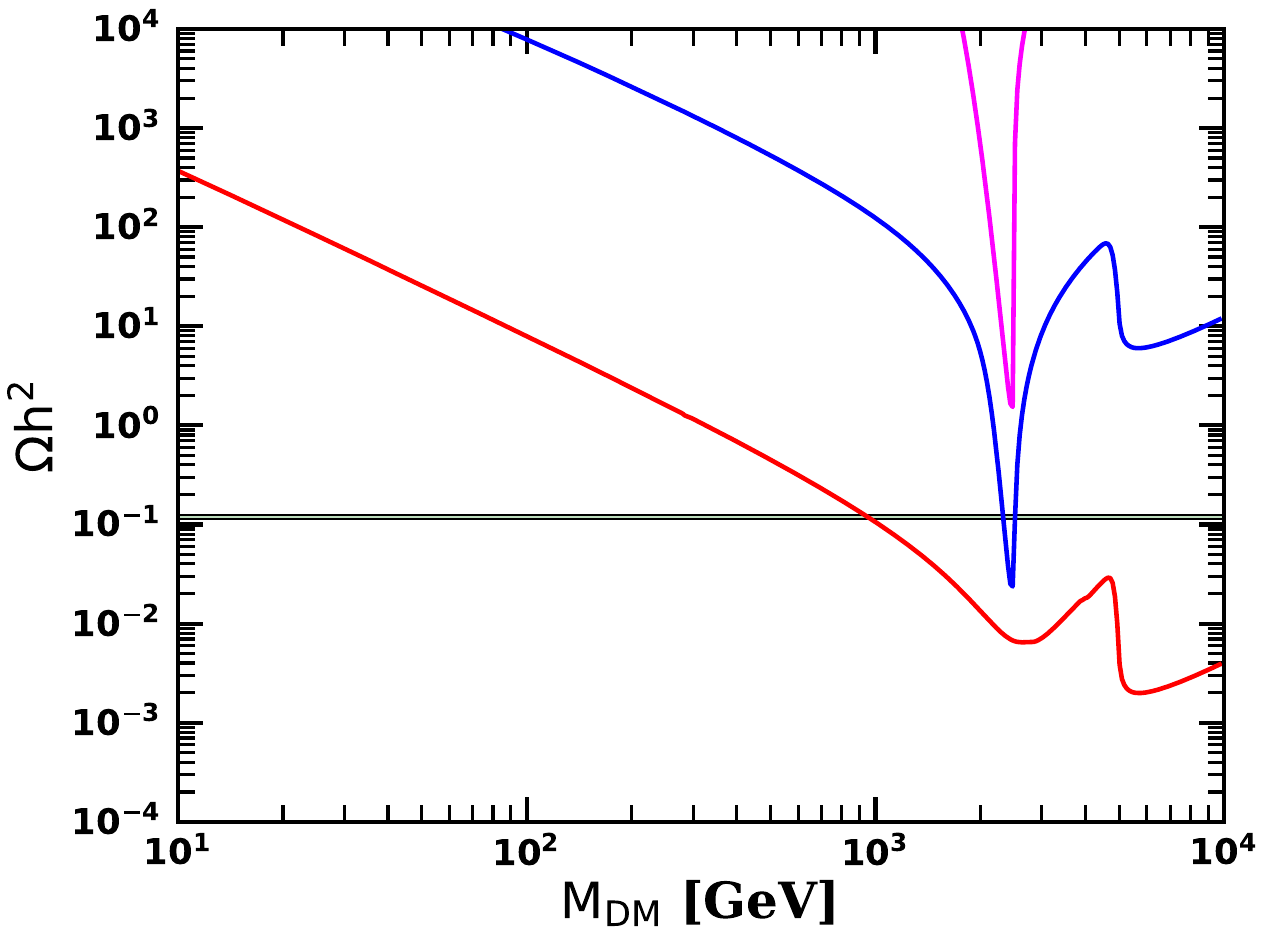}    \includegraphics[width=0.49\linewidth]{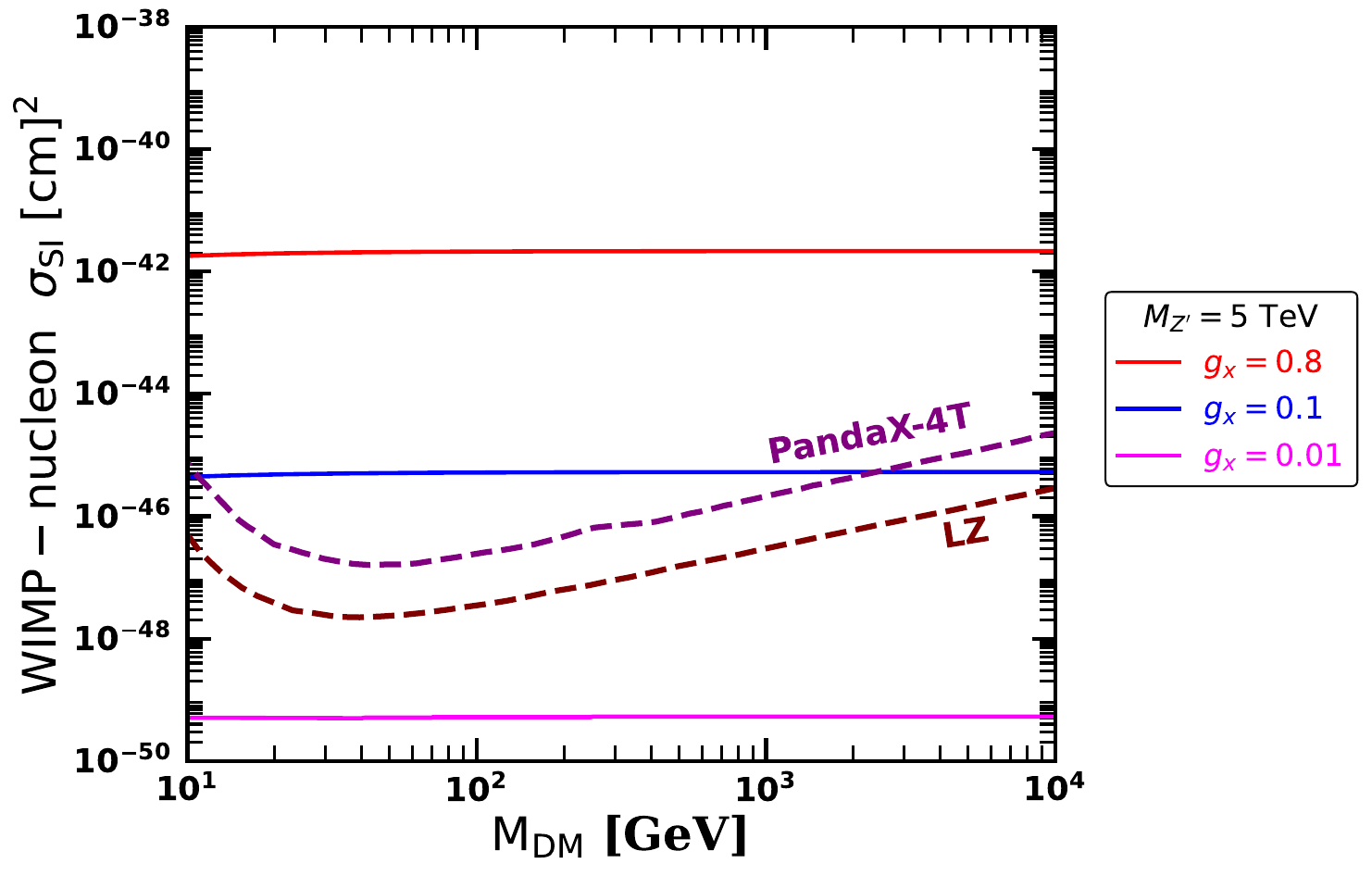}
    \includegraphics[width=0.425\linewidth]{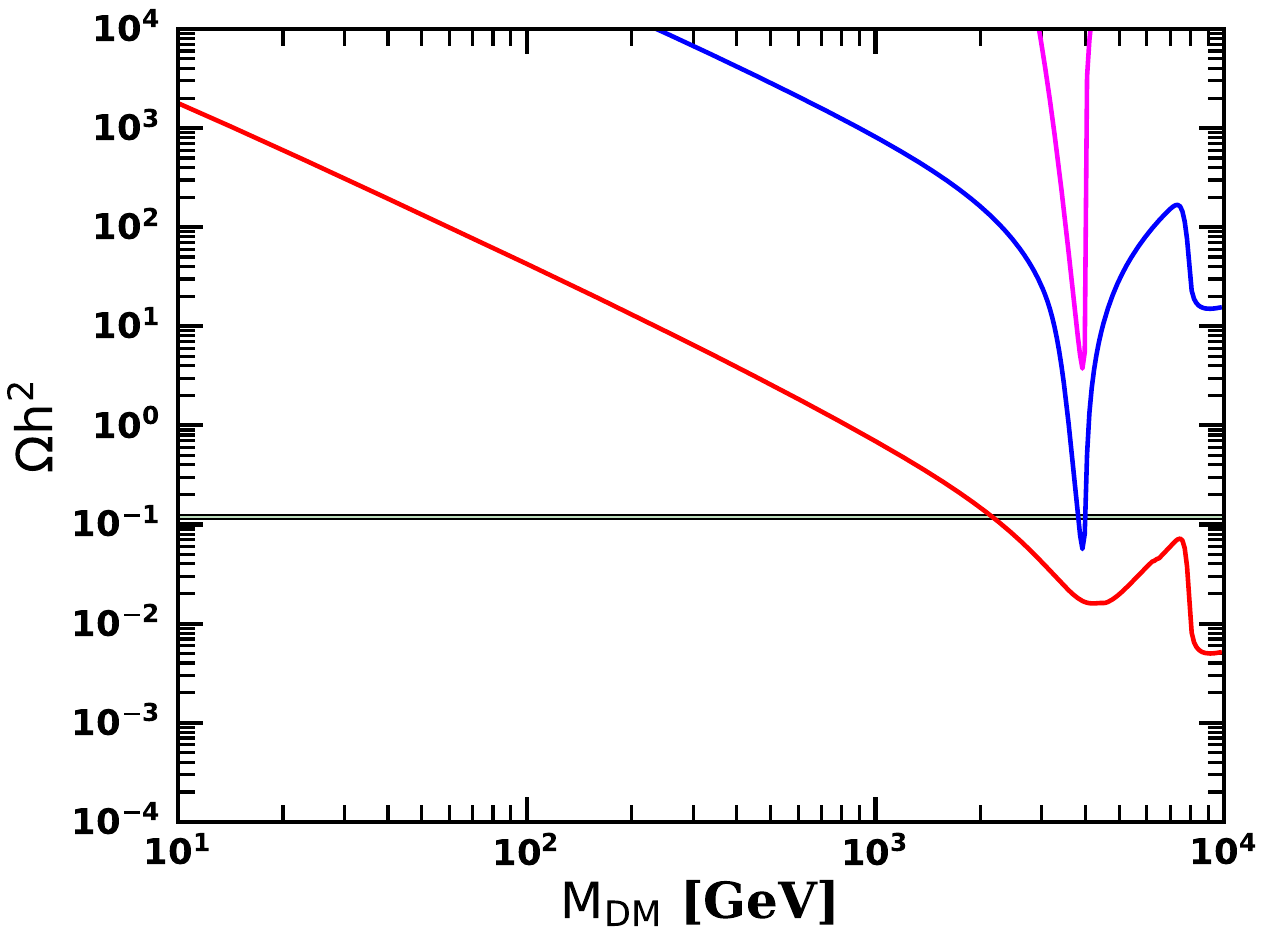}    \includegraphics[width=0.49\linewidth]{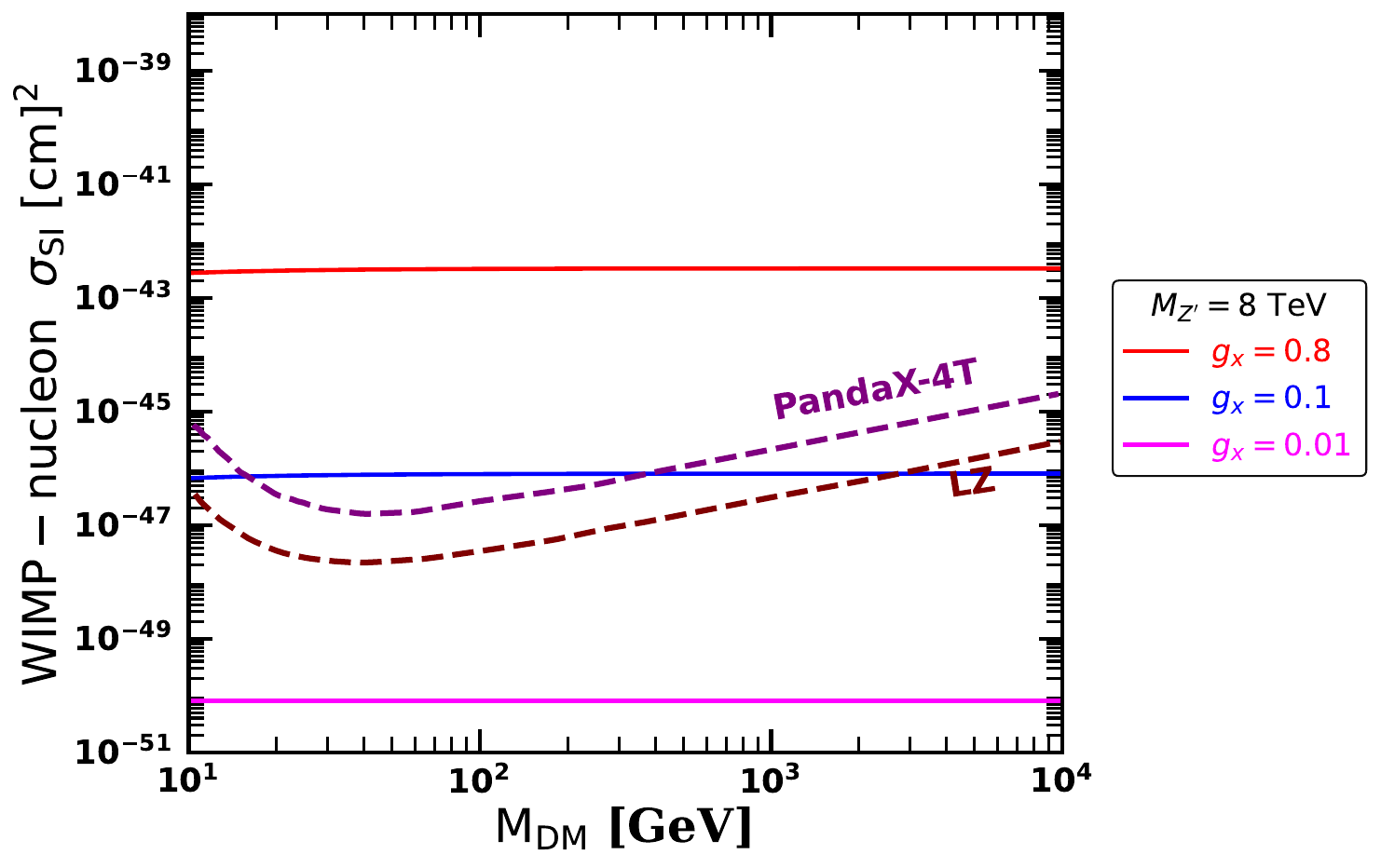}
    \caption{
    Relic density (left panels) and spin-independent WIMP-nucleon scattering cross section (right panels) as a function of the DM mass $M_{\rm DM}$ for the $S_1$-dominated singlet DM scenario. The different rows correspond to $Z'$ masses $M_{Z'} = 1,\ 5,\ 8$ TeV (from top to bottom). In each panel, the colored curves represent different values of the gauge coupling $g_x$. The horizontal green band on the left panel shows the $3\sigma$ allowed range for cold DM $0.1126 \le \Omega h^2 \le 0.1246$ \cite{Planck:2018vyg}.  The dashed purple and red lines in the right panels show the latest upper bounds from the PandaX-4T \cite{PandaX:2024qfu} and LZ \cite{LZ:2024zvo} collaborations.}
    \label{fig:OnlyZP_S1DM}
\end{figure}

%%%%
We first examine the role of pure $Z'$ mediated interactions in shaping the phenomenology of singlet DM $S_1$.
%%%%
These gauge interactions allow the DM particle to annihilate into fermions via $Z^\prime$ exchange, $\text{DM}~\text{DM}^{*} \to Z' \to \psi \bar{\psi}$.
The contributions of these channels depend primarily on the gauge coupling $g_x$ and the $Z^\prime$ mass $M_{Z^\prime}$. 
Consequently, in this pure gauge interaction scenario, the DM phenomenology is governed by only a few key parameters: $M_{\text{DM}}$, $g_x$, and $M_{Z^\prime}$.
%%%%%%%%
To illustrate the impact of these parameters on the DM relic density and direct detection prospects, we present the results in Fig.~\ref{fig:OnlyZP_S1DM} for the $S_1$ singlet-dominated scenario.
The left panels show the DM relic density $\Omega h^2$ as a function of the DM mass $M_{\text{DM}}$ for several benchmark values of the gauge coupling $g_x$.
%%%
The green band shows the $3\sigma$ allowed range for cold DM derived from the Planck satellite data, $0.1126 \le \Omega h^2 \le 0.1246$ \cite{Planck:2018vyg}.
%%%
The right panels display the corresponding spin-independent WIMP-nucleon cross section $\sigma_{\text{SI}}$ versus $M_{\text{DM}}$ for the same benchmark values.
%%%
The dashed purple and red lines correspond to the latest upper bounds from the PandaX-4T \cite{PandaX:2024qfu} and LZ \cite{LZ:2024zvo} collaborations.
%%%%
From top to bottom, the rows correspond to different fixed values of the $Z^\prime$ mass, specifically $M_{Z^\prime} = 1, 5,$ and $8~\text{TeV}$.
%%%
%%%%%
%%%%%

The qualitative behavior remains analogous across all considered benchmark scenarios. In particular, two distinct dips appear in the relic density curves.
%%%
The first dip corresponds to the $Z^\prime$ resonance region, where $M_{\text{DM}} \sim M_{Z^\prime}/2$, while the second dip occurs when the $\text{DM}~\text{DM}^{*} \to Z^\prime Z^\prime$ annihilation channel becomes kinematically accessible.
The relic density decreases as the DM mass approaches the $Z^\prime$ resonance region ($M_{\text{DM}} \sim M_{Z^\prime}/2$), where it reaches its minimum value. It subsequently increases as the DM mass moves away from the resonance, featuring a second small dip once the $\text{DM}~\text{DM}^{*} \to Z^\prime Z^\prime$ channel opens up.
%%%%
The right panel of Fig.~\ref{fig:OnlyZP_S1DM}, shows the corresponding direct detection cross section. Since quarks carry purely vector charges under $U(1)_{B-L}$, the resulting interactions are vectorial, leading to spin-independent scattering. The corresponding cross section in the zero-momentum transfer limit is given by \cite{DelNobile:2021wmp,DeRomeri:2025nkx},
%%%%
\begin{equation}\label{Eq:DD:SpinI_Zp_Crc}
    \sigma_{\rm SI} =  \frac{\mu^{2}}{\pi} \left(\frac{Q_{\rm N}\,Q_{DM}\, g_{x}^{2}}{M_{Z'}^{2}}\right)^{2}\,,
\end{equation}
%%%%
where $\mu = M_{\rm N} M_{\rm DM}/(M_{\rm N} + M_{\rm DM})$ is the reduced mass of the nucleon-DM system. The quantity $Q_{\rm N} = \sum_{q} Q_q\, \mathcal{N}_{q}$ denotes the effective $U(1)_{B-L}$ charge of the nucleon, where $\mathcal{N}_{q}$ is the number of valence quarks in the nucleon, and $Q_{\rm DM}$ is the $U(1)_{B-L}$ charge of the DM candidate.
For the DM mass of order $\mathcal{O}(\rm GeV)$ or higher, the reduced mass $\mu$ approaches the nucleon mass and becomes essentially independent of the DM mass.
Thus, $\sigma_{\rm SI}$ is primarily governed by the gauge coupling $g_x$ and the mediator mass $M_{Z^\prime}$.
%%%%%%%%%%%
Increasing the $g_x$ reduces the relic density while enhancing the $\sigma_{\rm SI}$. Conversely, increasing the $M_{Z'}$ enhances the relic density but suppresses the $\sigma_{\rm SI}$. This behavior can be seen by comparing the top and bottom panels of Fig.~\ref{fig:OnlyZP_S1DM}.
For relatively small gauge coupling $g_x \sim 0.1$ and below, the relic density tends to be significantly higher. However, for the larger gauge couplings ($g_x \sim 0.8$), the observed relic density can be satisfied over a broad DM mass range extending from a few hundred GeV to the TeV scale. However, such large couplings are severely restricted by collider bounds (see Sec. \ref{sec:Zprime}). In addition, $\sigma_{\rm SI}$ corresponding to such large coupling is also excluded by the direct detection constraints, as illustrated in the right panels.
The qualitative behavior of the $S_2$-dominated scenario remains analogous to the $S_1$ case and is presented in App.~\ref{App:S2_Dom}.
%%%%

\begin{figure}[!h]
    \centering
        \includegraphics[width=0.43\linewidth]{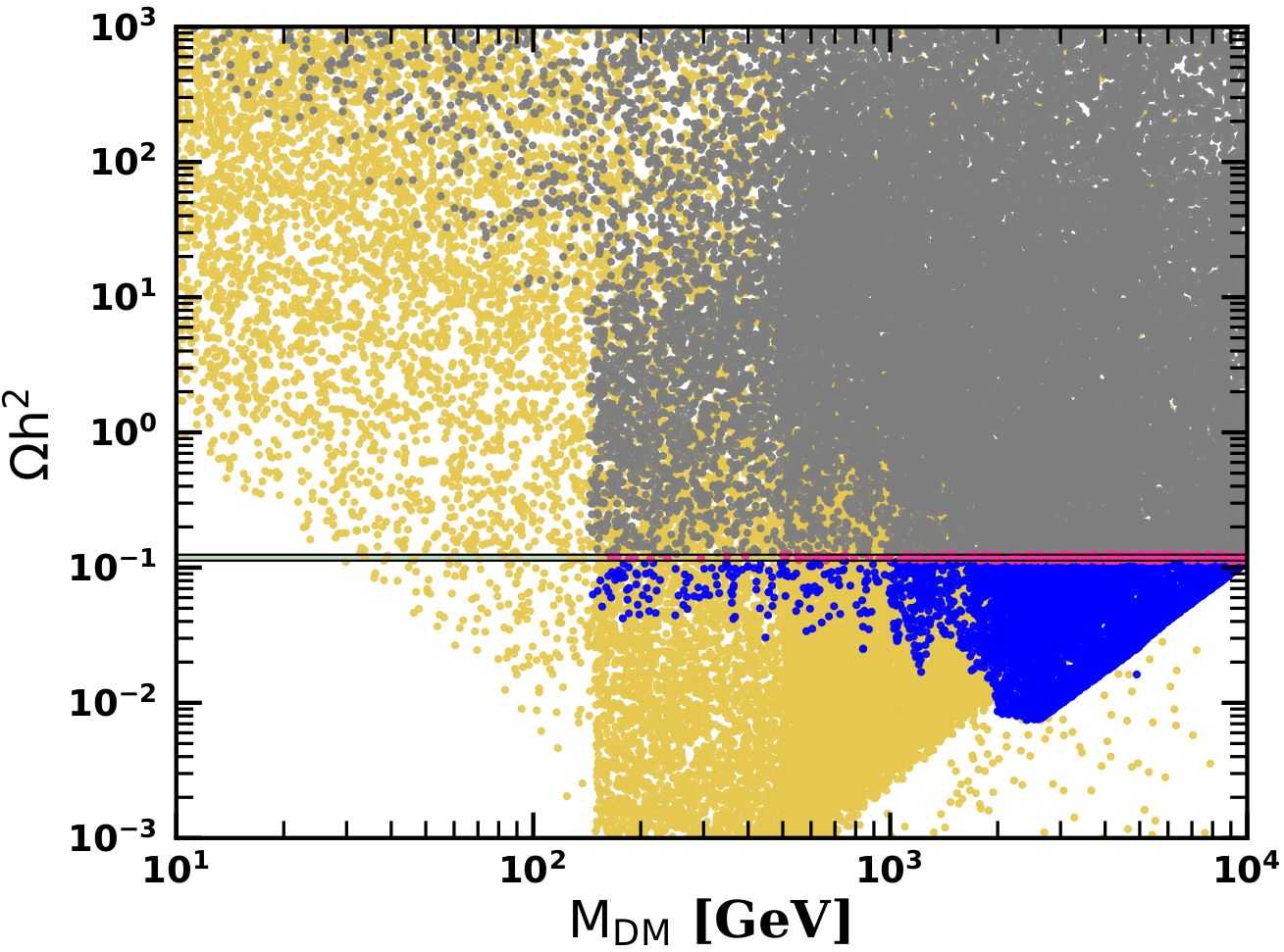}
        \includegraphics[width=0.56\linewidth]{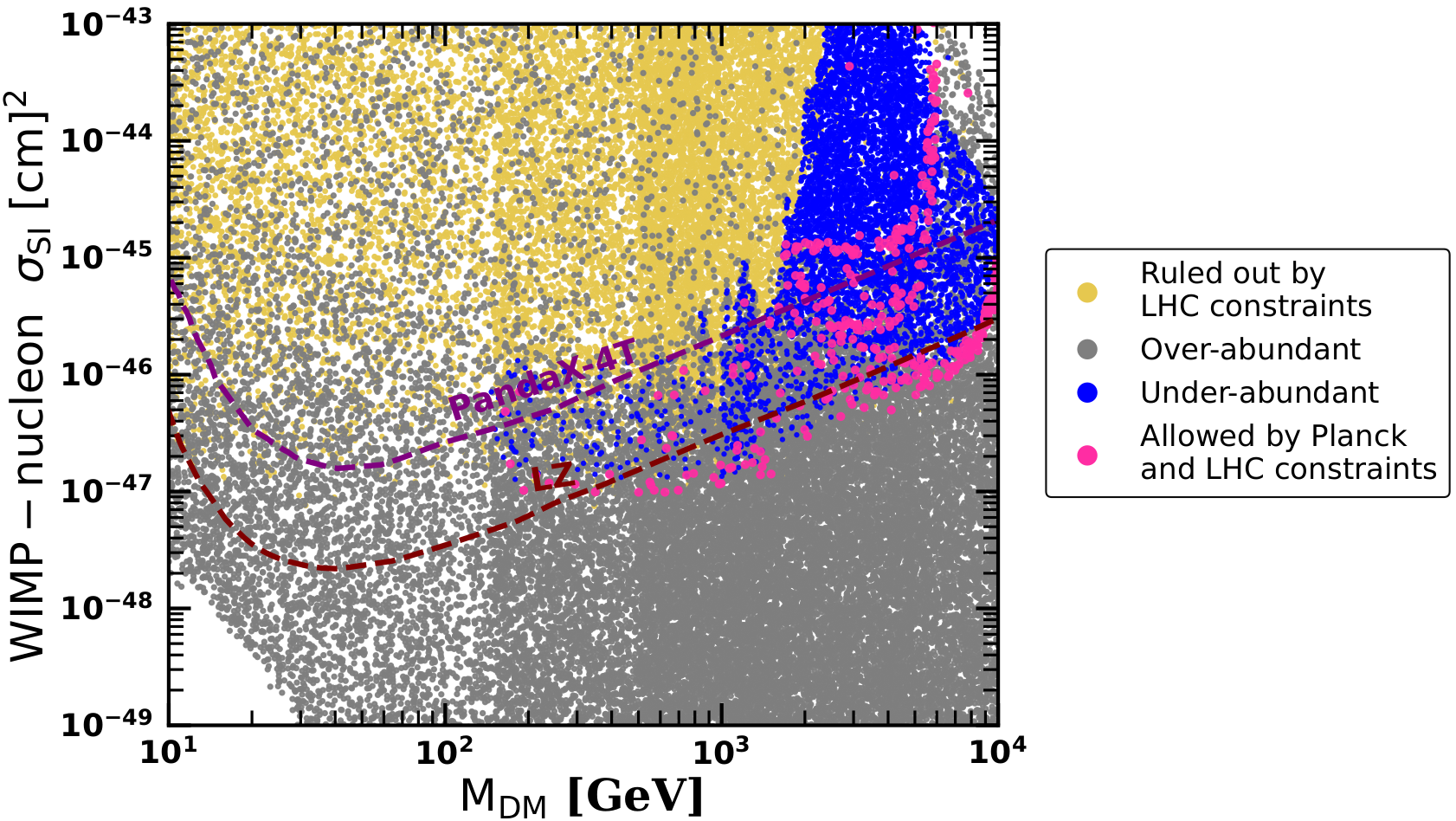}
    \caption{Left: DM relic density $\Omega h^2$; Right: spin-independent WIMP–nucleon cross section $\sigma_{\rm SI}$ as functions of $M_{\rm DM}$ for the mixed $S_1$–$S_2$ singlet DM scenario . Yellow points are excluded by LHC bounds on $M_{Z'}/g_x$ from dilepton searches. Gray (blue) points correspond to over-abundant (under-abundant) relic density. Magenta points satisfy collider constraints and lie within the $3\sigma$ allowed range for cold DM.
 }
    \label{fig:ZP_Dom_Scatter}
\end{figure}

Having discussed the qualitative features of the singlet DM candidate $S_1$ for representative benchmark scenarios with only $Z'$ mediated interactions, we now extend our analysis to the generic parameter space of the model. We continue to focus exclusively on the $Z'$ portal while considering the dark sector containing both the singlet scalars, $S_1$ and $S_2$.
We present this scenario in Fig. \ref{fig:ZP_Dom_Scatter}. 
The left panel shows the DM relic density as a function of the DM mass $M_{\text{DM}}$ for the mixed $S_1$-$S_2$ singlet DM scenario.\footnote{The lightest state among $S_1$ and $S_2$ serves as the DM candidate.}
%%%
%%%%%%%%%%%%%%%%%%
%
The gauge parameters are scanned over the ranges, $g_x \in [10^{-4}, 1]$ and $M_{Z^\prime} \in [0.3, 10]~\text{TeV}$. Here, we also incorporate the ATLAS constraints on $M_{Z'}/g_{x}$ from dilepton searches as discussed in Sec.~\ref{sec:Zprime}. 
%%%
Most of the parameter space yielding a low relic density with large $g_x$ and a light $Z'$ boson is excluded by collider bounds, as indicated by the yellow points. 
The blue and gray points satisfy the collider limits but result in under-abundant and over-abundant DM relic densities, respectively. 
Finally, the magenta points simultaneously satisfy the collider constraints and fall within the $3\sigma$ allowed range for cold DM. 
%%%%
From Fig. \ref{fig:ZP_Dom_Scatter}, it is evident that a sizable region of the parameter space simultaneously satisfies the observed DM relic abundance and remains consistent with current collider and direct detection constraints. Notably, a significant fraction of the allowed parameter space lies close to the present LZ exclusion limit. Therefore, this scenario offers promising prospects for future direct detection experiments \cite{DARWIN:2016hyl,PANDA-X:2024dlo,XLZD:2024nsu}, which will be capable of probing a substantial portion of the currently viable parameter space.

%%%%%%%%%%%%%%%%%%%%%
%%%%%%%%%%%%%%%%%%%%%
%%%%%%%%%%%%%%%%%%%%%

\subsubsection{Singlet DM with scalar-dominated interactions}

Having discussed pure $Z'$ interactions, we now discuss scalar-dominated interactions. 
%%%
These interactions resemble those of the Higgs portal scenario. However, due to the rich scalar sector, the behavior is different from the simple Higgs portal scenario due to additional annihilation and co-annihilation channels.
%%%
We consider the $S_1$-dominated singlet DM scenario, corresponding to the limit $m_{s_1}^2 \ll m_{s_2}^2$.
%%%
In this regime, along with the SM Higgs ($H_1 \equiv H_{\text{SM}}$) mediated channels, there are additional annihilation channels mediated by the heavier CP-even scalar $H_2$. 
%%%
Mixing between these two scalars opens up further annihilation channels, $\text{DM} ~\text{DM}^{*} \to H_2 \to \text{SM} ~\text{SM}$. 
This mixing is governed by the VEV of the singlet scalar, $v_{\chi}$, and the coupling $\lambda_{H \chi}$ (see Eq.~\ref{Eq:App:Higgs_Mixing}).
%%%
The other parameter that primarily determines the strength of these annihilation channels is the DM coupling to $H_2$, denoted as $\lambda_{\chi s_1}$.
%%%%
These additional channels can also contribute to the direct detection cross section, which is governed by the coupling $\lambda_{H s_1}$.
%%%%
%%%%
%%
\begin{figure}[!h]
    \centering
        \includegraphics[width=0.39\linewidth]{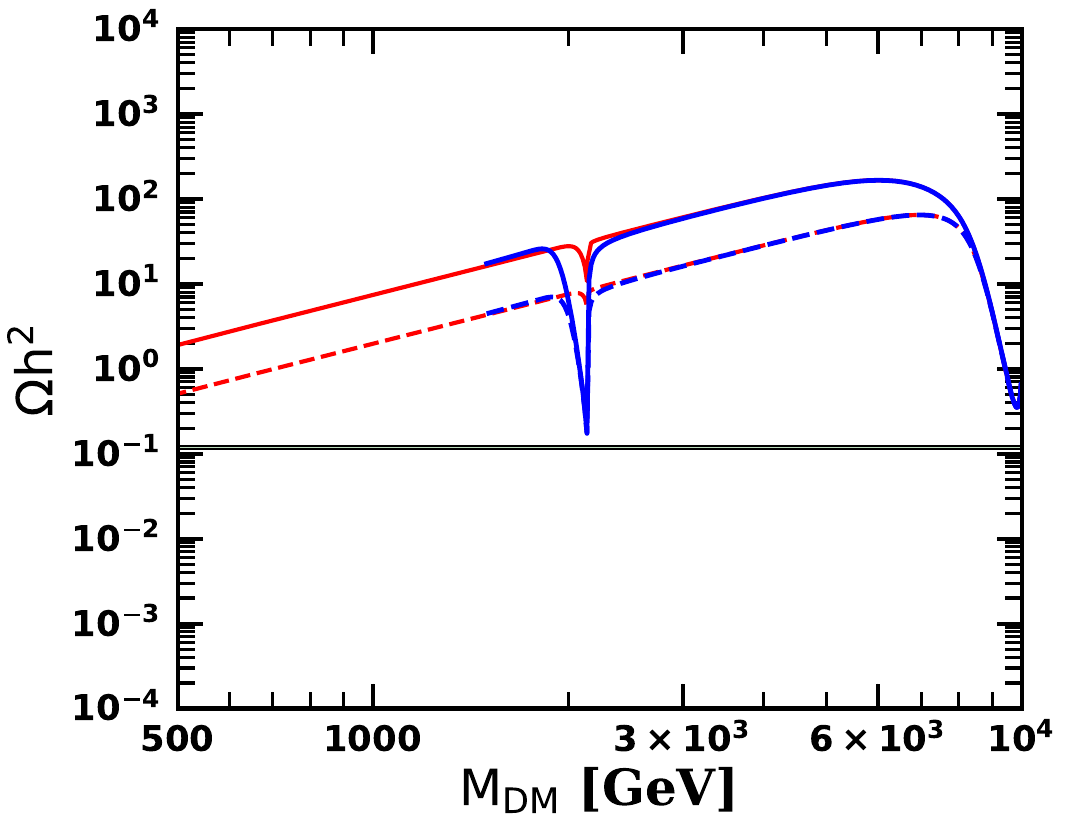}
        \includegraphics[width=0.59\linewidth]{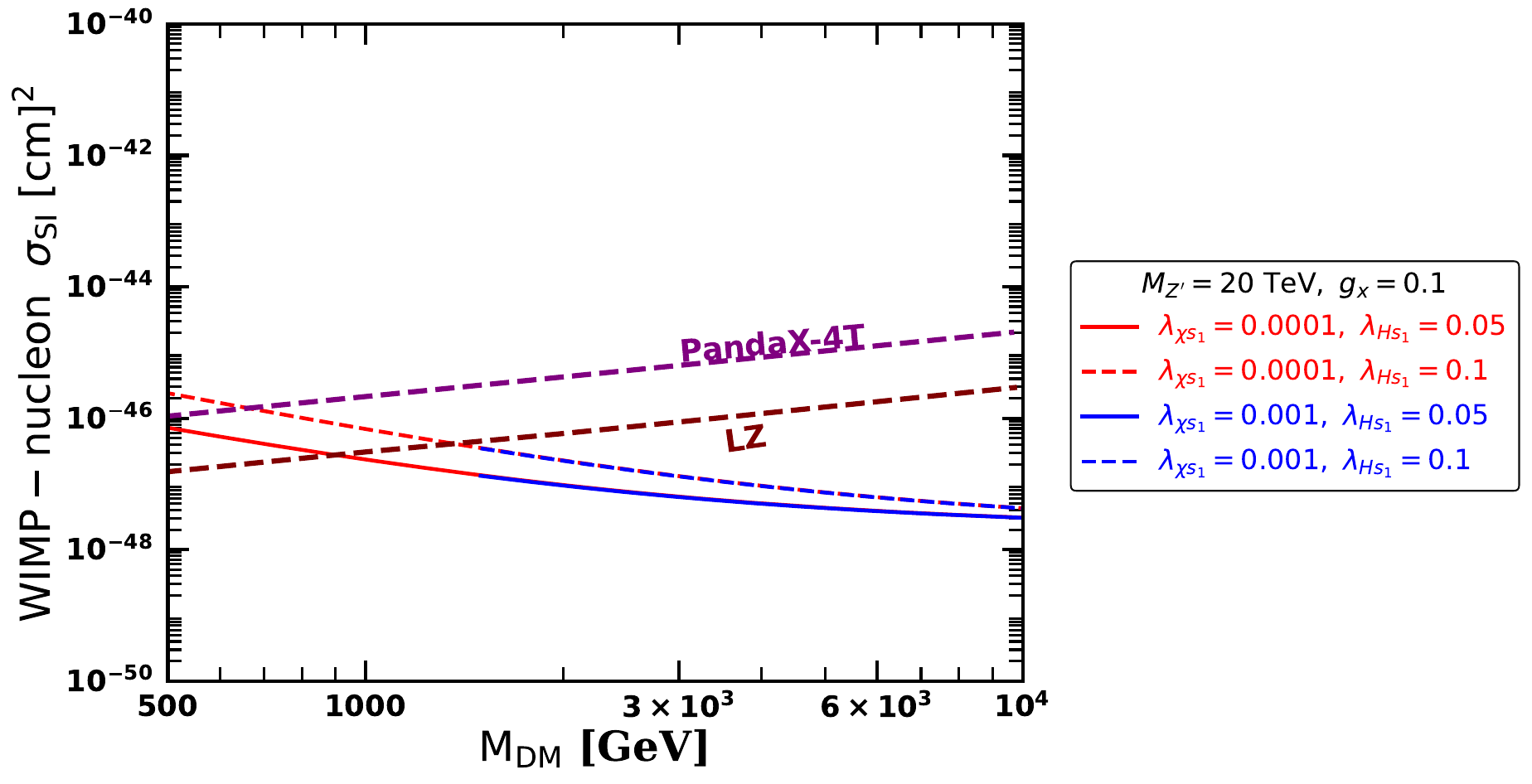}
    \caption{DM $\Omega h^2$ (left) and  $\sigma_{\text{SI}}$ (right) versus $M_{\text{DM}}$ for the $S_1$ singlet DM scenario under scalar dominated interactions. The curves represent different combinations of the scalar couplings $\lambda_{\chi s_1}$ and $\lambda_{H s_1}$ with fixed $M_{Z^\prime} = 20~\text{TeV}$, $g_x =0.1$ and $M_{H_2} \approx 4.2~\text{TeV}$. See text for details.}
    \label{fig:Scalar_Dom_S1_Low}
\end{figure}
%%%%%%%%%%%%%%%%%%

The effects of these $H_2$ mediated annihilation channels are shown in Fig. \ref{fig:Scalar_Dom_S1_Low}.
The left (right) panel of Fig.~\ref{fig:Scalar_Dom_S1_Low}, shows $\Omega h^2$ ($\sigma_{\rm SI}$) as a function of $M_{\text{DM}}$ for the $S_1$ singlet DM scenario under scalar-dominated interactions.
To work in the scalar-dominated regime, we set a heavy mass $M_{Z^\prime} = 20~\text{TeV}$ with $g_x = 0.1$, which subsequently fixes the singlet VEV $v_{\chi}$.
The heavy scalar mass is fixed at $M_{H_2} \approx 4.2~\text{TeV}$. 
To illustrate the distinct impact of these scalar parameters, we choose specific combinations of benchmark values for $\lambda_{\chi s_1}$ and $\lambda_{H s_1}$. 
%%%
The red curves correspond to a fixed $\lambda_{\chi s_1} = 10^{-4}$ with $\lambda_{H s_1} = 0.05$ (solid) and $0.1$ (dashed). The blue curves correspond to a larger fixed coupling $\lambda_{\chi s_1} = 10^{-3}$ with $\lambda_{H s_1} = 0.05$ (solid) and $0.01$ (dashed). 
%%%
Notably, because $\lambda_{\chi s_1}$ contributes directly to the DM mass matrix, the chosen benchmark for the blue curves does not allow $M_{\rm DM} \lesssim 1.5~\text{TeV}$.
%%%
In the left panel, the first dip corresponds to the $H_2$ resonance, while the second dip at the high mass tail arises from the $Z^\prime$ resonance. 
As evident from the left panel of Fig. \ref{fig:Scalar_Dom_S1_Low}, since $\lambda_{\chi s_1}$ primarily governs the DM interaction with the heavy scalar $H_2$, increasing its value significantly deepens the $H_2$ resonance profile ($M_{\text{DM}} \sim M_{H_2}/2$) without altering the curve behavior elsewhere or affecting $\sigma_{\rm SI}$. 
In contrast, the SM Higgs-mediated contributions are governed by $\lambda_{H s_1}$. As a result, increasing its value uniformly suppresses the $\Omega h^2$ across the mass spectrum while simultaneously increasing $\sigma_{\rm SI}$, as shown in the right panel.
Furthermore, the $\sigma_{\rm SI}$ for scalar-mediated interactions scales as $\sigma_{\rm SI} \propto 1/M_{\rm DM}^2$ \cite{DelNobile:2021wmp,He:2008qm}. Consequently, increasing the DM mass $M_{\rm DM}$ suppresses $\sigma_{\rm SI}$, which appears as a noticeable downward slope in the right panels.

As evident from Fig.~\ref{fig:Scalar_Dom_S1_Low}, beyond the SM Higgs resonance region, $H_2$-mediated annihilation channels may yield the observed DM relic abundance while satisfying direct detection limits near the $H_2$ resonance.
%%%%
%%
\begin{figure}[!h]
    \centering
        \includegraphics[width=0.405\linewidth]{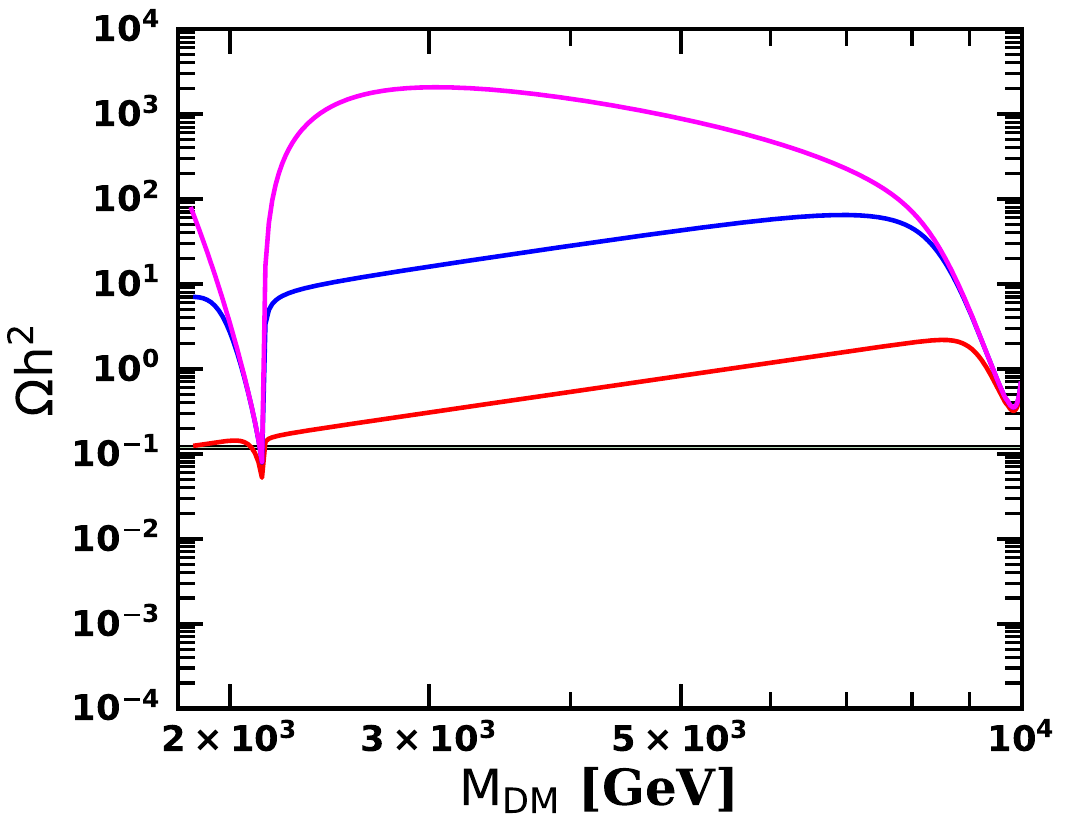}
        \includegraphics[width=0.56\linewidth]{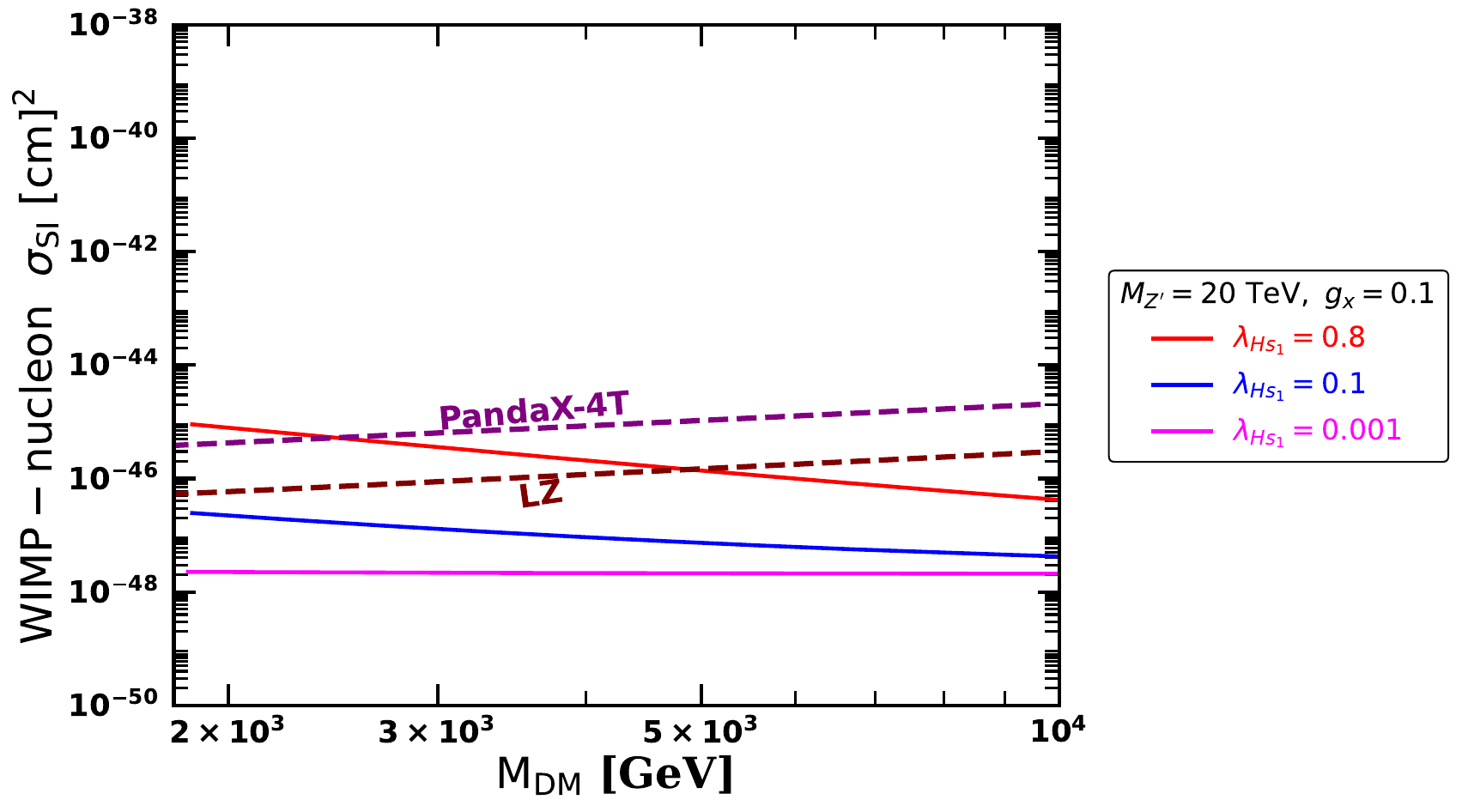}
    \caption{DM $\Omega h^2$ (left) and  $\sigma_{\text{SI}}$ (right) as functions of $M_{\rm DM}$ for different benchmark values of the scalar couplings $\lambda_{\chi s_1}$ and $\lambda_{H s_1}$ in the $S_1$-dominated singlet DM scenario. See text for further details. 
}
    \label{fig:Scalar_Dom_S1}
\end{figure}
%%%%%%%%%%%%%%%%%%
%%
This behavior near the $H_2$ resonance region is illustrated in Fig.~\ref{fig:Scalar_Dom_S1}. 
Adopting the same scalar benchmark values, we fix $M_{H_2} = 4.2~\text{TeV}$.
%%%
The benchmark values for $\lambda_{\chi s_1}$ are chosen such that $M_{\text{DM}}$ can scan down to at least $1.8~\text{TeV}$, thereby fully capturing the $H_2$ resonance region ($M_{\text{DM}} \sim M_{H_2}/2$). 
The Higgs-DM coupling $\lambda_{H s_1}$ is varied across three distinct benchmark values, represented by different colors. 
For the smallest benchmark value ($\lambda_{H s_1} = 0.001$), the SM Higgs-mediated channels become sufficiently suppressed; consequently, the flat line in the direct detection plot reflects the residual contributions from both $Z^\prime$ and $H_2$ exchange.
%%%
The observed relic density can be satisfied near the $H_2$ resonance across all benchmarks, with direct detection bounds simultaneously respected for $\lambda_{H s_1} \lesssim 0.1$.
In Appendix~\ref{App:S2_Dom}, we present the corresponding analysis for the $S_2$-dominated singlet DM scenario, where the qualitative features remain consistent with those shown in Fig.~\ref{fig:Scalar_Dom_S1}.

\begin{figure}[!h]
    \centering        \includegraphics[width=0.43\linewidth]{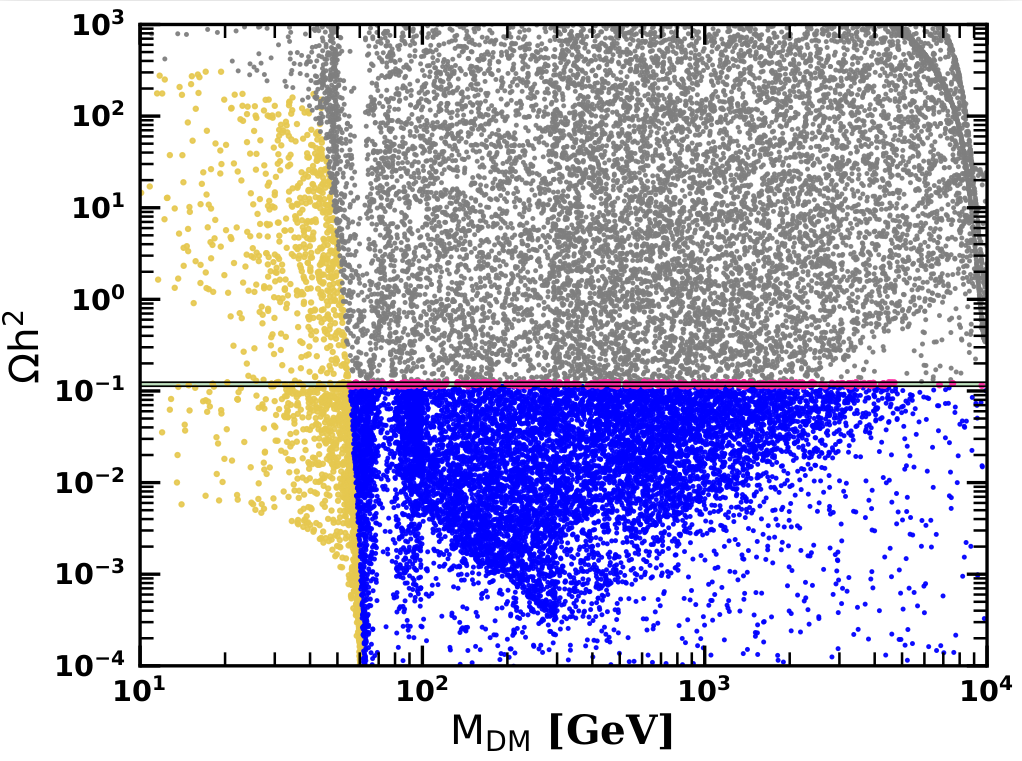}
      \includegraphics[width=0.56\linewidth]{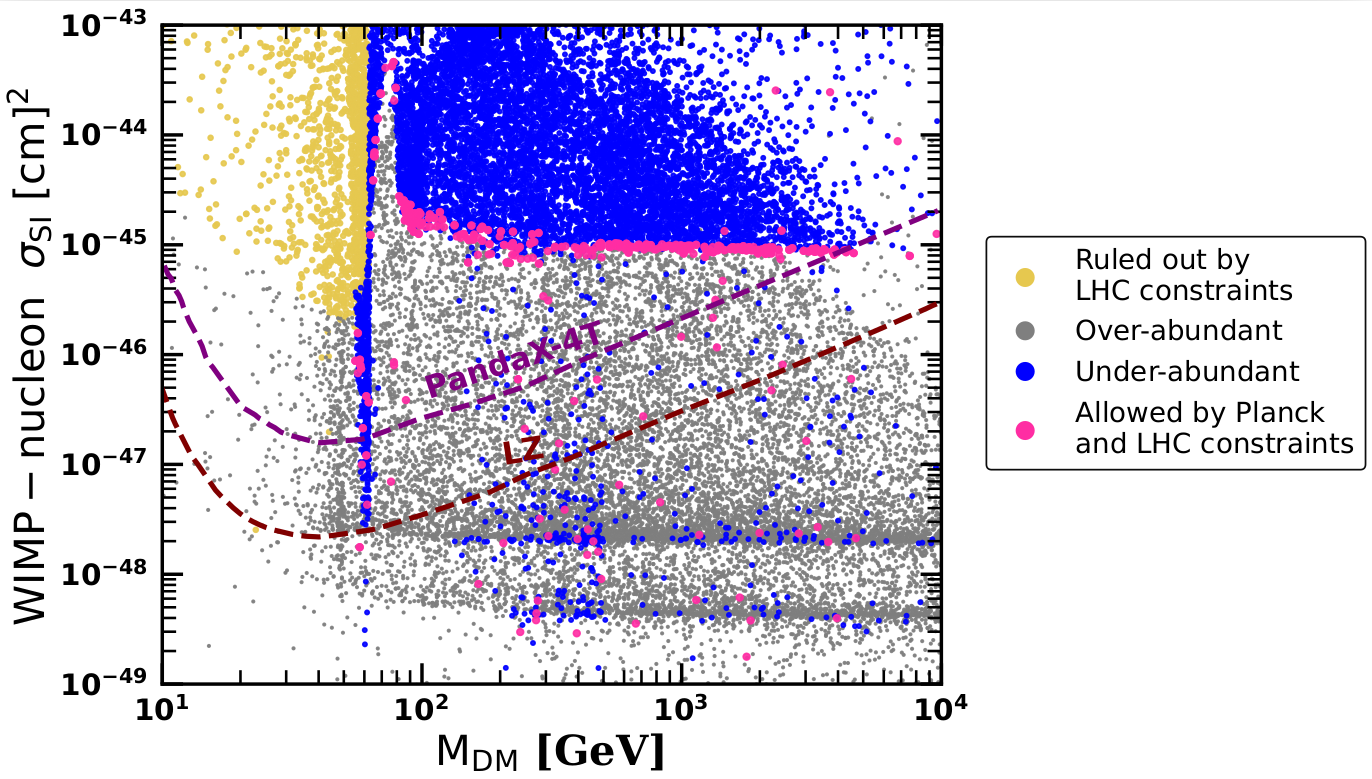}
    \caption{DM $\Omega h^2$ (left) and  $\sigma_{\text{SI}}$ (right) as functions of $M_{\rm DM}$, showing the full parameter scan for the mixed $S_1$-$S_2$ singlet DM scenario. Color coding is the same as Fig. \ref{fig:ZP_Dom_Scatter}. See text for further details. }
    \label{fig:Scalar_Dom_Scatter}
\end{figure}
%%%%%%%%%%%%%%%%%%

In addition to the new $H_2$-mediated annihilation channels, co-annihilation processes also play an important role. Having established the qualitative features of the $H_2$-mediated annihilation channels, we now turn to the exploration of the generic parameter space incorporating both annihilation and co-annihilation channels.
To illustrate the full parameter space of scalar singlet DM with scalar-dominated interactions in this model, we present in Fig.~\ref{fig:Scalar_Dom_Scatter} the variation of $\Omega h^2$ ($\sigma_{\rm SI}$) with $M_{\rm DM}$ for the mixed $S_1$-$S_2$ DM scenario.
In generating Fig.~\ref{fig:Scalar_Dom_Scatter}, all relevant scalar quartic couplings are scanned over the range $[10^{-4}, \sqrt{4\pi}]$, while the trilinear coupling $\kappa$ is varied within $[10^{-4}, 100]$. The gauge parameters are fixed at $g_x = 0.1$ and $M_{Z'} = 20~\text{TeV}$.
%%%
The yellow points are ruled out by current LHC constraints on invisible Higgs decays \cite{ATLAS:2023tkt}. 
Blue and gray points satisfy collider limits but correspond to under-abundant and over-abundant relic density, respectively, while magenta points satisfy all collider constraints and lie within the $3\sigma$ allowed range for cold DM.
From Fig.~\ref{fig:Scalar_Dom_Scatter}, it is evident that the DM parameter space spans a broad mass range, enabled by the presence of additional annihilation and co-annihilation channels.

\subsubsection{Singlet DM with mixed scalar and gauge interactions}

\begin{figure}[!h]
    \centering
        \includegraphics[width=0.405\linewidth]{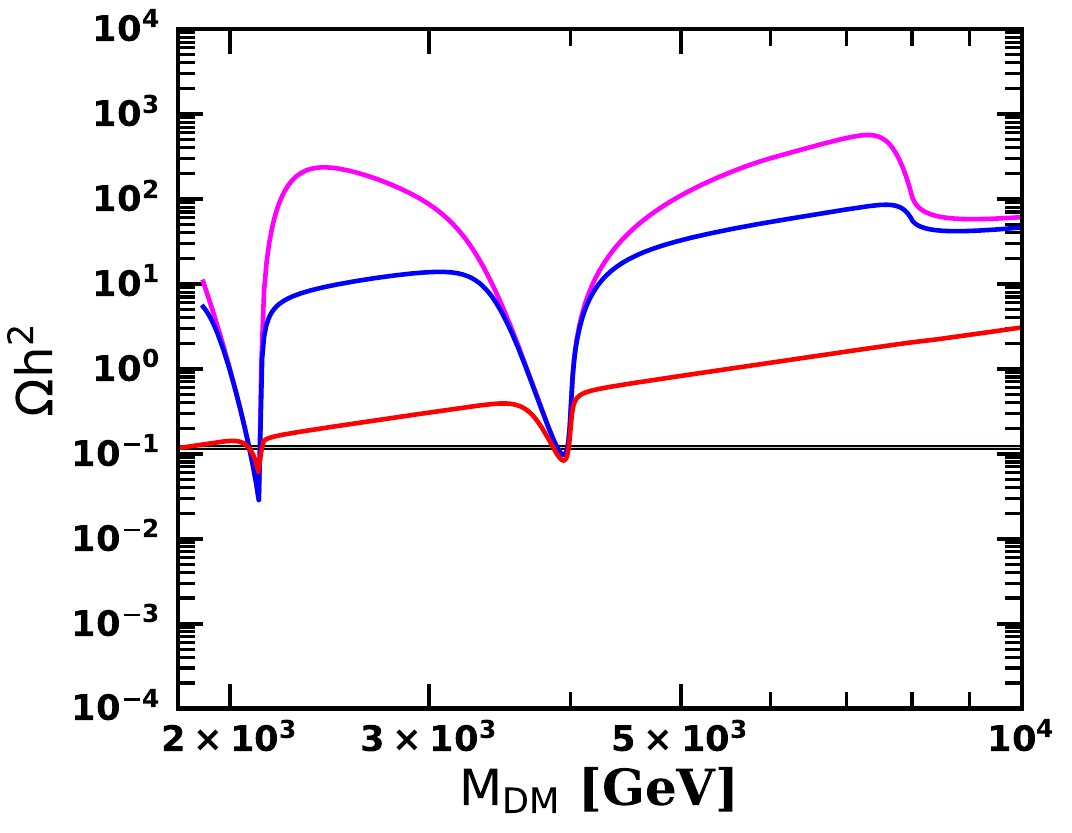}
        \includegraphics[width=0.56\linewidth]{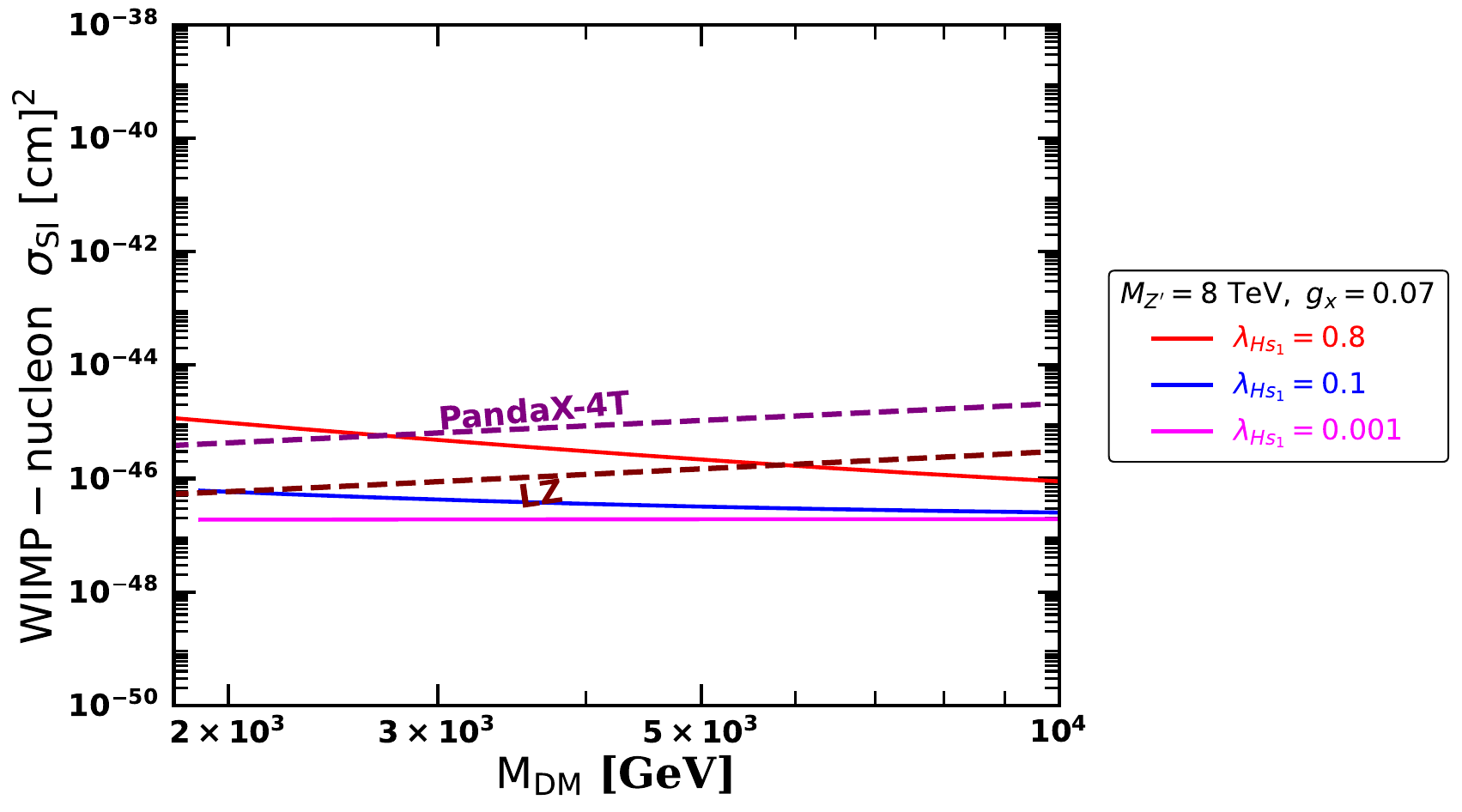}

        \includegraphics[width=0.405\linewidth]{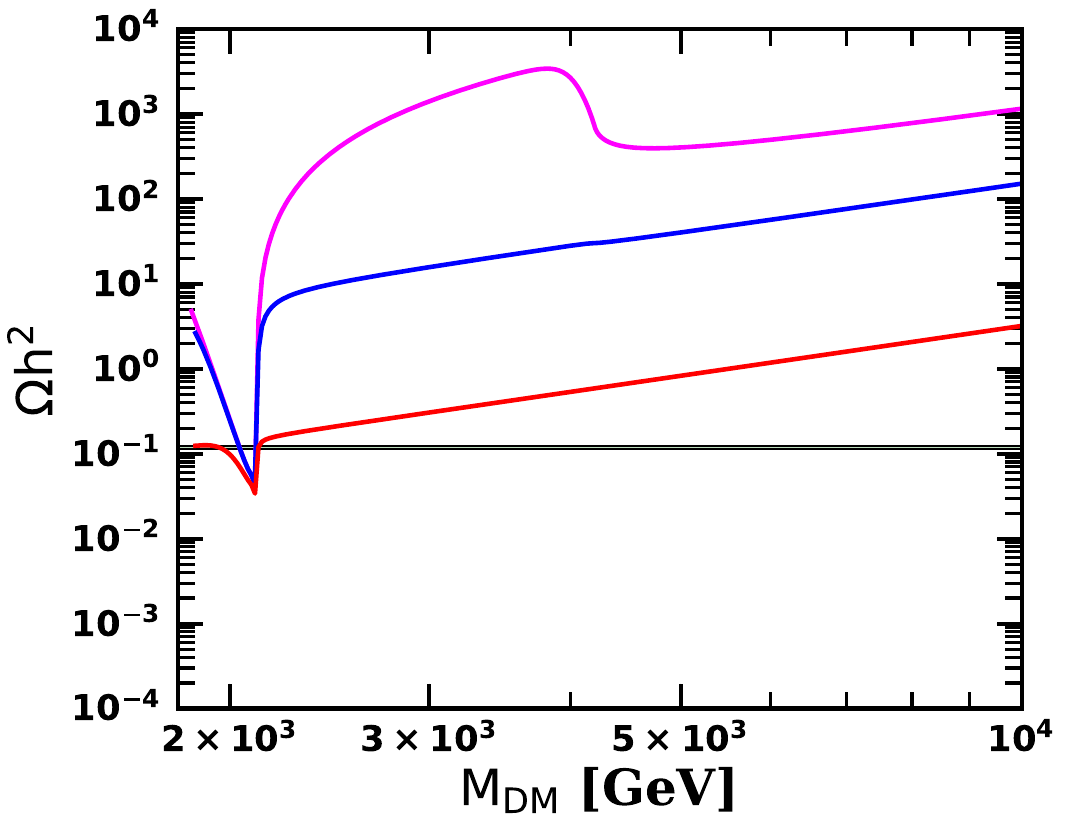}
        \includegraphics[width=0.56\linewidth]{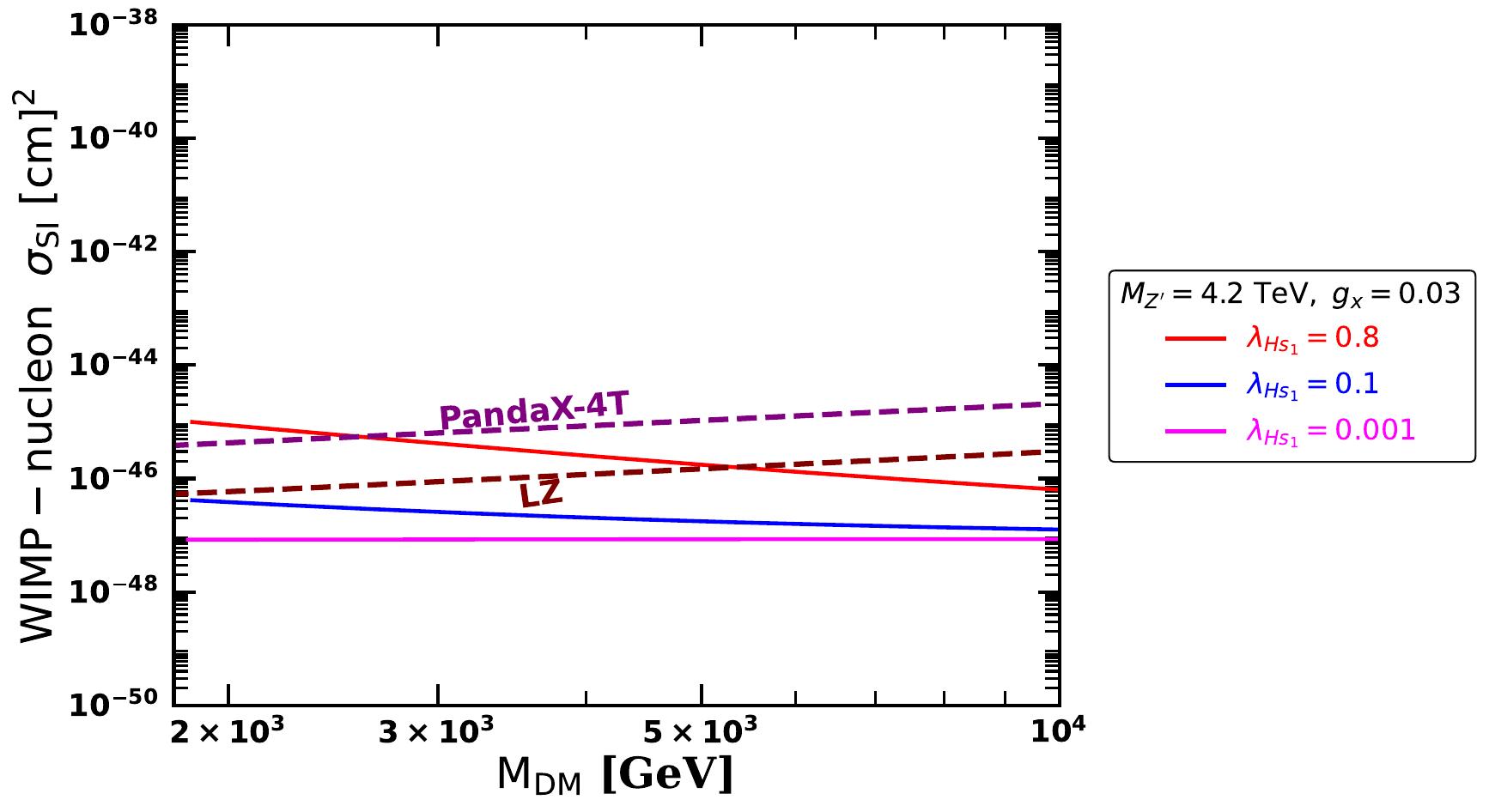}
    \caption{Same as Fig.~\ref{fig:Scalar_Dom_S1}, but including the effects of both $H_2$ and $Z^\prime$ resonances. Top and bottom panels correspond to different choices of $M_{Z'}$ and $g_x$.}
    \label{fig:ZP_Scalar_Dom_S1}
\end{figure}

Having analyzed the $Z^\prime$ and scalar portals independently, we now examine their combined impact to obtain a complete picture of the parameter space. We begin by considering the effect of the $Z^\prime$ portal on the $S_1$-dominated singlet DM scenario discussed in the previous section for the case of pure scalar interactions (see Fig. \ref{fig:Scalar_Dom_S1}). The heavy Higgs ($H_2$) mass and its mixing with $H_1$ are kept identical to those used earlier. 
%%%
To study the interplay with the $Z^\prime$ portal, we consider a relatively light $Z^\prime$ boson with a sizable gauge coupling.
In Fig.~\ref{fig:ZP_Scalar_Dom_S1}, the left and right panels display $\Omega h^2$ and $\sigma_{\rm SI}$, respectively, as functions of $M_{\rm DM}$ for the $S_1$-dominated scenario. 
From top to bottom, the rows correspond to different fixed values of the $Z^\prime$ mass and gauge coupling. 
In the first row, the $Z^\prime$ mass is fixed to be $8~\text{TeV}$.
In the second row, it is taken to be $4.2~\text{TeV}$ so that both the $H_2$ and $Z^\prime$ resonances overlap at the same DM mass position ($M_{\text{DM}} \sim M_{H_2}/2 = M_{Z^\prime}/2$).
The benchmark values of $\lambda_{\chi s_1}$ are again chosen to allow $M_{\text{DM}}$ as low as $1.8~\text{TeV}$, fully capturing the $H_2$ resonance region ($M_{\text{DM}} \sim M_{H_2}/2$).
Notably, because the singlet VEV $v_{\chi}$ is lower in these configurations than in the previous benchmark, the value for $\lambda_{\chi s_1}$ is slightly larger in these cases.
%%%%%%%%%%%%%%%%%%%%%%%%%%%%%%%%
%%%%%%%%%%%%%%%%%%
As seen from the top panels of Fig.~\ref{fig:ZP_Scalar_Dom_S1}, both the $H_2$ and $Z^\prime$ resonance peaks become a bit broader than in the previous scenarios.
In the bottom panels, where both the $H_2$ and $Z^\prime$ masses align, the resonances lie at the same position; consequently, the combined resonance peak broadens up further, though not significantly.
Across both configurations, the presence of the active $Z^\prime$ portal slightly improves the viable DM parameter space.
\begin{figure}[!h]
    \centering
        \includegraphics[width=0.43\linewidth]{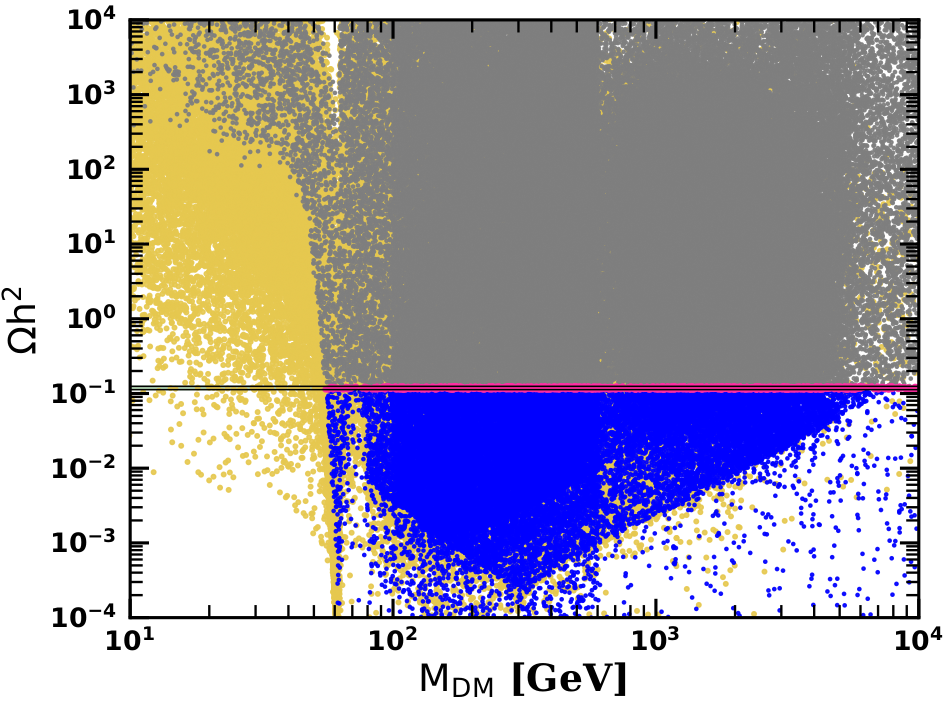}
        \includegraphics[width=0.56\linewidth]{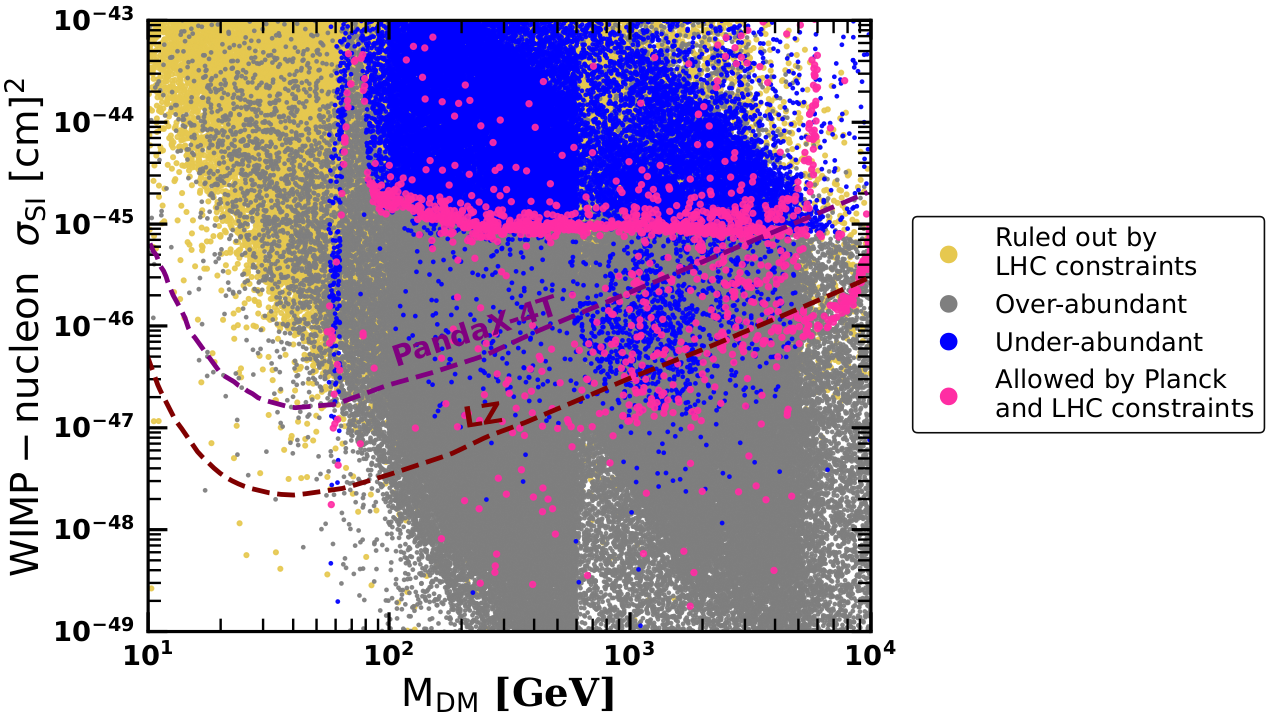}
        
    \caption{Same as Fig.~\ref{fig:Scalar_Dom_Scatter}, but including the effects of both scalar and $Z^\prime$ portals for the singlet-dominated mixed $S_1$–$S_2$ DM scenario. Color coding follows the same convention. See text for further details.}
    \label{fig:ZP_Scalar_Dom}
\end{figure}
%%%%%%%%%%%%%%%%%%
To conclude, we present in Fig.~\ref{fig:ZP_Scalar_Dom} the full parameter space where both the scalar and $Z^\prime$ portals contribute to the DM relic abundance and direct detection, for the singlet-dominated mixed $S_1$–$S_2$ DM scenario in our model. In generating Fig.~\ref{fig:ZP_Scalar_Dom}, all relevant scalar quartic couplings are scanned over the range $[10^{-4}, \sqrt{4\pi}]$, while the trilinear coupling $\kappa$ is varied within $[10^{-4}, 100]$. The gauge parameters are taken in the ranges $g_x \in [10^{-4}, 1]$ and $M_{Z'} \in [0.3, 20]~\text{TeV}$.
From Fig.~\ref{fig:ZP_Scalar_Dom}, it is evident that the inclusion of both $Z^\prime$ and scalar portals makes the viable DM parameter broad, extending from the vicinity of the SM Higgs resonance to the TeV scale.

\FloatBarrier

\subsection{Fermionic DM Scenario}
\begin{figure}[!h]
    \centering
        \includegraphics[width=0.43\linewidth]{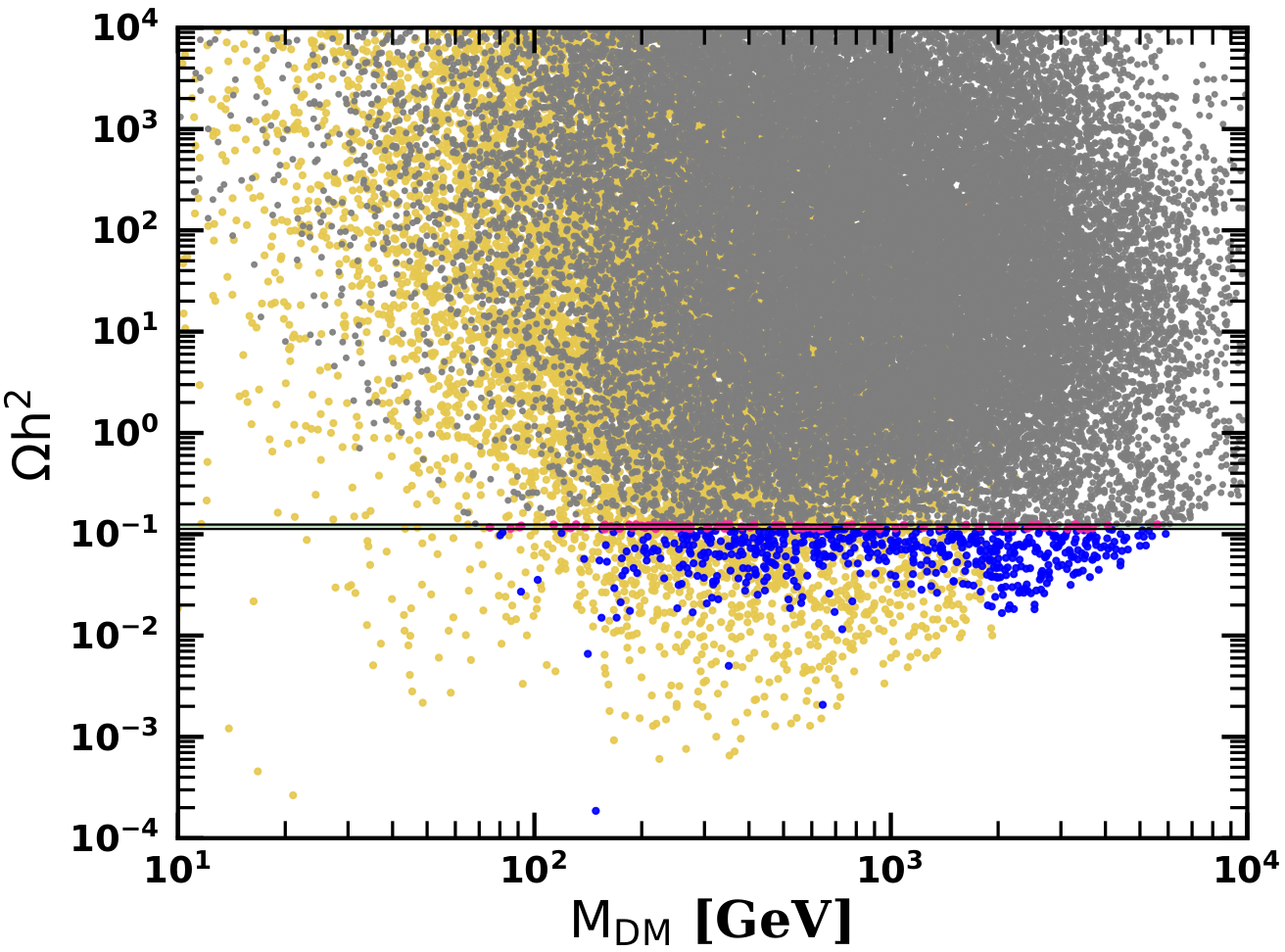}
        \includegraphics[width=0.56\linewidth]{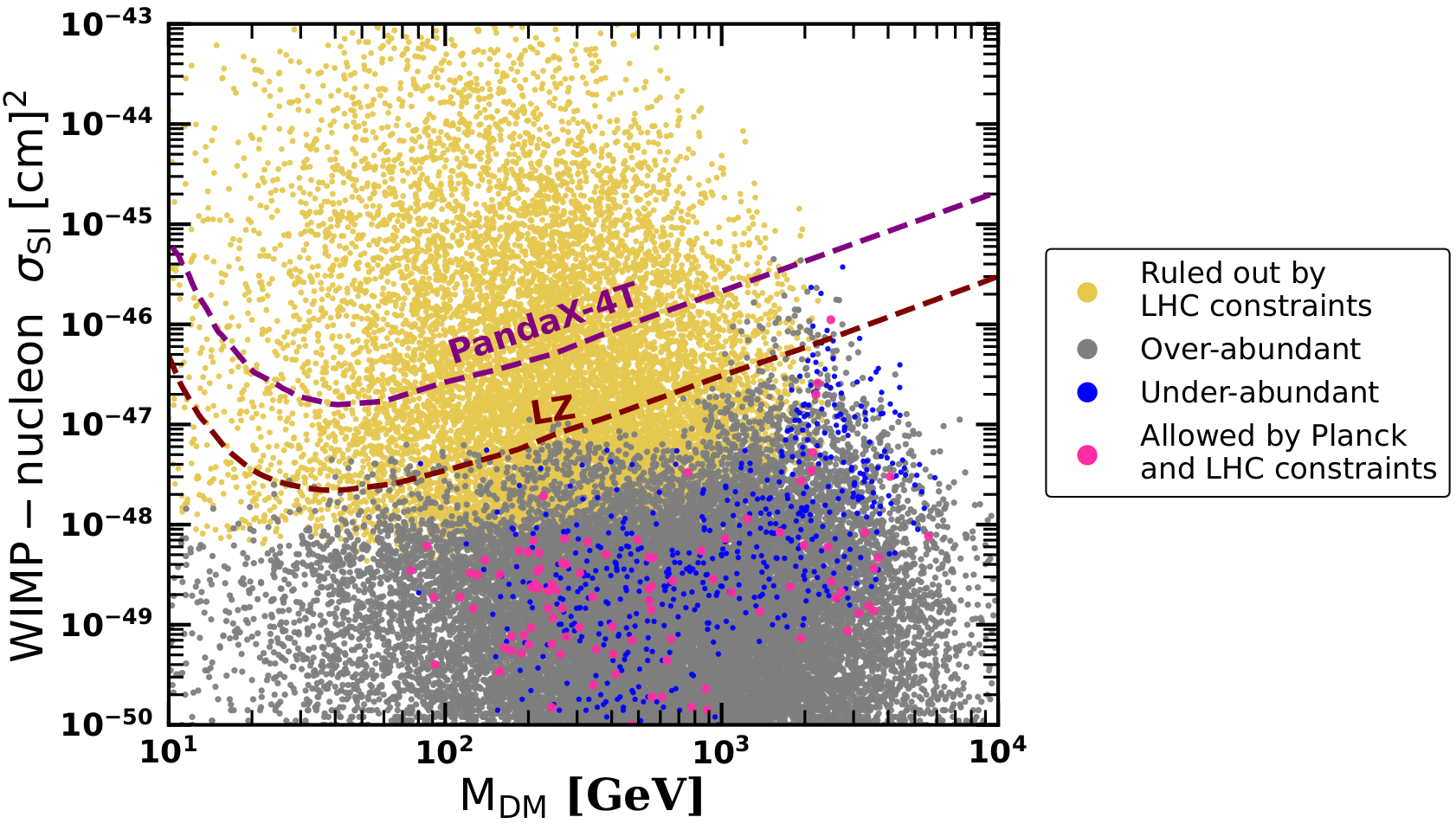}
    \caption{ Relic density (left) and spin-independent direct detection cross section (right) as functions of $M_{\rm DM}$. The color coding are same as Fig. \ref{fig:ZP_Scalar_Dom_S1}.  }
    \label{fig:ZP_Fermion_DM_Scatter}
\end{figure}

Finally, we consider the possibility of fermionic DM in this model.
The $Z_6$ odd fermion $f \equiv (f_L, f_R)$ serves as a viable fermionic DM candidate. We take $f$ to be the lightest particle in the $Z_6$ odd sector $(M_{f} < M_{\zeta_{p}}, M_{\eta^{\pm}})$, thereby ensuring its stability. 
%%
%%%
In Fig.~\ref{fig:ZP_Fermion_DM_Scatter}, we present $\Omega h^2$ ($\sigma_{\rm SI}$) in the left (right) panel as a function of the DM mass $M_{\rm DM}$. The yellow points are excluded by collider constraints from the LHC, as discussed in Sec.~\ref{sec:Zprime}. The gray and blue points satisfy collider constraints but correspond to over-abundant and under-abundant regions, respectively. The magenta points are consistent with both Planck and LHC data.

%%%%
%%
%%%%%%%%%%%%%%%%%%

The fermionic DM candidate exhibits both $Z'$-mediated and scalar-mediated annihilation as well as co-annihilation channels, which contribute to its relic abundance. In contrast, direct detection is predominantly governed by $Z'$ mediated interactions, as the fermionic DM does not couple to quarks at tree level via scalar mediators. 
Consequently, in the zero-momentum transfer limit, the relevant direct detection cross section reduces to the spin-independent result as provided in Eq.~\ref{Eq:DD:SpinI_Zp_Crc}.
%%%
From Fig.~\ref{fig:ZP_Fermion_DM_Scatter}, it is evident that the fermionic DM candidate also reproduces the observed relic density, as measured by Planck, while simultaneously evading current direct detection bounds over a wide range of DM masses.

%%%%%%%%%%%%%%%%%%%%%%%%%%%%%%%%%%%%%%%%%%%%%%%%%%%%%%%%%%%%%%%
\FloatBarrier
\section{Conclusion} \label{sec:conc}

In this work, we proposed a simple Dirac scoto-seesaw framework that simultaneously explains the neutrino mass generation and DM stability. Owing to the anomaly-free chiral $U(1)_{B-L}$ charge assignment $(-4,-4,5)$ of $\nu_{R_i}$, the model naturally realizes the observed two mass-squared differences. The two $-4$ charged $\nu_{R}$ generate a tree level seesaw contribution responsible for $\Delta m^2_{\rm atm}$, whereas the $5$ charged state contributes only through the scotogenic loop, leading to $\Delta m^2_{\rm sol}$. The spontaneous breaking of $U(1)_{B-L}$ via the VEV of the newly added singlet scalar $\chi$ leaves a residual $Z_6$ symmetry that guarantees DM stability.
In addition, we introduce fermions $N \equiv (N_L,N_R)$ and $f \equiv (f_L,f_R)$, together with a doublet scalar $\eta$ and singlet scalars $S_1,S_2$. The fields $(f,\eta,S_1,S_2)$ are odd under the residual $Z_6$ symmetry, whereas all other particles are even. As a result, the lightest $Z_6$ odd state is stable and serves as a viable DM candidate. 

In the neutrino sector, we study two realizations of the framework: case-I with one generation of $N$, which yields two massive neutrinos, and case-II with two generations of $N$, where all three neutrinos acquire masses.
Our analysis shows that both NO and IO are allowed in case-I, while only NO remains viable in case-II. The BF values correspond to $\chi^2_{\rm min}=1.475~(9.033)$ for NO (IO) in case-I and $\chi^2_{\rm min}=0.536$ for NO in case-II. The presence of one massless neutrino in case-I leads to sharp predictions for $\sum m_i$ and $m_\beta$. 
In contrast, case-II features three massive neutrinos and exhibits the correlations among $\sum m_i$, $m_\beta$, and the neutrino masses. Consequently, both $m_\beta$ and $\sum m_i$ can attain larger values than in case-I.
The preferred BF value of $\delta_{\rm CP}$ is close to CP conservation for NO and maximal CP violation for IO in case-I, while case-II favors nearly maximal CP violation.

Furthermore, the gauged $U(1)_{B-L}$ symmetry gives rise to a neutral gauge boson $Z'$, accessible at collider experiments. Using recent ATLAS dilepton searches, we find that the corresponding mass limits are weaker than those in the conventional vector $B-L$ scenario. This can be attributed to the chiral charge assignment $(-4,-4,5)$ of $\nu_{R_i}$, which enhances the invisible decay width of $Z'$ and consequently suppresses its dilepton branching fraction.
Turning to the dark sector, the scalar DM candidate $\zeta_1$ originates from the mixing of the $Z_6$ odd scalars $(\eta,S_1,S_2)$. While the doublet-like regime is already known to satisfy DM constraints via co-annihilation channels, we focus on the singlet-dominated scenario. We show that the $Z'$ and scalar portals significantly enrich the DM phenomenology compared to the minimal singlet scalar DM framework by opening additional annihilation and co-annihilation channels. As a result, the observed relic abundance can be achieved while satisfying current collider and direct detection constraints over a broad mass range, extending well beyond the Higgs resonance region.  
Finally, we show that the fermionic DM candidate $f$ also reproduces the observed relic density while evading current direct detection bounds over a wide DM mass range.
Notably, part of the viable parameter space remains within reach of future direct detection experiments for both singlet scalar and fermionic DM.

In summary, the simplest Dirac scoto-seesaw framework built upon the anomaly-free $(-4,-4,5)$ chiral charge assignment naturally explains the origin of the two neutrino mass scales and realizes a viable scalar and fermionic DM scenario. Extending the framework with flavor symmetries may lead to additional neutrino sector correlations and remains an interesting avenue for future study.

\section*{Acknowledgments}
\noindent
 SKK and RK are supported by the National Research Foundation of Korea under Grant NRF-2023R1A2C100609111. The work of HKP is supported by the Prime Minister Research Fellowship (ID: 0401969). HKP also acknowledges the hospitality of SeoulTech University during his visit, where part of this work was completed.
%%%%%%%%%%%%%
\FloatBarrier
\appendix
%%%%%%%%%%%%

\section{Scalar Sector and Mass Spectrum} \label{sec:scalar}

In this section, we present the scalar potential of the model and derive the corresponding scalar mass spectrum. The scalar sector consists of the $SU(2)_L$ doublets $H$ and $\eta$, together with the $SU(2)_L$ singlet scalars $\chi$, $S_1$, and $S_2$.
Following the charge assignment of scalars provided in Table \ref{tab:field_content}, the scalar potential of the model is given as follows
\begin{align}\label{eq:scalar_Potential}
   % \begin{split}
         \mathcal{V}_{\mathtt{S}} &=  m_{H}^2 H^{\dagger}H + \frac{\lambda_1}{2} (H^{\dagger}H)^{2} + m_{\eta}^2 \eta^{\dagger}\eta + \frac{\lambda_2}{2} (\eta^{\dagger}\eta)^{2}+ \lambda_{3}(H^{\dagger}H)(\eta^{\dagger}\eta)+ \lambda_{4} (H^{\dagger} \eta)(\eta^{\dagger}H) \nonumber \\
        & + m_{\chi}^2 (\chi^{*} \chi) + \frac{\lambda_{\chi}}{2} (\chi^{*} \chi)^2  + m_{s_i}^2 (S_{i}^{*} S_{i}) + \frac{\lambda_{s_{i}}}{2} (S_{i}^{*} S_{i})^2 + \lambda_{H\chi} (H^{\dagger}H) (\chi^{*} \chi) + \lambda_{\eta \chi} (\eta^{\dagger}\eta)(\chi^{*} \chi) \nonumber \\& + \lambda_{H s_i} (H^{\dagger}H)(S_{i}^{*} S_{i}) + \lambda_{\eta s_i} (\eta^{\dagger}\eta)(S_{i}^{*} S_{i}) + \lambda_{\chi s_i}(\chi^{*} \chi)(S_{i}^{*} S_{i})+ \lambda_{s_1 s_2} (S_{1}^{*} S_{1}) (S_{2}^{*} S_{2}) \nonumber  \\
        &  + ( \lambda_6 H^{\dagger} \eta \chi^{*} S_2^{*} + \kappa S_2 \chi^{*} S_{1}^{*} + \text{H.c.} )\,.
   % \end{split}
\end{align}
 Among the scalar fields, only $\eta$, $S_1$, and $S_2$ participate in the loop and they are odd under the residual $Z_6$ symmetry. Whereas $H$ and $\chi$ transform trivially under $Z_6$. The residual $Z_6$ symmetry emerges from the spontaneous breaking of the $U(1)_{B-L}$ gauge symmetry through the VEV of $\chi$. Simultaneously, the electroweak symmetry is broken by the VEV of $H$. In contrast, the remaining scalar fields do not acquire VEVs, ensuring the preservation of the residual $Z_6$ symmetry. Consequently, the lightest $Z_6$ odd particle is stable and can serve as a viable DM candidate.

Upon spontaneous symmetry breaking, the fields can be written as
%%%%
\begin{equation}\label{eq:Field_Expansion}
    H = \begin{pmatrix}
        G^{+} \\ \frac{v_{H} + h + iG^{0}}{\sqrt{2}} 
    \end{pmatrix},~     
    \chi = \frac{1}{\sqrt{2}} (v_{\chi} + R_{\chi} + i I_{\chi})\,,~ \eta = \begin{pmatrix}
        \eta^{+} \\ \eta_{0} 
    \end{pmatrix},
\end{equation}
%%%%
By solving the minimization conditions, the mass parameters $m_H^2$ and $m_{\chi}^2$ can be expressed as
%%%%
\begin{equation}\label{eq:Tadpole}
\begin{split}
    & 2 m_{H}^{2} + \lambda_{1}v_{H}^2 +  \lambda_{H \chi} v_{\chi}^{2} = 0 \\
    & 2 m_{\chi}^{2}  + \lambda_{\chi} v_{\chi}^{2} + \lambda_{H \chi} v_{H}^2 = 0 \,.
\end{split}    
\end{equation}
We now determine the scalar mass spectrum of the model. Since $H$ and $\chi$ are even under the residual $Z_6$ symmetry, their components mix with one another. Likewise, the $Z_6$ odd scalars $\eta$, $S_1$, and $S_2$ mix among themselves. Following electroweak symmetry breaking, $G^{\pm}$ are absorbed as the Goldstone bosons associated with the $W^{\pm}$ gauge bosons, while the physical charged scalar $\eta{^\pm}$ acquires a mass given by
%%%%
\begin{equation}
    M_{\eta^{\pm}}^{2} = m_{\eta}^{2} + \frac{1}{2} ( v_{H}^{2} \lambda_{3} + v_{\chi}^{2} \lambda_{\eta \chi})\,.
\end{equation}
%%%%
The CP even scalars $h$ and $R_{\chi}$ mix with each other, leading to the following mass matrix in the $(h,R_\chi)^T$ basis,
%%%%
\begin{equation}\label{Eq:App:Higgs_Mixing}
    \mathcal{M}^{2}_{H} = \begin{pmatrix}
        \lambda_{1}v_{H}^2 & v_{H} v_{\chi} \lambda_{H\chi} \\
        v_{H} v_{\chi} \lambda_{H\chi} & \lambda_{\chi}v_{\chi}^2
    \end{pmatrix}\, = 
    \begin{pmatrix}
        A & C  \\
        C & B
    \end{pmatrix}\,.
\end{equation}
%%%%
The mass eigenvalues of light and heavy mass eigenstates as
%%%%%%%%%%%%%%%%%%%%%%%%%%%%%%%%%%%%%%%%%%%%%%%%%
\begin{align}
M^{2}_{H_{1}} & =\frac{1}{2}\left[A+B-\sqrt{(A-B)^2+4C^2}\right], \\
M^{2}_{H_{2}} & =\frac{1}{2}\left[A+B+\sqrt{(A-B)^2+4C^2}\right].
\end{align}
We follow the convention $M_{H_{1}} < M_{H_{2}}$ and have identified $H_{1}$ as the SM Higgs, with mass $M_{H_{1}}=125$~GeV. The two mass eigenstates $H_{1}, H_{2}$ are related with the $(h, R_{\chi})$ fields through the following rotation matrix as
%%%%%%%%%%%%%%%%%%%%%%%%%%%%%%%%%%%%%%%%%%%%%%%%%%%%%%%%%
\begin{equation}
\begin{bmatrix}
H_{1} \\
H_{2}  \end{bmatrix} = \begin{bmatrix}
\cos\alpha & -\sin\alpha \\
\sin\alpha & \cos\alpha
\end{bmatrix} \begin{bmatrix}
h \\
R_{\chi}
\end{bmatrix}, \,\, \text{with}\,\, \tan 2\alpha=\frac{2C}{B-A}.
\end{equation} 
%%%%%%%%%%%%%%%%%%%%%%%
%%%%%%%%%%%%%%%%%%%%%%%

In the $Z_6$ odd scalars $(\eta_0,S_1,S_2)$ mix with one another, giving rise to the mass matrix
%%%
\begin{equation}  \mathcal{M^{\mathrm{2}}_{\text{DM}}} = \begin{pmatrix}
        m_{\eta}^{2} +\frac{v_{H}^2}{2} (\lambda_{3} + \lambda_{4}) + \lambda_{\eta \chi} \frac{v_{\chi}^{2}}{2} &  0 & ~~~~\frac{v_{H} v_{\chi}}{2} \lambda_{6} \\
        0 & m_{s_{1}}^{2} + \frac{1}{2}( \lambda_{Hs_1} v_{H}^{2} + \lambda_{\chi s_1 }v_{\chi}^{2}) & ~~~~\kappa \frac{v_{\chi}}{\sqrt{2}} \\
        \frac{v_{H} v_{\chi}}{2} \lambda_{6} & \kappa \frac{v_{\chi}}{\sqrt{2}} ~~~ & m_{s_{2}}^{2} + \frac{1}{2}( \lambda_{Hs_2} v_{H}^{2} + \lambda_{\chi s_2 }v_{\chi}^{2}) 
    \end{pmatrix}
\end{equation}
%%%
This mass matrix is rotated to the mass basis by an orthogonal matrix $\mathcal{O}$. The mass eigenstates $\zeta_{p}$ and the gauge eigenstates $H^{D} = (\eta_{{0}}, S_1, S_2)$ are related as
$\zeta_{p} = \mathcal{O}[p,q] H_{q}^{D}$, with $p,q = 1,2,3$.
%%%

In the gauge sector, since the SM Higgs doublet is not charged under the $U(1)_{B-L}$ gauge group, there is no tree level mass mixing between the $Z$ and $Z^\prime$ bosons.
Furthermore, we assume the kinetic mixing between the hypercharge $U(1)_Y$ and $U(1)_{B-L}$ gauge fields to be negligibly small and explicitly set it to zero.
Consequently, the physical gauge boson masses are given by,
\begin{equation}
    M_{Z}^{2} = \frac{v_{H}^2}{4} (g^2 + g'^2),~ M_{W}^{2} = \frac{v_{H}^2 g^2}{4},~M_{Z'}=3 v_{\chi} g_{x}\,,
\end{equation}
where $g'$, $g$ and $g_x$ are hypercharge, $SU(2)_{L}$ and $U(1)_{B-L}$ gauge couplings, respectively.

%%%%%

\section{$S_2$ dominated singlet DM Scenario}\label{App:S2_Dom}

In this section, we present the results for the $S_2$-dominated singlet DM scenario. For a direct comparison with the $S_1$-dominated case discussed in the main text, we adopt the same benchmark values as used in sec. \ref{sec:dm}.
%%%%
%%%%%
\begin{figure}[!h]
    \centering
    \includegraphics[width=0.425\linewidth]{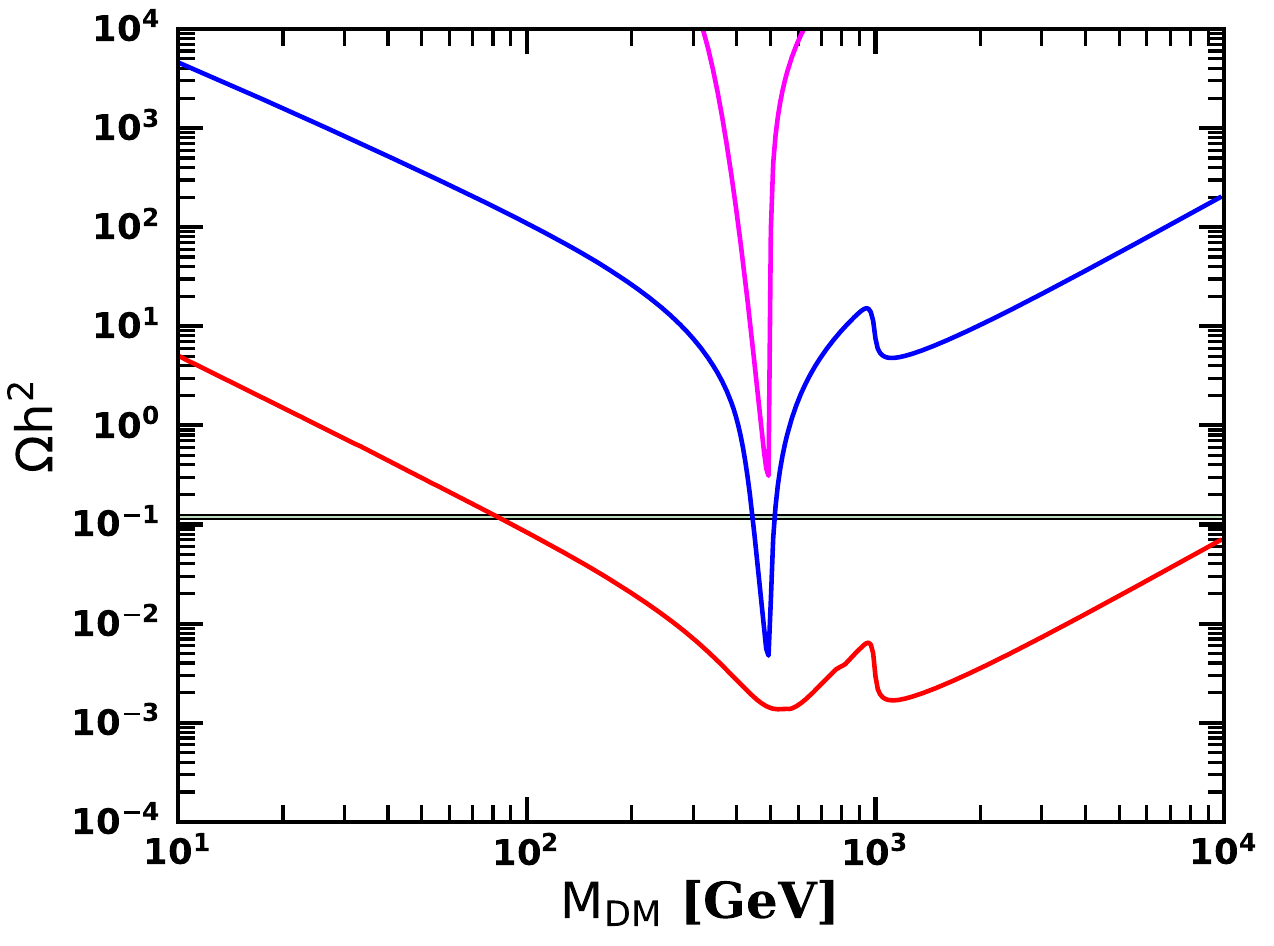}    \includegraphics[width=0.49\linewidth]{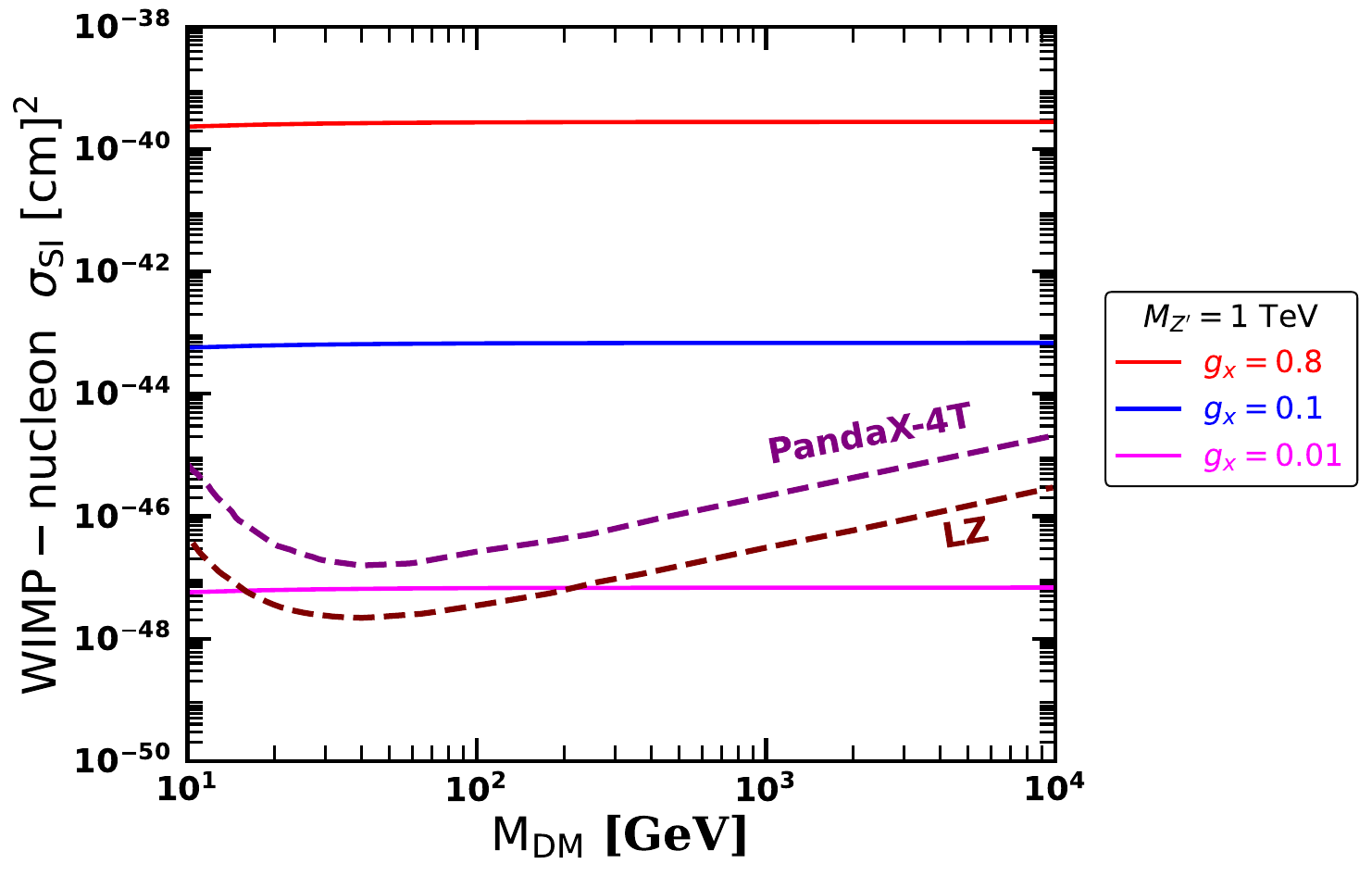}
    \includegraphics[width=0.425\linewidth]{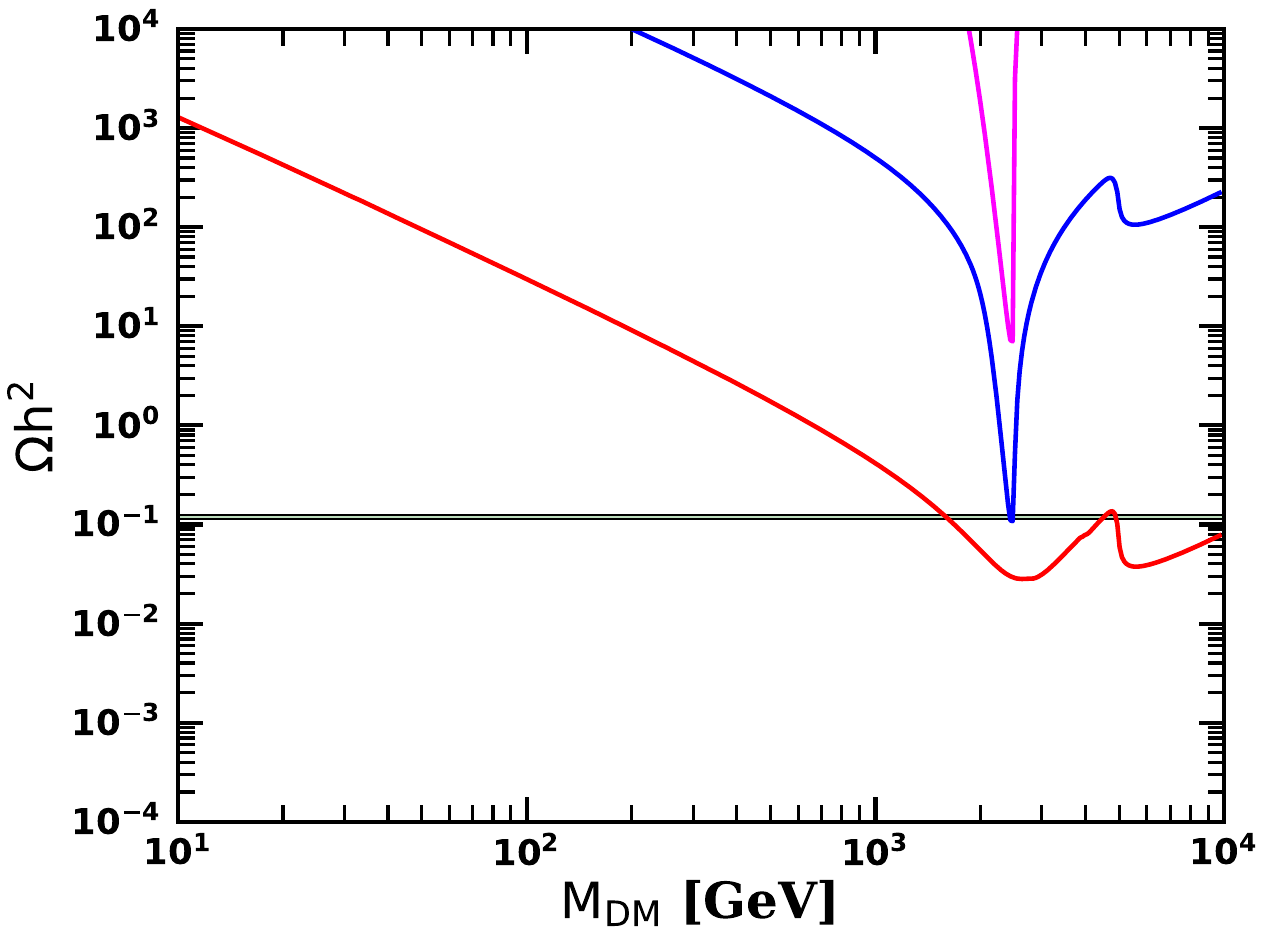}    \includegraphics[width=0.49\linewidth]{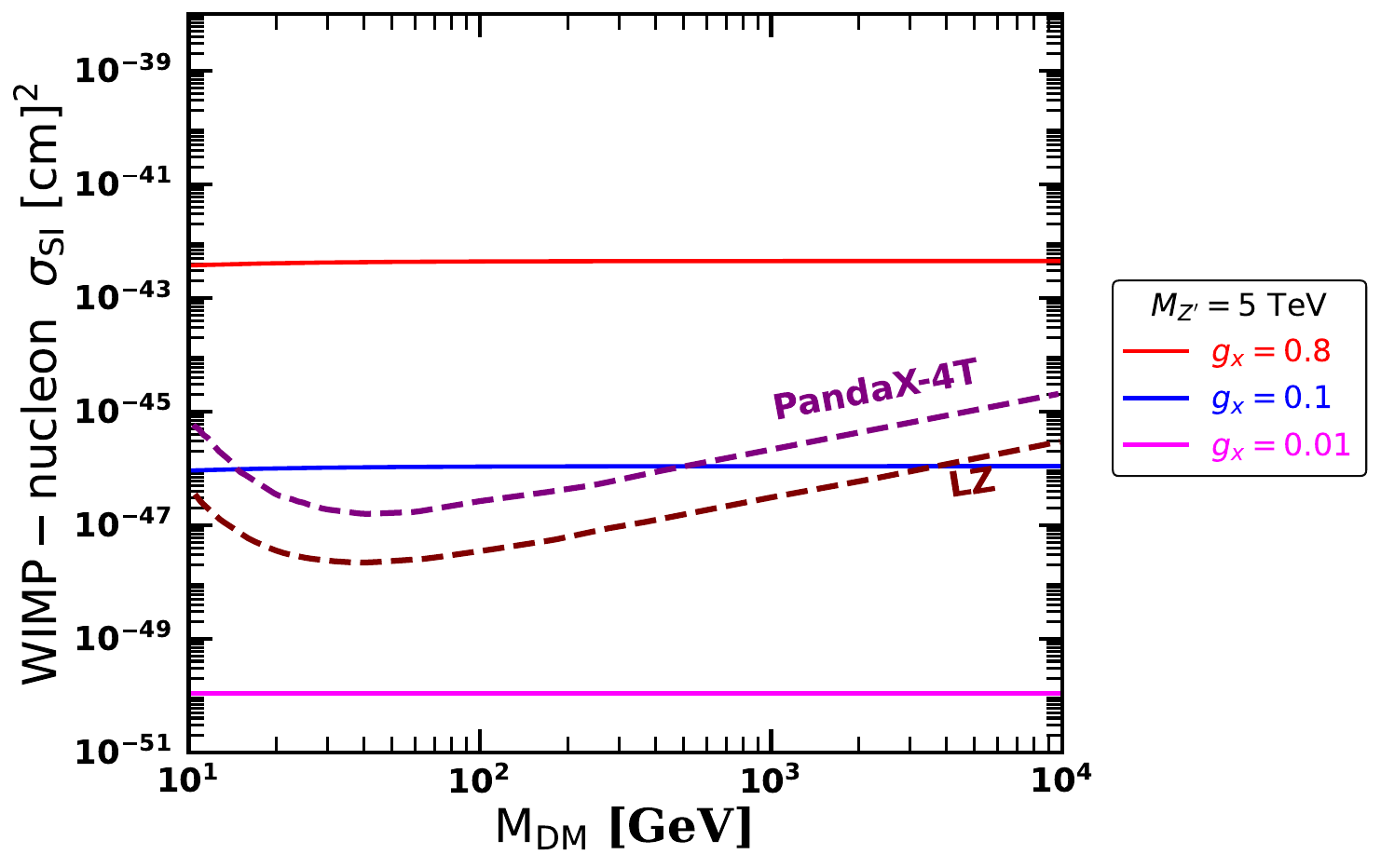}
    \includegraphics[width=0.425\linewidth]{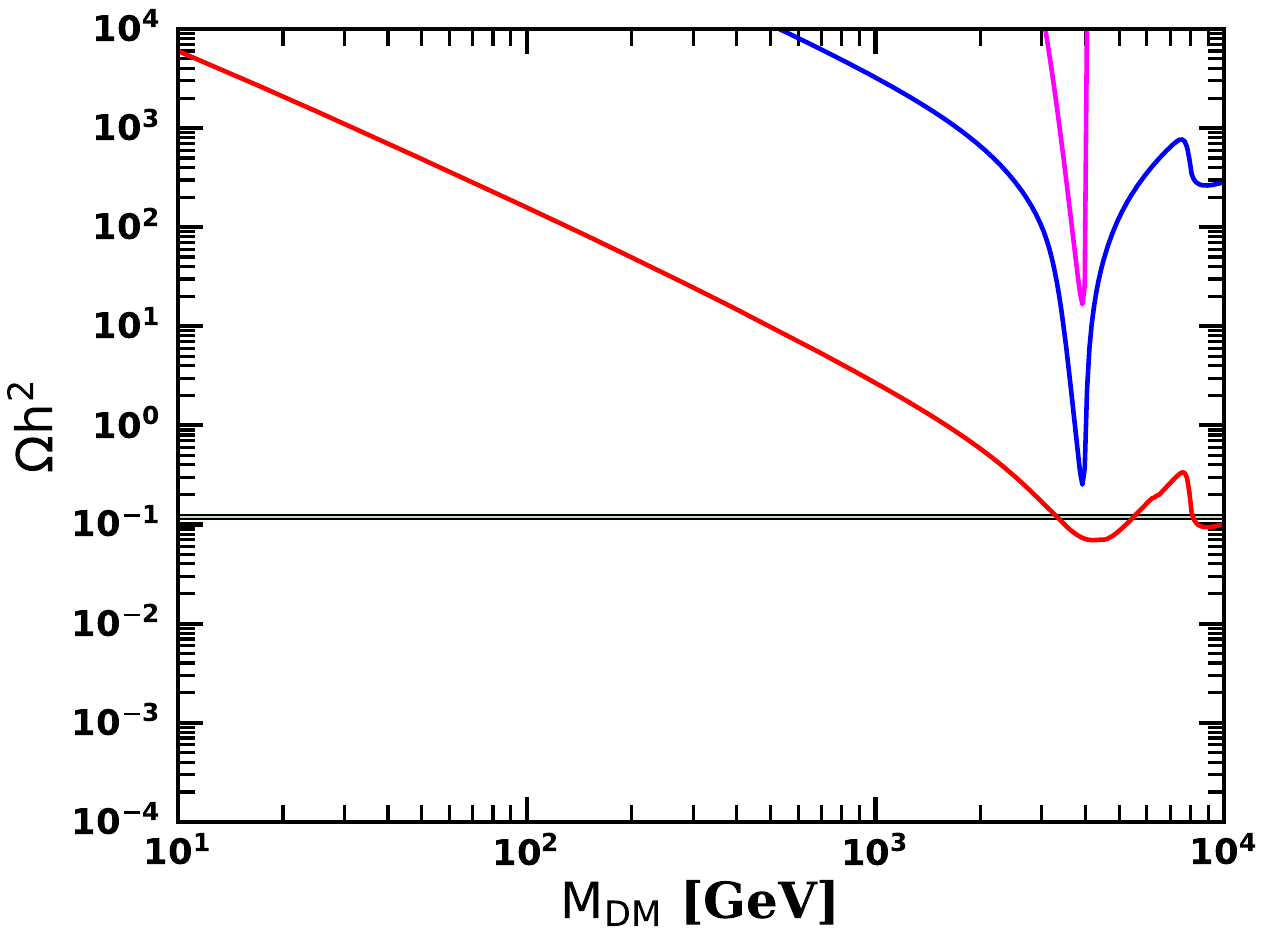}    \includegraphics[width=0.49\linewidth]{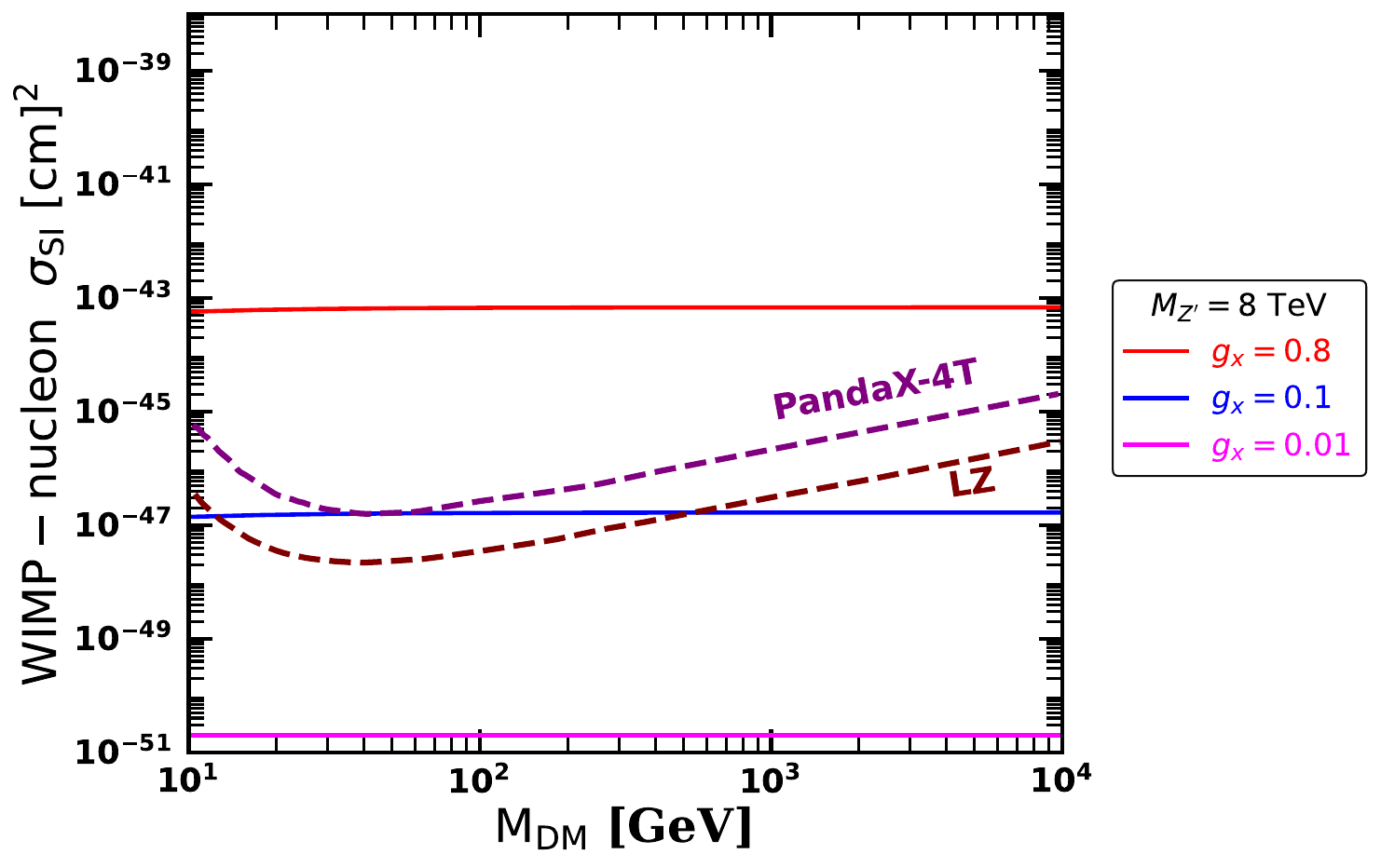}
    \caption{Relic density (left panels) and spin-independent WIMP-nucleon scattering cross
section (right panels) as a function of the DM mass $M_{\text{DM}}$ for the $S_2$-dominated singlet DM
scenario. Benchmark cases are same as Fig. \ref{fig:OnlyZP_S1DM}.}
    \label{fig:OnlyZP_S2DM}
\end{figure}
%%%%%
We begin with the $S_2$-dominated singlet DM scenario with pure gauge interactions.
In Fig.~\ref{fig:OnlyZP_S2DM}, we show the same quantities as in Fig.~\ref{fig:OnlyZP_S1DM} but for the $S_2$-dominated singlet DM case. The qualitative behavior remains similar. The only notable difference is that, in the $S_2$ case, the relic density is relatively larger while the spin-independent direct detection cross section is relatively smaller. This is due to the smaller $U(1)_{B-L}$ charge of $S_2$ compared to $S_1$.
%%%

%%
\begin{figure}[!h]
    \centering
        \includegraphics[width=0.405\linewidth]{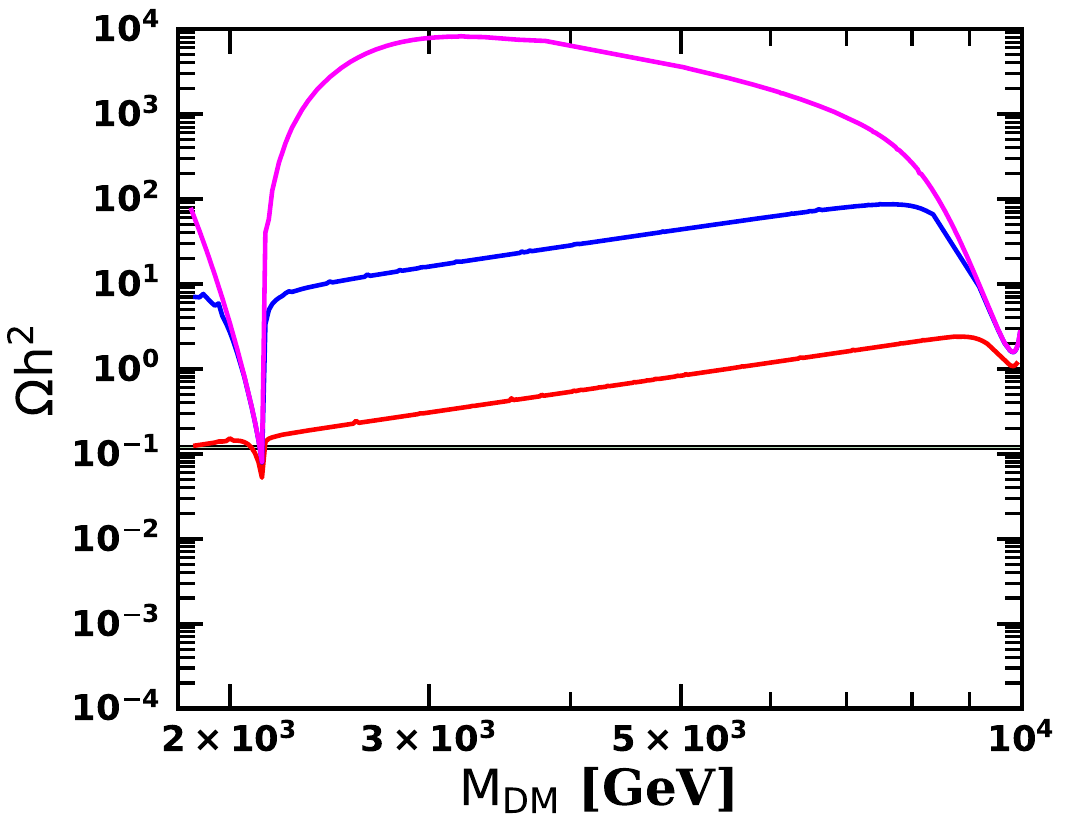}
        \includegraphics[width=0.56\linewidth]{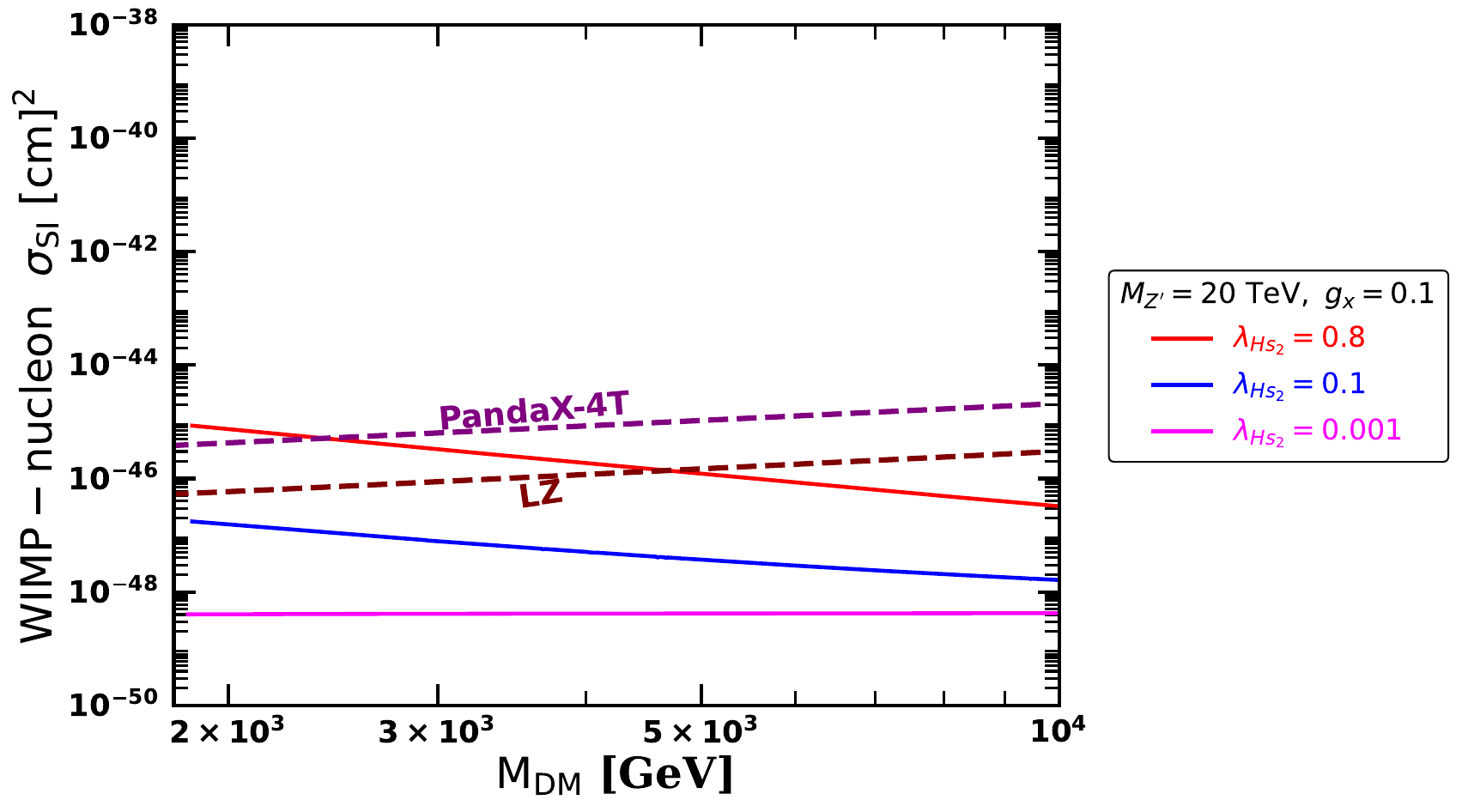}
    \caption{Relic density (left panels) and spin-independent WIMP-nucleon scattering cross section (right panels) as a function of the DM mass $M_{\text{DM}}$ for the $S_2$-dominated singlet DM scenario with scalar-dominated interactions. Benchmark cases are same as Fig. \ref{fig:Scalar_Dom_S1_Low}.}
    \label{fig:Scalar_Dom_S2}
\end{figure}
%%%%%%%%%%%%%%%%%%

The behavior of the $S_2$-dominated singlet DM scenario with scalar-dominated interactions is shown in Fig.~\ref{fig:Scalar_Dom_S2}.
All parameter benchmarks and relevant couplings for this scenario are chosen to be identical to those in the $S_1$-dominated case presented in Fig. \ref{fig:Scalar_Dom_S1}.
As illustrated, the qualitative behavior remains identical to that in Fig.~\ref{fig:Scalar_Dom_S1}. 
%%%
Because the scalar sectors are identical, the position of the $H_2$ resonance is unchanged. 
However, the relic density at the $Z^\prime$ resonance dip is slightly higher than in the previous scenario because $S_2$ carries a smaller $U(1)_{B-L}$ charge than $S_1$. 
%%%
This suppressed gauge interaction also manifests in the direct detection plots: for the benchmark with the smallest Higgs-DM coupling ($\lambda_{H S_2} = 0.001$), where the cross section is primarily dominated by $Z^\prime$ exchange, the direct detection cross section is lower than in the $S_1$ case. 
As before, both the observed relic density and direct detection limits can be simultaneously satisfied for $\lambda_{H S_2} \lesssim 0.1$.

\begin{figure}[!h]
    \centering
        \includegraphics[width=0.405\linewidth]{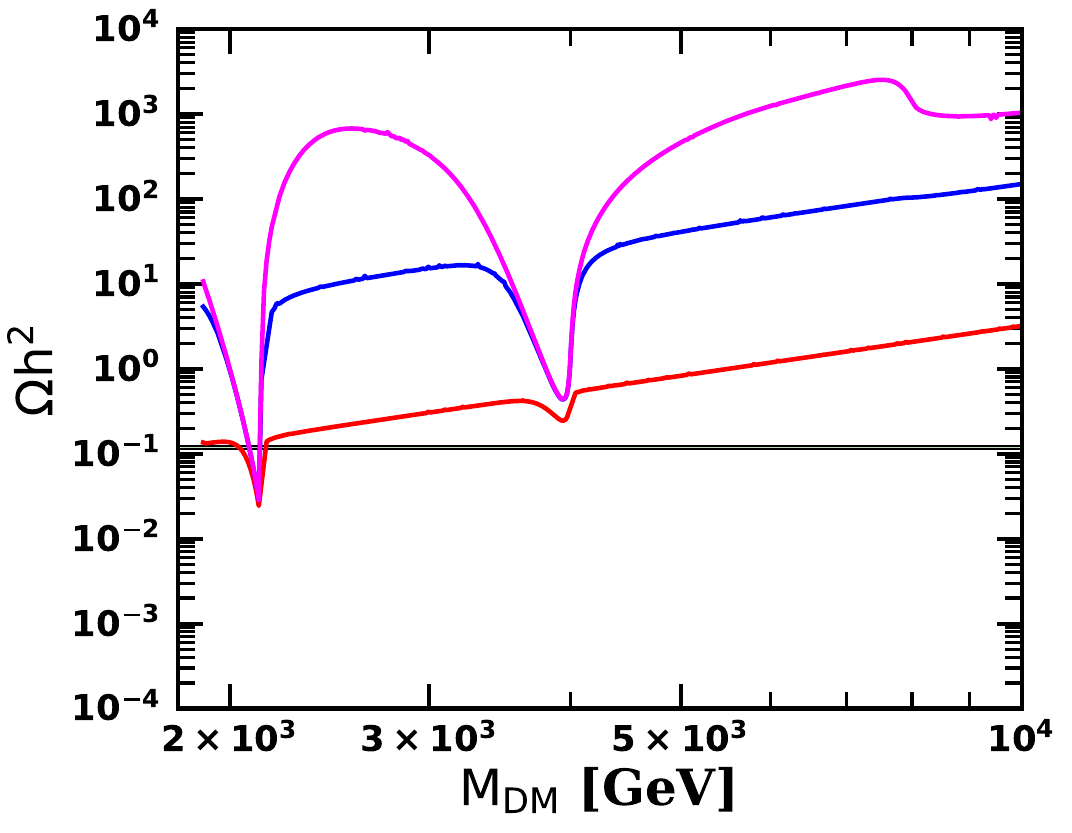}
       \includegraphics[width=0.56\linewidth]{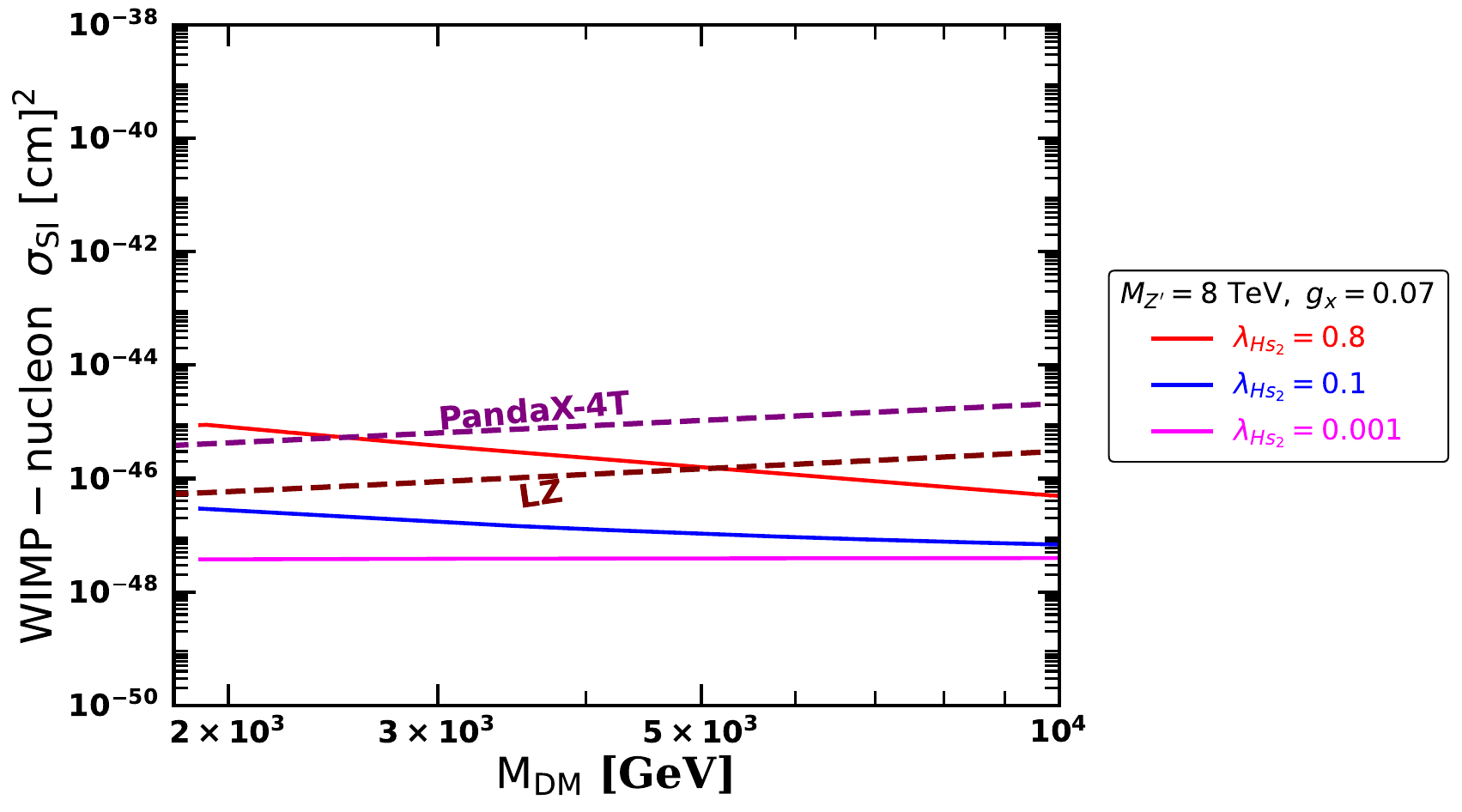}

   \includegraphics[width=0.405\linewidth]{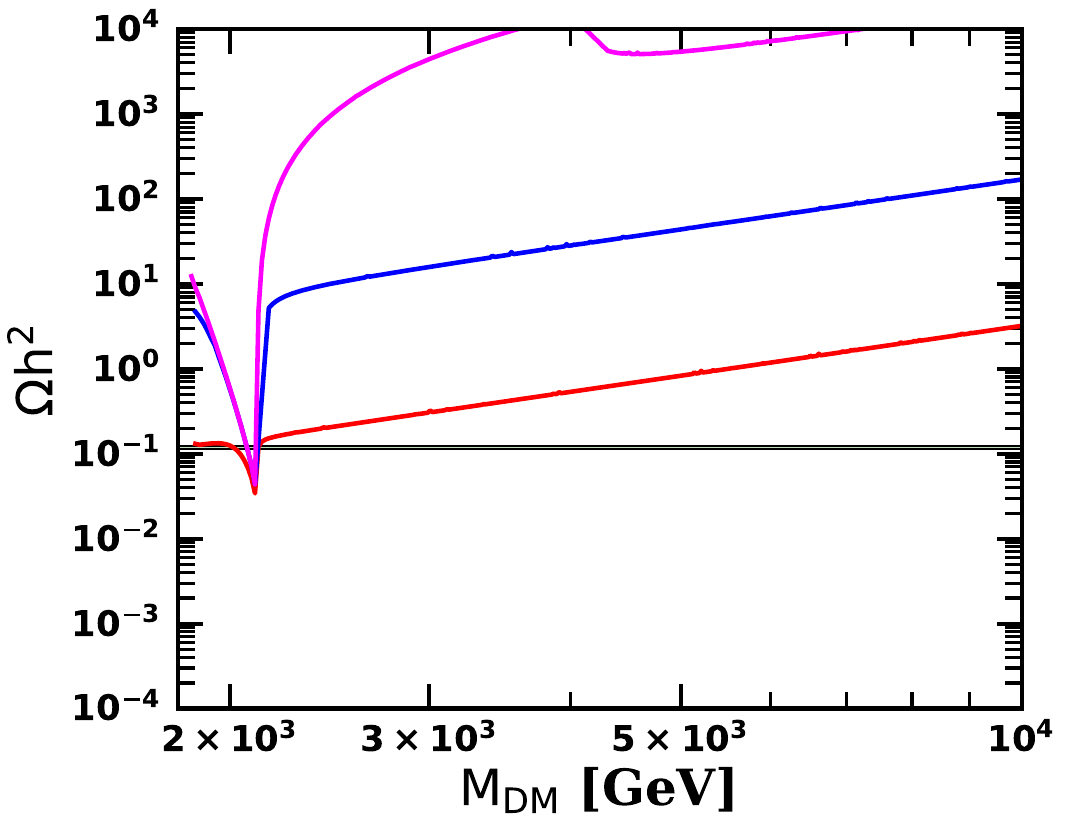}
       \includegraphics[width=0.56\linewidth]{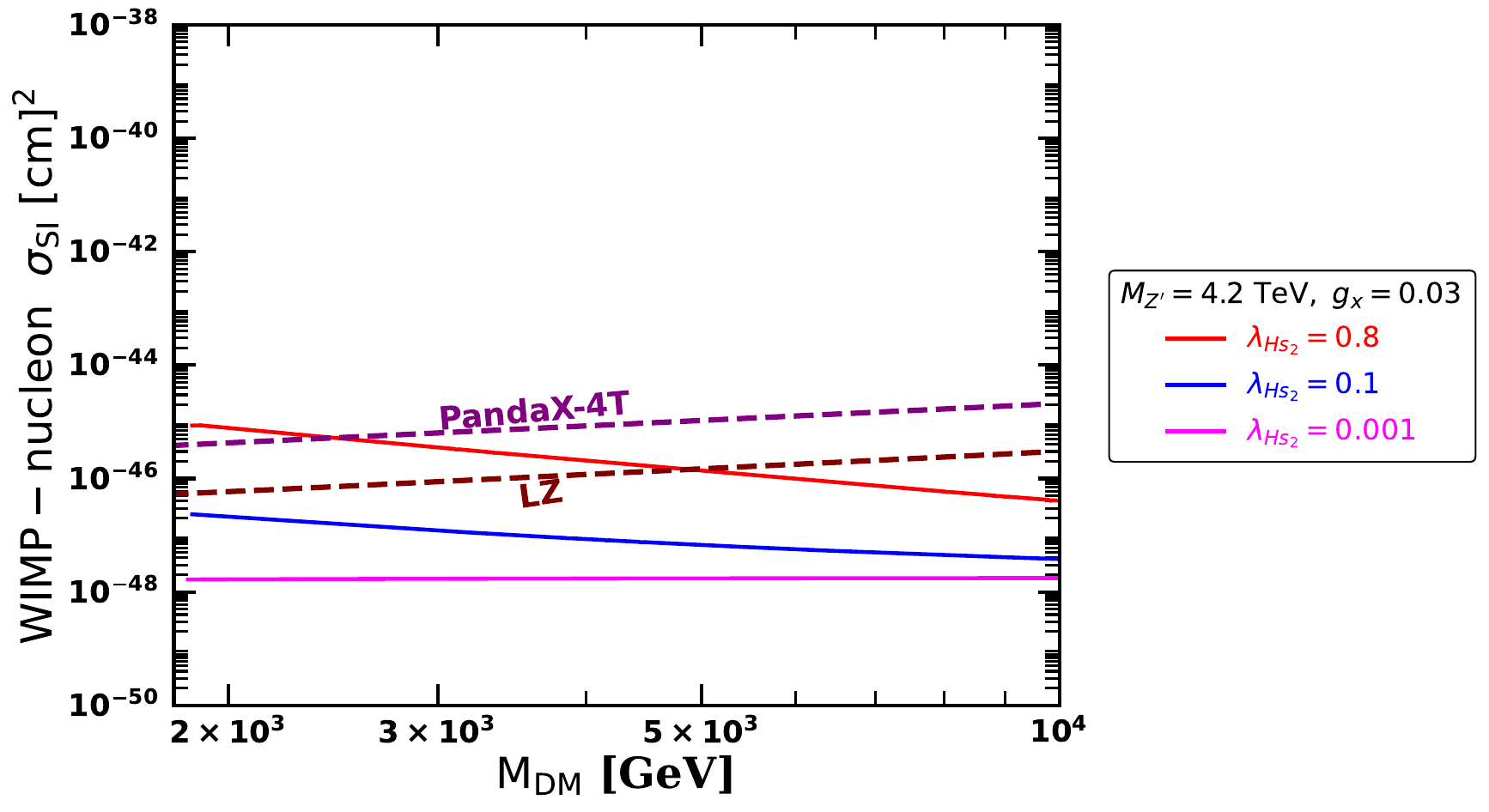}
    \caption{Same as Fig.~\ref{fig:ZP_Scalar_Dom_S1}, but for the $S_2$-dominated singlet DM scenario. See text for details.}
    \label{fig:ZP_Scalar_Dom_S2}
\end{figure}

The effect of the $Z^\prime$-mediated annihilation channels on Fig.~\ref{fig:Scalar_Dom_S2} is shown in Fig.~\ref{fig:ZP_Scalar_Dom_S2}. The benchmark choices are kept identical to those of the $S_1$-dominated scenario in Fig.~\ref{fig:ZP_Scalar_Dom_S1}. Again, the only noticeable difference arises in the $Z^\prime$ resonance dip, which is slightly higher in this case due to the smaller $U(1)_{B-L}$ charge of $S_2$ compared to $S_1$.

%%%%%%%%%%%%%%%%%%
\FloatBarrier

%%%%%%%%%%%%%%%%%%%%
\bibliographystyle{utphys}
\bibliography{references}
\end{document}